\documentclass[12pt,preprint]{aastex}

\usepackage{morefloats}
\usepackage{color}

\shorttitle{Stellar Evolution and the Effects of Variable Composition on Habitable Zones}
\shortauthors{Truitt et al.}

\begin{document}

\title{Expanding the Catalog: Considering the Importance of \\ Carbon, Magnesium, and Neon in the Evolution of Stars \\ and Habitable Zones}

\author{Amanda Truitt\altaffilmark{1} \& Patrick A. Young\altaffilmark{1}}
\altaffiltext{1}{School of Earth and Space Exploration, Arizona State University, Tempe, AZ 85287}

\newcommand{\Msol}{\mbox{$\rm{M_{\odot}\ }$}}
\newcommand{\msol}{\mbox{$\rm{M_{\odot}\ }$}}
\newcommand{\Rsol}{\mbox{$\rm{R_{\odot}\ }$}}
\newcommand{\rsol}{\mbox{$\rm{R_{\odot}\ }$}}
\newcommand{\sol}{\mbox{$\rm{_{\odot}\ }$}}
\newcommand{\Ni}{\mbox{$^{56}$Ni}}
  
\begin{abstract}    

Building on previous work, we have expanded our catalog of evolutionary models for stars with variable composition; here we present models for stars of mass 0.5 - 1.2 \Msol, at scaled metallicities of 0.1 - 1.5 Z\sol, and specific C/Fe, Mg/Fe, and Ne/Fe values of 0.58 - 1.72 C/Fe\sol, 0.54 - 1.84 Mg/Fe\sol and 0.5 - 2.0 Ne/Fe\sol, respectively. We include a spread in abundance values for carbon and magnesium based on observations of their variability in nearby stars; we choose an arbitrary spread in neon abundance values commensurate with the range seen in other low Z elements due to the difficult nature of obtaining precise measurements of neon abundances in stars. As indicated by the results of \citet{truitt15}, it is essential that we understand how differences in individual elemental abundances, and not just the total scaled metallicity, can measurably impact a star's evolutionary lifetime and other physical characteristics. In that work we found that oxygen abundances significantly impacted the stellar evolution; carbon, magnesium, and neon are potentially important elements to individually consider due to their relatively high (but also variable) abundances in stars. We present 528 new stellar main sequence models, and we calculate the time-dependent evolution of the associated habitable zone boundaries for each based on mass, temperature, and luminosity. We also reintroduce the 2 Gyr ``Continuously Habitable Zone" (CHZ$_{2}$) as a useful tool to help gauge the habitability potential for a given planetary system. 

\end{abstract}


\section{INTRODUCTION}

We are working to understand how stars of different mass and composition evolve, and how stellar evolution directly influences the location of the habitable zone (HZ) around a star. Most of the prevailing research on exoplanet habitability focuses on the notion that the HZ is simply the range of distances from a star over which liquid water could exist on the surface of a terrestrial planet (e.g. \citet{kwr93}). Since the radial position of the HZ is determined primarily by the host star's luminosity and spectral characteristics (which also serve as boundary conditions for planetary atmosphere calculations), it is extremely important to understand as much as we can about the broad range of potential exoplanet host stars that exist. Evaluating the potential for liquid water on the surface of a planet requires a deep understanding of the link between stars and the circumstellar environment.

Previous work has been done on stellar evolution with considerations of variations in abundances, including the elements we are interested in here (e.g. \citet{dot07,vberg12,beom16}). However, our work is an important consideration for several reasons. First of all, the referenced groups use solar abundance mixtures from \citet{gs98}, whereas we use abundance values from \citet{lod10}. We use a different stellar evolution code as well, though we have implemented OPAL opacities as has the previous work. \citet{dot07} uses artificially enhanced abundances for elements relative to solar abundance value, whereas our ranges of variation in abundance values stem from real observations (except for neon), discussed in \S 2. Nonetheless, our results are relatively consistent with those presented, specifically that enhancing elements like carbon (at constant solar metallicity value) can actually cause the mean opacity to decrease. We do see a turnover in main sequence (MS) lifetimes for those models that we have simultaneously enriched in both metallicity and individual carbon abundance, and we confirm that enhanced carbon decreases the MS lifetime at enhanced abundance values. The work of \citet{vberg12} and \citet{beom16} are less directly relevant, in that they discuss abundance values taken from low metallicity stars in globular clusters; we use abundance ranges from nearby FGK field (disk) stars, so there is not a great deal of overlap. The premise that elemental variation exists in stars, however, is an important one to consider.

We reiterate the pressing need to thoroughly represent the large variation that exists for potential exoplanet host stars, based both on the specific chemical composition as well as the individual detailed evolutionary history (addressed in our previous paper, \citet{truitt15}; hereafter T15). Though other groups have done excellent work on the evolution of HZs as a function of a star's overall scaled metallicity \citep{valle14,oishi16}, we argue that it is important to consider the specific elemental abundance ratios of stars if we want to make any comprehensive assessments about the habitability potential of a particular system. The work by \citet{valle14} in particular is extremely relevant to this work, in that they have evaluated several characteristics for considering the general habitability of a particular stellar system, and also consider the boundaries for a 4 Gyr continuously habitable zone. This is similar to the discussion we introduced in T15, though we use a 2 Gyr continuously habitable zone instead (see \S 3.3). The results presented in their paper are robust, and our data confirm the relationships between mass, metallicity, and HZ distances that they report. Specifically, our results are consistent with a decreasing HZ distance with increasing metal content and/or decreasing mass of the host star; we also confirm that low mass and/or higher metal stars spend a larger amount of time on the MS and could potentially offer a higher degree of habitability potential. The major important distinction we make with our work is not that we use a different stellar evolution code and HZ prescriptions, but that we consider how variations in specific elemental abundance values (and not only scaled metallicity relative to solar value) significantly impact stellar evolution.

In the current environment, with the almost-constant discovery (e.g. \citet{zieg16}) and statistical confirmation (e.g. \citet{morton16}) of new exoplanets, it is imperative that we as a scientific community have an efficient and consistent way to narrow down the search for potentially habitable exoplanets. If we can define boundary conditions based on certain stellar physical parameters, we will be better equipped to assess whether a planet discovered in a star's HZ is actually a worthwhile candidate to perform follow-up observations for characterization, utilizing the kind of missions recommended in the most recent Decadal Review of Astronomy and Astrophysics: transmission spectroscopy with James Webb Space Telescope (e.g. \citet{bi16}), or direct detection with a coronagraph, interferometer, or starshade (e.g. \citet{turn12}).

Following T15, here we expand our investigation into the effects of variations to the elemental abundance ratios in stars. Specifically, we consider carbon and magnesium, since they are important players in the overall stellar evolution (e.g. \citet{ser16}). We also discuss the contributions of neon (and briefly, nitrogen); however, we don't know the extent of variability in these two elements in real stars due to the lack of observational abundance determinations. The discussion of Ne and N is based on speculation that these elements could potentially vary by a factor of two relative to solar abundances (i.e. 0.5 Ne/Fe\sol would be the depleted value, while 2.0 Ne/Fe\sol is enriched), a similar scale to other elements nearby on the periodic table. Neon is more important than nitrogen to the evolution in terms of providing opacity, the main effect of different elemental abundances. We have made our entire catalog of stellar evolution tracks available as an online database\footnote{http://bahamut.sese.asu.edu/$\sim$payoung/AST\_522/Evolutionary\_Tracks\_Database.html}, with an included interactive interpolation tool; it is designed for use by the astrobiology and exoplanet communities to characterize the evolution of stars and HZs for any real planetary candidates of interest. In this paper, we describe our choice of parameter space and the stellar evolution code in \S 2, our interpretation of the results of the models in \S 3, and our conclusions in \S 4.

\section{METHODS}
\subsection{Parameter Space \label{sec.ps}}

Here we present an extended grid of stellar models suitable for the prediction of HZ locations. In T15 we discussed the importance of mass, metallicity, and oxygen abundance to the stellar evolution. In this paper, we focus on the variation observed in carbon and magnesium abundances, which also produce a measurable effect in the stellar evolution (albeit smaller than the effect observed for variations in oxygen) and which also exhibit substantially variable abundance ratios in neighboring stars \citep{neves09,mish08,take07,young14,pagano15}. We also include discussion on the practicality of considering neon's contribution to stellar evolution, though the range of abundance values we quote are not based directly on observational data. In this work, ratios without brackets (e.g. C/Fe) indicate the linear absolute abundance ratio in terms of mass fraction, while a bracketed ratio denotes the log of the atom number relative to the solar abundance value for that same element. The latter is the conventional [C/Fe] given by
\begin{equation}
[C/Fe] = log_{10} \frac{(C/Fe)}{(C/Fe)\sol}
\end{equation}
We primarily quote linear ratios relative to solar (i.e. C/Fe = 1.72 C/Fe\sol) since the range of abundance ratios is small enough to not require logarithmic notation. We use mass fraction as this is the conventional usage for stellar evolution calculations.

We again consider the major contributors to stellar evolution: mass, metallicity (Z), and individual specific elemental abundances. Variations in Z alone are made with a fixed abundance pattern that is uniformly scaled, while the spread in carbon and magnesium values we use reflects the actual observed variations in abundance ratios in nearby stars \citep{ram07,bond06,bond08,gonz10,hink14,young14}. One exception is that the range in neon values we use does not result from observed neon abundances in stars; rather, we vary neon relative to solar to create a range of values that we might reasonably expect to see in stars if neon could be measured more accurately. Changes in C/Fe\sol, Mg/Fe\sol, and Ne/Fe\sol at each metallicity are made by changing the absolute abundance of each element while holding all other metal abundances constant. The relative abundances of hydrogen and helium are adjusted in compensation to ensure the sum of mass fractions = 1.

Beyond the original grid for oxygen (discussed in T15) that comprised a total of 376 models, we now introduce an additional 240 models for each carbon and magnesium. Also, for the purposes of this work, we've produced a smaller grid of 48 models for neon that includes only end-member cases of interest, resulting in a total addition of 528 new models. The grids for C, Mg, and Ne still encompass stars of mass 0.5 - 1.2 M\sol at each 0.1 M\sol (which includes spectral types from approximately M0 - F0 at solar metallicity), overall scaled metallicity values of 0.1 - 1.5 Z\sol at each 0.1 Z\sol, and now abundance values of C, Mg, and Ne ranging from 0.58 - 1.72 C/Fe\sol, 0.54 - 1.84 Mg/Fe\sol, and 0.5 - 2.0 Ne/Fe\sol.

\subsection{TYCHO}

The models included in our catalog were simulated using the stellar evolution code TYCHO \citep{yoar05}. As detailed in T15, TYCHO outputs information on stellar surface quantities for each time-step of a star's evolution, which we then use to calculate the inner and outer radii of the HZ as a function of the star's age. New OPAL opacity tables \citep{ir96,rn02} were generated at the specific abundance values needed for each enriched and depleted C/Fe, Mg/Fe, and Ne/Fe value to match the desired composition of the stellar model. The TYCHO evolutionary tracks are used as input to our HZ calculator (CHAD) which is easily upgradable to incorporate improved HZ predictions as they become available.

We have recently implemented improved low temperature ($\sim$2400 K) opacity tables in TYCHO, and we are now able to more accurately simulate evolutionary tracks, particularly for very low mass stars. The new low temperature opacities are based on \citet{f05,sbfa09} and include dust grain opacity. Ultimately, it will be extremely important to include M-stars in our catalog due to the high probability that they may host a habitable world \citep{borucki10,borucki11,bat13}. We will explore the ramifications of variable stellar composition in a grid of M-stars in a future paper. We have recalculated the original oxygen grid that was discussed in T15; we provide updated oxygen values alongside data for carbon, magnesium, and neon for certain parameters of interest.

\section{RESULTS}

As we examined at length in T15, the main factors that influence the time evolution of the classical HZ are the host star's luminosity ($L$) and effective temperature ($T_{eff}$), their rates of change, and the stellar MS lifetime. TYCHO evolutionary tracks are used to estimate the extent of the HZ at each point in the stellar evolution. For these estimates we follow the prescriptions of \citet{kopp13,kopp14}, which proceed from \citet{sel07} and \citet{kwr93}. These prescriptions parameterize the orbital radii of the HZ as a function of $L$ and $T_{eff}$, which facilitates the translation from stellar evolution tracks to HZ distance estimations. We reconfirm that mass and scaled metallicity influence these factors considerably. Following from \citet{ylp12} and T15, wherein the focus was variability in the oxygen abundance (ranging from 0.44 to 2.28 O/Fe\sol), we now examine the outcome of varying the the abundance ratios of C/Fe\sol, Mg/Fe\sol, and Ne/Fe\sol; these are other elements that are relatively significant to the stellar evolution over the entire range of mass and metallicity represented in our grid.

\subsection{Stellar Properties and Main Sequence Lifetimes}

Table~\ref{tab1} shows the MS lifetimes (in Gyr) for standard and end member abundance values for all elements of interest (carbon, updated oxygen values, magnesium, and neon), as well as end member metallicity values, for all masses in our grid. When considering how a star's specific chemical composition translates to its MS lifetime, we would expect that a star with higher metallicity (or enriched elemental abundances) would live longer than a star of the same mass with lower overall opacity. Surprisingly, this is not what we see for some of the carbon models in our grid. Upon close inspection of the listed table values (particularly for the 1.5 Z\sol cases), an unexpected trend emerges; specifically, it appears that some of the depleted carbon cases (0.58 C/Fe\sol) actually have $longer$ MS lifetimes than the associated enriched carbon cases (1.72 C/Fe\sol). With further examination of the lifetimes given for the other elements, it is clear that the MS lifetimes do not exhibit the same inverted lifetime expectancies for these models as they do for some of the carbon cases.

In order to understand the puzzling behavior of the carbon models, we have examined two possibilities. First, since discrepancies in the expected stellar ages are sufficiently small compared to the overall calculated MS lifetimes, numerical uncertainties in the code that determine where TYCHO terminates the MS may be larger than the variability that we actually measure for the MS lifetimes. TYCHO determines the Terminal Age Main Sequence (TAMS) by stopping the code when the abundance of hydrogen in the innermost model zone drops below 1 part in 10$^{6}$. Rezoning in TYCHO is adaptive, so minor differences in the size of the innermost zones and diffusion/convection across those zones can cause small (i.e. $<$ 1\%) variation in the output value of the TAMS. Second, because of compositional normalization that is applied when creating opacity tables, the depleted carbon (and magnesium) models start out with slightly more hydrogen to ensure that the total mass fraction = 1, which may allow for ``extra'' MS lifetime if that hydrogen becomes available for core burning. For the highest mass stars in our sample that develop convective cores, the extent of the convective core changes slightly due to the change in electron fraction (the convective core is high enough in temperature to be dominated by electron scattering opacity) and the energy generation by the CNO cycle with a different amount of catalysts. Additional carbon also shifts the position of the second peak in the opacity vs. temperature relationship in the OPAL tables, impacting the location of the convection zone base. Each of these are very small effects; it turns out that the variations in lifetime from carbon are also relatively small. Given that the effect is seen preferentially at higher metallicity and higher mass, the dominant effects are a combination of slightly increased hydrogen mass fraction and central zoning, which enhance convective transport and the CNO catalysts that play a role in the more massive stars.

To further interpret how differences in the stellar models are caused by variations in the specific elemental abundance values, notably carbon, it is also important that we distinguish between the effects from the stellar opacity and the effects from nuclear reactions. Our models (at solar metallicity) encompass the mass range wherein the transition from pp-chain to CNO-dominated hydrogen burning occurs ($\leq$ 1.1 M\sol). The energy generated by the CNO cycle depends on the mass fractions of hydrogen and total CNO catalysts $X_{p}$*$X_{CNO}$. For models of the same mass but at much lower opacity (i.e. Z = 0.1 Zsol) and relatively higher temperatures, it might seem that the transition between pp-chain and CNO burning should therefore shift to a lower mass value. However, since CNO burning is extremely temperature sensitive and the transition is dependent on the amount of CNO catalysts ($X_{p}$/$X_{CNO}$)$^{1/12.1}$ \citep{arn05}, the lower metallicity stars in our sample don't actually transition to CNO burning at a significantly lower mass. The same consideration can be made for changes in the opacity due to enhanced or depleted specific elemental abundances. Changes to carbon in particular will alter the amount of energy generated via the CNO cycle, so the effect is more pronounced in the highest mass stars in our sample. Overall, however, the contribution from carbon is relatively insignificant.

\begin{figure}
\centering
\includegraphics[width=0.4\textwidth]{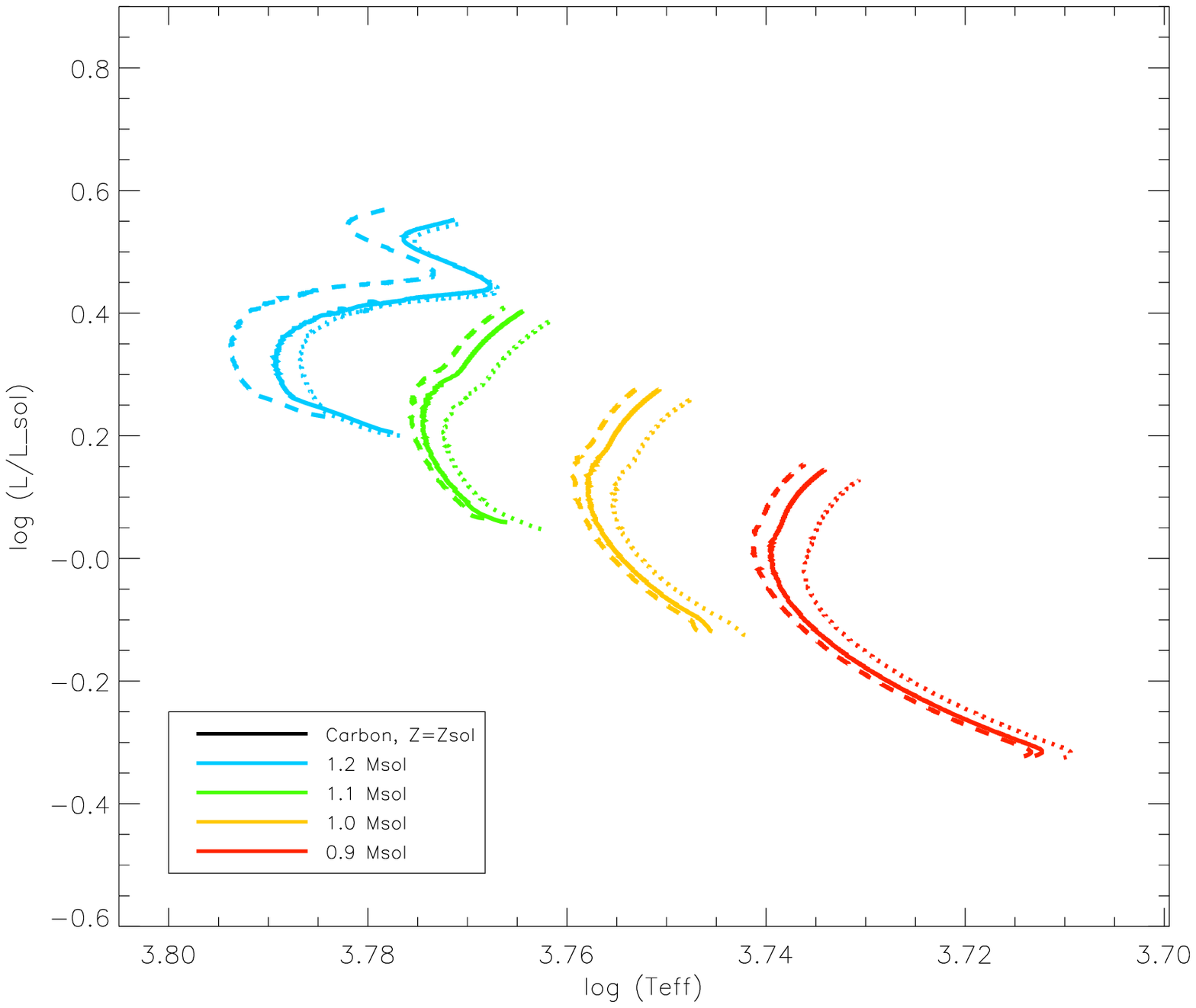}
\includegraphics[width=0.4\textwidth]{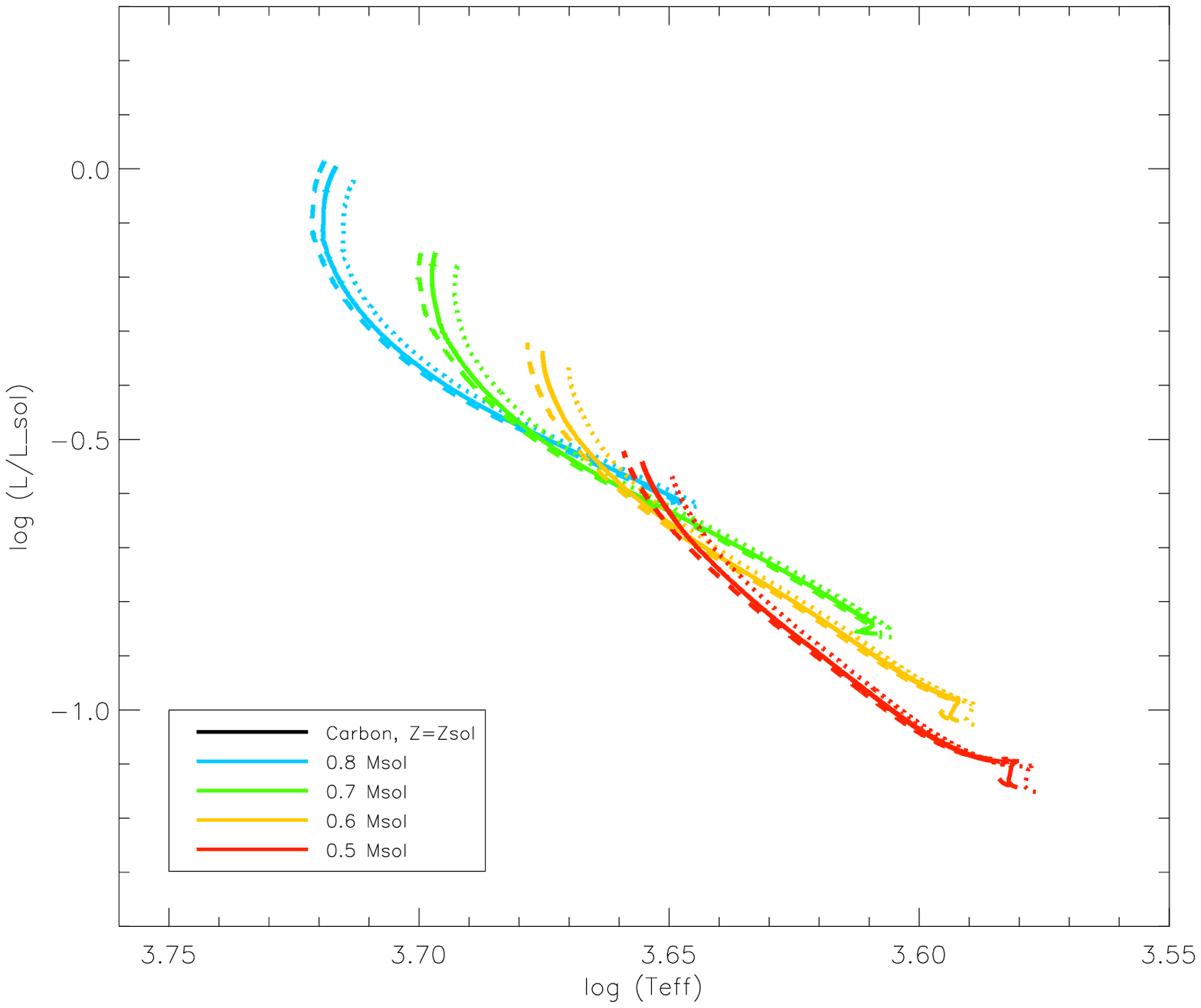}
\includegraphics[width=0.4\textwidth]{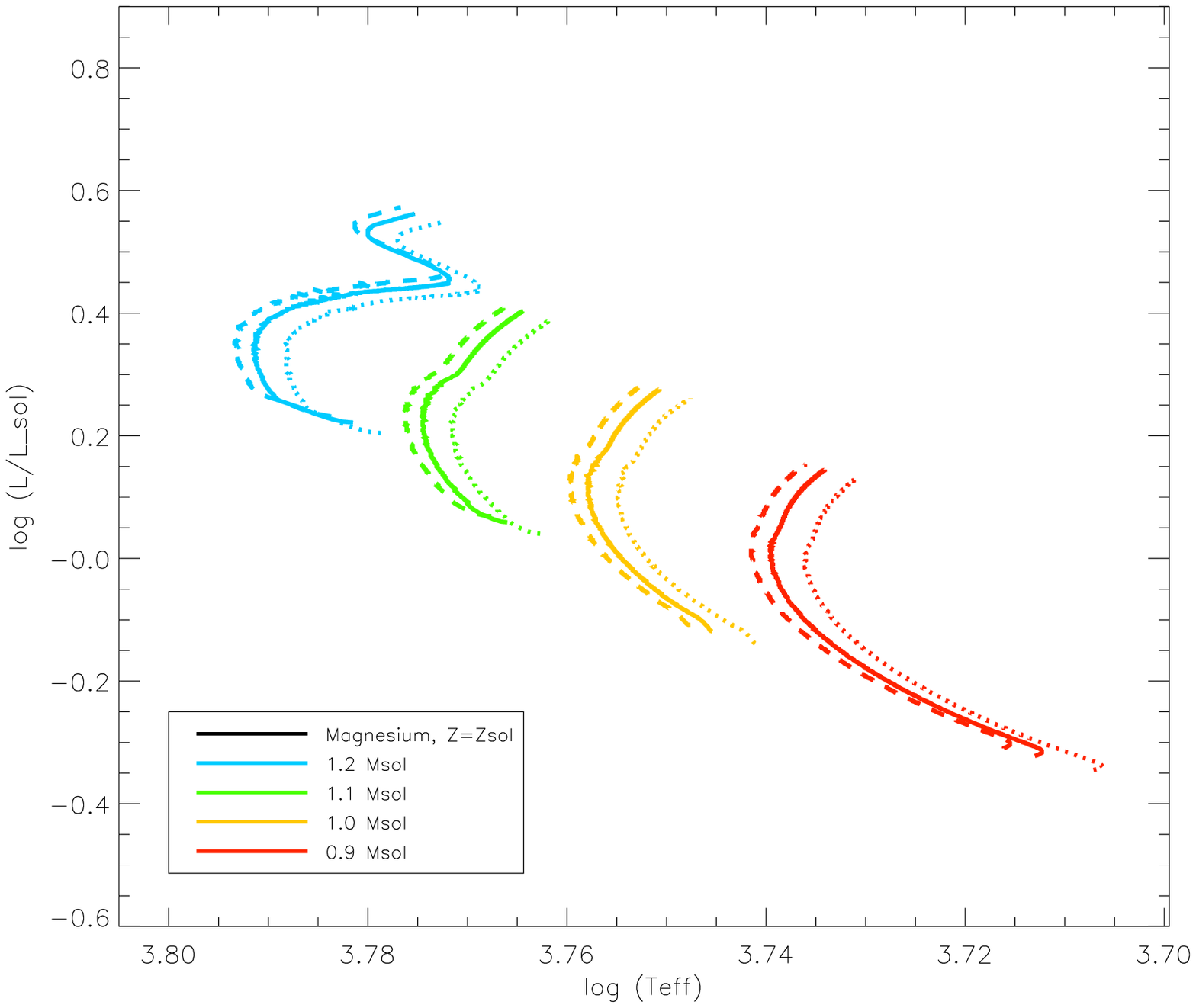}
\includegraphics[width=0.4\textwidth]{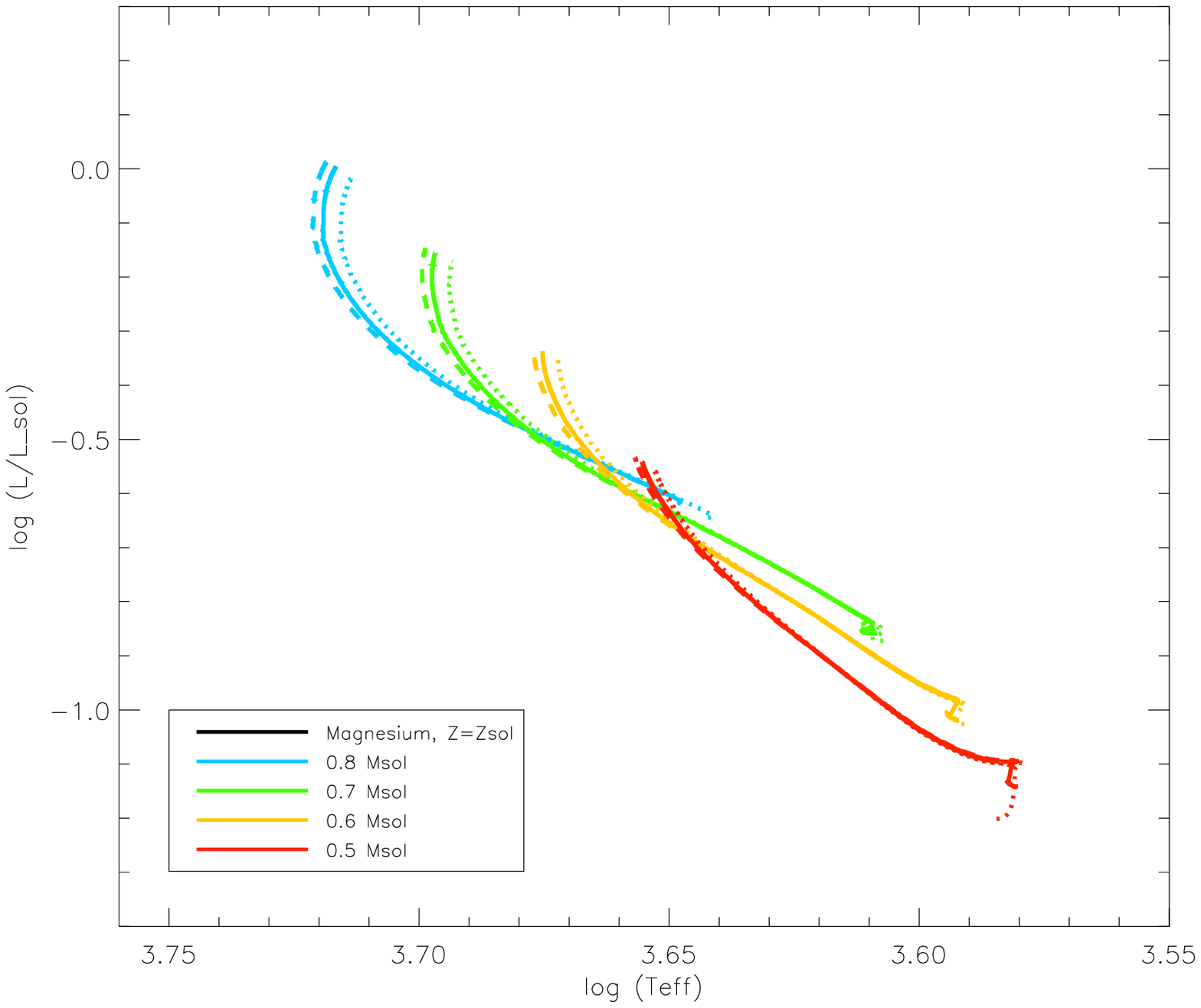}
\includegraphics[width=0.4\textwidth]{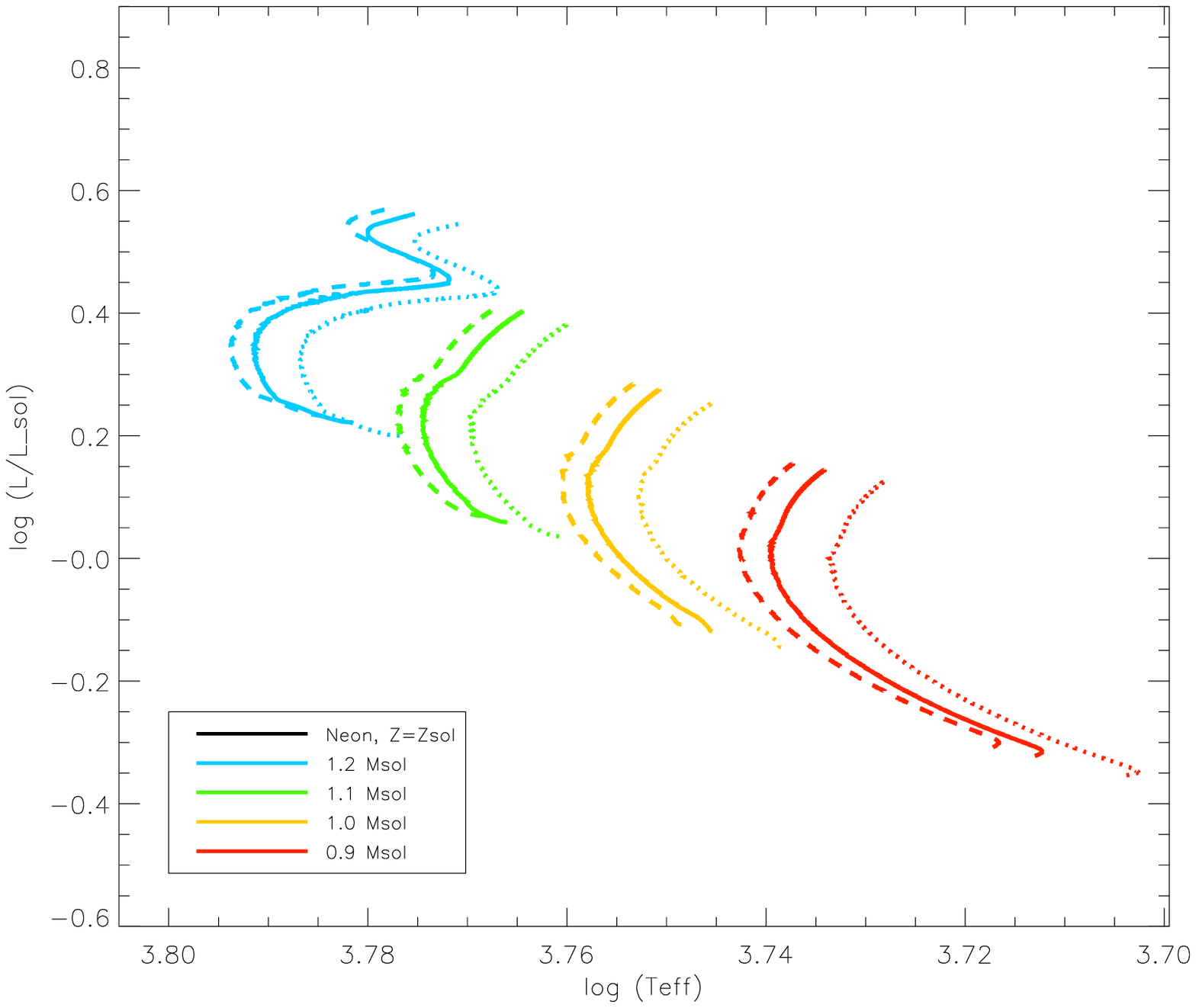}
\includegraphics[width=0.4\textwidth]{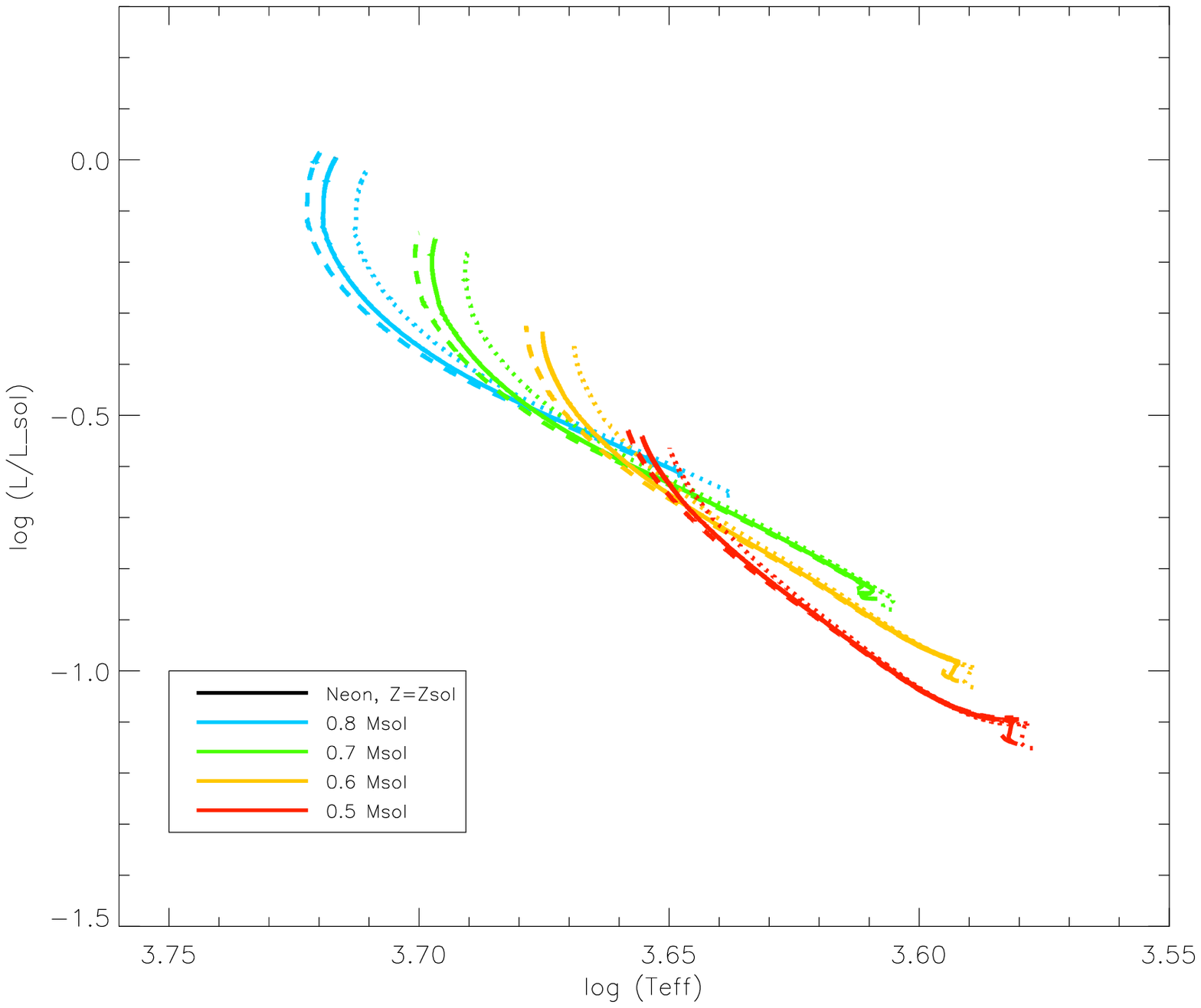}
\caption{HRD, Evolutionary tracks from ZAMS to TAMS for all masses. The left column shows masses 0.9 - 1.2 M\sol, while the right column is 0.5 - 0.8 M\sol. Each row represents a different elemental composition, where the depleted abundances are dashed, the solar abundance value is solid, and the enriched abundances are dotted. The top row is for carbon (0.58 - 1.72 C/Fe\sol), the middle row is for magnesium (0.54 - 1.84 Mg/Fe\sol), and the bottom row is for neon (0.5 - 2.0 Ne/Fe\sol). All abundance values are at Z = Z\sol. The rightward-most dotted line in each row is for the 0.5 M\sol star with the enriched abundance value, while the leftward-most dashed line is for the 1.2 M\sol star with the depleted abundance value.\label{hrd}}
\end{figure}

Figure~\ref{hrd} shows the Hertzsprung-Russell Diagrams (HRDs) for evolutionary tracks from ZAMS (Zero Age Main Sequence) to TAMS for all masses in our grid. The top row is for carbon, where C/Fe = 0.58 C/Fe\sol (dashed), 1.0 C/Fe\sol (solid), and 1.72 C/Fe\sol (dotted), all at Z = Z\sol. The middle and bottom rows (respectively) show the similar HRDs for magnesium and neon, where Mg/Fe = 0.54 Mg/Fe\sol and Ne/Fe = 0.5 Ne/Fe\sol (dashed), 1.0 Mg/Fe\sol and 1.0 Ne/Fe\sol (solid), and 1.84 Mg/Fe\sol and 2.0 Ne/Fe\sol (dotted), again at Z = Z\sol. The rightward-most dotted lines are for the lowest mass star with enriched elemental values, while the leftward-most dashed line is for the highest mass star with depleted values.

For the higher mass models (the left-hand column of Figure~\ref{hrd}) we see evidence of the Kelvin-Helmholtz mechanism (KH ``jag"), wherein a star nearing the end of its MS lifetime begins to cool and compress due to decreased internal pressure from the end of core hydrogen burning. This compression reheats the core, causing the observed fluctuations in $L$ and $T_{eff}$. A detailed scrutiny of the figures reveals a slight crossover that occurs in the late MS for both carbon and magnesium, for the depleted (dashed line) cases relative to standard (solid line) cases. The crossover occurs due to the slightly larger core in the high C models; thus, the shift in the KH-jags for these models on the HR diagram is physical, from variability that exists in the interior structures of the stars. The total abundance of carbon in stars is, generally, a factor of several higher than for that of magnesium (e.g. \citet{lod10}); however, the abundance range of carbon (from 0.58 - 1.72 C/Fe\sol) is smaller than that of magnesium (from 0.54 - 1.84 Mg/Fe\sol), and magnesium contributes more opacity per gram in the stellar interior than carbon does (e.g. \citet{morse40}). Thus, magnesium actually makes a bigger difference to the evolution relative to its abundance in stars. Oxygen is not only much more abundant than carbon, but also has a high contribution to the opacity.

Table~\ref{tab2} shows $\Delta$($L$/$L_{ZAMS}$) at each mass and end-member composition for all elements. As expected, the change in luminosity over the MS is largest for less enriched compositions except in the case of the higher metallicity, higher mass stars, where the shape of the K-H jag obscures the trend. Table~\ref{tab3} similarly shows $\Delta$$T_{eff}$ at each mass and end-member composition for all elements. The lowest mass, lowest opacity models all exhibit the largest change in temperature over the course of their MS lifetimes, even though they don't live quite as long as higher opacity stars at the same mass. Interestingly, even though we see the highest $\Delta$T values for depleted magnesium (0.54 Mg/Fe\sol at 0.1 Z\sol), the largest change in $L$ actually occurs for the depleted oxygen model (0.44 O/Fe\sol, though also at 0.1 Z\sol).

This work constitutes a sound argument for considering the contributions of neon (and, to some extent, nitrogen) to the stellar evolution. Neon would definitely be an important player in the evolution based on its opacity contributions per unit mass (similar to that of magnesium). It is difficult to assign the appropriate abundance ratio ranges for modeling, as it is challenging to measure neon in stars with much certainty, although work has been done to measure neon abundances from the X-ray spectra of cool stars \citep{dt05}. For the purposes of this work, we have assigned an artificial range of neon abundances (enriched and depleted by factor of two from the solar neon abundance, similar to the range of other low Z elements) which we can use to estimate contributions to the stellar evolution. Nitrogen is also not easily measurable in stars, but can probably be safely neglected; it is similar in opacity per gram to carbon, but relatively less abundant in stars, by a factor of about 4 in the Sun \citep{hkt04}. Thus, its contribution to the stellar evolution is likely negligible even though it is more abundant than either magnesium or neon. One exception to this would be if nitrogen is actually observed to be widely variable in stars with future measurements; if the abundance values vary a great deal more between individual stars than other elements, it could be an important consideration.

Now consider the rate of change of the luminosity (Table~\ref{tab4}) for all masses in our grid at end-member compositions. It is especially useful to look at the change of luminosity per Gyr, because some of the models undergo a larger change in $L$ than do the higher opacity models at the same mass, but potentially over longer or shorter MS lifetimes. This could have different implications for whether the change in luminosity with time is greater or smaller for low opacity models (or if it varies), and whether that occurs during the second half of the star's MS lifetime. With few exceptions, the low opacity models at each mass and elemental composition change more in $L$ per Gyr than their counterparts at higher opacities. Additionally, and as expected, it's clear that the higher mass models experience a significantly larger change in $L$ over the course of their MS lifetimes.

As we understand how the luminosity changes over time (the rate of change, as well as the total change), we see that the range of orbits in the HZ at different points in the MS evolution can vary substantially. Table~\ref{tab5} shows the fraction (listed as percentages) of orbital radii that only enter the habitable zone $after$ the midpoint of the MS for each star. The results indicate that up to a half of all orbits that are in the HZ only become habitable in the second half of the host star's MS lifetime. The effect is more pronounced at higher mass and enriched composition, at each element of interest. When considering the potential for detectability, it is wise to avoid planets that have only recently entered the HZ of the host star; not only would we potentially circumvent the issue of cold starts (discussed in \S 3.3), but we also assume that life requires enough time spent in ``habitable'' conditions before it would yield detectable biosignatures. This is a somewhat narrow assumption that depends on specific habitability considerations; indeed, \citet{silva16} introduces an alternative ``atmospheric mass habitable zone for complex life" with an inner edge that is not affected by the uncertainties inherent to the calculation of the runaway greenhouse limit.

\begin{deluxetable}{lcccccccc}
\tabletypesize{\scriptsize}
\tablecaption{MS lifetimes (Gyr) for each mass and end-member composition for all elements.\label{tab1}}
\tablehead{
  \colhead{Composition}
& \colhead{0.5 M\sol}
& \colhead{0.6 M\sol}
& \colhead{0.7 M\sol}
& \colhead{0.8 M\sol}
& \colhead{0.9 M\sol}
& \colhead{1.0 M\sol}
& \colhead{1.1 M\sol}
& \colhead{1.2 M\sol}
}
\startdata
0.1 $Z\sol$, 0.58 C/Fe\sol  &  69.067  &  40.446  &  22.575  &  13.475  &  8.618  &  5.825  &  4.099  &  2.960 \\
0.1 $Z\sol$, C/Fe\sol  &  70.579  &  41.228  &  23.089  &  13.741  &  8.778  &  5.914  &  4.154  &  2.996 \\
0.1 $Z\sol$, 1.72 C/Fe\sol  &  73.251  &  43.010  &  24.037  &  14.245  &  9.072  &  6.094  &  4.274  &  3.073 \\
$Z\sol$, 0.58 C/Fe\sol  &  109.038  &  72.578  &  45.839  &  26.770  &  16.185  &  10.408  &  6.916  &  5.060 \\
$Z\sol$, C/Fe\sol  &  109.490  &  72.762  &  45.853  &  26.691  &  16.123  &  10.363  &  6.902  &  5.029 \\
$Z\sol$, 1.72 C/Fe\sol  &  110.992  &  73.605  &  46.274  &  26.849  &  16.141  &  10.364  &  6.903  &  5.091 \\
1.5 $Z\sol$, 0.58 C/Fe\sol  &  114.774  &  77.401  &  50.089  &  30.009  &  18.158  &  11.651  &  7.716  &  5.634 \\
1.5 $Z\sol$, C/Fe\sol  &  114.328  &  76.838  &  49.495  &  29.513  &  17.831  &  11.459  &  7.608  &  5.578 \\
1.5 $Z\sol$, 1.72 C/Fe\sol  &  113.697  &  75.978  &  48.530  &  28.672  &  17.291  &  11.107  &  7.391  &  5.571 \\\\
0.1 $Z\sol$, 0.44 O/Fe\sol  &  65.047  &  37.352  &  20.954  &  12.586  &  8.095  &  5.496  &  3.878  &  2.823 \\
0.1 $Z\sol$, O/Fe\sol  &  70.579  &  41.228  &  23.089  &  13.741  &  8.778  &  5.914  &  4.154  &  2.996 \\
0.1 $Z\sol$, 2.28 O/Fe\sol  &  82.565  &  50.281  &  28.579  &  16.742  &  10.492  &  6.974  &  4.848  &  3.481 \\
$Z\sol$, 0.44 O/Fe\sol  & 102.238  &  66.968  &  41.196  &  23.657  &  14.494  &  9.397  &  6.322  &  4.580 \\
$Z\sol$, O/Fe\sol  &  109.490  &  72.762  &  45.853  &  26.691  &  16.123  &  10.363  &  6.902  &  5.029 \\
$Z\sol$, 2.28 O/Fe\sol  &  120.713  &  82.074  &  52.680  &  31.429  &  18.586  &  11.789  &  7.776  &  5.826 \\
1.5 $Z\sol$, 0.44 O/Fe\sol  &  109.250  &  72.825  &  46.198  &  27.129  &  16.495  &  10.687  &  7.162  &  5.214 \\
1.5 $Z\sol$, O/Fe\sol  &  114.328  &  76.838  &  49.495  &  29.513  &  17.831  &  11.459  &  7.608  &  5.578 \\
1.5 $Z\sol$, 2.28 O/Fe\sol  &  118.072  &  79.284  &  51.197  &  30.807  &  18.400  &  11.671  &  8.197  &  5.984 \\\\
0.1 $Z\sol$, 0.54 Mg/Fe\sol  &  69.918  &  40.721  &  22.791  &  13.584  &  8.678  &  5.855  &  4.121  &  2.973 \\
0.1 $Z\sol$, Mg/Fe\sol  &  70.579  &  41.228  &  23.089  &  13.741  &  8.778  &  5.914  &  4.154  &  2.996 \\
0.1 $Z\sol$, 1.84 Mg/Fe\sol  &  71.798  &  42.171  &  23.638  &  14.040  &  8.957  &  6.029  &  4.235  &  3.052 \\
$Z\sol$, 0.54 Mg/Fe\sol  & 107.956  &  71.461  &  44.730  &  25.910  &  15.691  &  10.105  &  6.728  &  4.939 \\
$Z\sol$, Mg/Fe\sol  &  109.490  &  72.762  &  45.853  &  26.691  &  16.123  &  10.363  &  6.902  &  5.029 \\
$Z\sol$, 1.84 Mg/Fe\sol  &  112.245  &  75.150  &  47.931  &  28.215  &  16.934  &  10.864  &  7.221  &  5.265 \\
1.5 $Z\sol$, 0.54 Mg/Fe\sol  &  112.890  &  75.632  &  48.427  &  28.702  &  17.382  &  11.179  &  7.427  &  5.490 \\
1.5 $Z\sol$, Mg/Fe\sol  &  114.328  &  76.838  &  49.495  &  29.513  &  17.831  &  11.459  &  7.608  &  5.578 \\
1.5 $Z\sol$, 1.84 Mg/Fe\sol  &  116.918  &  78.997  &  51.409  &  31.181  &  18.702  &  11.988  &  7.948  &  5.789 \\\\
0.1 $Z\sol$, 0.5 Ne/Fe\sol  &  69.107  &  40.151  &  22.476  &  13.407  &  8.577  &  5.792  &  4.078  &  2.940 \\
0.1 $Z\sol$, Ne/Fe\sol  &  70.579  &  41.228  &  23.089  &  13.741  &  8.778  &  5.914  &  4.154  &  2.996 \\
0.1 $Z\sol$, 2.0 Ne/Fe\sol  &  73.572  &  43.452  &  24.364  &  14.432  &  9.185  &  6.171  &  4.329  &  3.123 \\
$Z\sol$, 0.5 Ne/Fe\sol  &  107.476  &  71.110  &  44.497  &  25.802  &  15.670  &  10.111  &  6.743  &  4.953 \\
$Z\sol$, Ne/Fe\sol  &  109.490  &  72.762  &  45.853  &  26.691  &  16.123  &  10.363  &  6.902  &  5.029 \\
$Z\sol$, 2.0 Ne/Fe\sol  &  114.068  &  76.425  &  48.860  &  28.793  &  17.194  &  10.990  &  7.289  &  5.314 \\
1.5 $Z\sol$, 0.5 Ne/Fe\sol  &  112.578  &  75.448  &  48.339  &  28.694  &  17.391  &  11.202  &  7.453  &  5.522 \\
1.5 $Z\sol$, Ne/Fe\sol  &  114.328  &  76.838  &  49.495  &  29.513  &  17.831  &  11.459  &  7.608  &  5.578 \\
1.5 $Z\sol$, 2.0 Ne/Fe\sol  &  117.587  &  79.362  &  51.609  &  31.251  &  18.688  &  11.945  &  7.904  &  5.751 \\
\enddata
\end{deluxetable}

\begin{deluxetable}{lllllllll}
\tabletypesize{\scriptsize}
\tablecaption{$\Delta$($L$/$L_{ZAMS}$) for each mass and end-member composition for all elements.\label{tab2}}
\tablehead{
  \colhead{Composition}
& \colhead{0.5 M\sol}
& \colhead{0.6 M\sol}
& \colhead{0.7 M\sol}
& \colhead{0.8 M\sol}
& \colhead{0.9 M\sol}
& \colhead{1.0 M\sol}
& \colhead{1.1 M\sol}
& \colhead{1.2 M\sol}
}
\startdata
0.1 $Z\sol$, 0.58 C/Fe\sol  &  5.717  &  6.004  &  4.363  &  3.013  &  2.388  &  1.955  &  1.618  &  1.373 \\
0.1 $Z\sol$, C/Fe\sol  &  5.679  &  5.815  &  4.368  &  2.981  &  2.357  &  1.918  &  1.584  &  1.341 \\
0.1 $Z\sol$, 1.72 C/Fe\sol  &  5.655  &  5.813  &  4.397  &  2.944  &  2.312  &  1.883  &  1.549  &  1.302 \\
$Z\sol$, 0.58 C/Fe\sol  & 3.137  &  3.956  &  4.175  &  3.278  &  1.993  &  1.516  &  1.207  &  1.174 \\
$Z\sol$, C/Fe\sol  &  2.995  &  3.823  &  4.056  &  3.185  &  1.929  &  1.483  &  1.209  &  1.190 \\
$Z\sol$, 1.72 C/Fe\sol  &  2.821  &  3.638  &  3.882  &  3.062  &  1.840  &  1.433  &  1.200  &  1.224 \\
1.5 $Z\sol$, 0.58 C/Fe\sol  &  2.623  &  3.400  &  3.739  &  3.141  &  1.938  &  1.462  &  1.170  &  1.155 \\
1.5 $Z\sol$, C/Fe\sol  &  2.495  &  3.264  &  3.591  &  2.994  &  1.844  &  1.419  &  1.174  &  1.192 \\
1.5 $Z\sol$, 1.72 C/Fe\sol  &  2.398  &  3.098  &  3.394  &  2.777  &  1.702  &  1.350  &  1.157  &  1.241 \\\\
0.1 $Z\sol$, 0.44 O/Fe\sol  &  5.928  &  5.857  &  4.335  &  3.061  &  2.438  &  2.004  &  1.681  &  1.420 \\
0.1 $Z\sol$, O/Fe\sol  &  5.679  &  5.815  &  4.368  &  2.981  &  2.357  &  1.918  &  1.584  &  1.341 \\
0.1 $Z\sol$, 2.28 O/Fe\sol  &  5.224  &  5.734  &  4.733  &  3.097  &  2.300  &  1.847  &  1.512  &  1.246 \\
$Z\sol$, 0.44 O/Fe\sol  & 3.198  &  4.017  &  4.139  &  2.980  &  1.904  &  1.511  &  1.260  &  1.206 \\
$Z\sol$, O/Fe\sol  &  2.995  &  3.823  &  4.056  &  3.185  &  1.929  &  1.483  &  1.209  &  1.190 \\
$Z\sol$, 2.28 O/Fe\sol  &  2.679  &  3.498  &  3.829  &  3.295  &  1.957  &  1.425  &  1.108  &  1.1998 \\
1.5 $Z\sol$, 0.44 O/Fe\sol  &  2.679  &  3.467  &  3.740  &  2.973  &  1.838  &  1.451  &  1.226  &  1.214 \\
1.5 $Z\sol$, O/Fe\sol  &  2.495  &  3.264  &  3.591  &  2.994  &  1.844  &  1.419  &  1.174  &  1.192 \\
1.5 $Z\sol$, 2.28 O/Fe\sol  &  2.293  &  3.031  &  3.362  &  3.254  &  1.713  &  1.288  &  1.257  &  1.2002 \\\\
0.1 $Z\sol$, 0.54 Mg/Fe\sol  &  5.700  &  5.799  &  4.331  &  2.973  &  2.343  &  1.911  &  1.586  &  1.344 \\
0.1 $Z\sol$, Mg/Fe\sol  &  5.679  &  5.815  &  4.368  &  2.981  &  2.357  &  1.918  &  1.584  &  1.341 \\
0.1 $Z\sol$, 1.84 Mg/Fe\sol  &  5.637  &  5.834  &  4.423  &  3.004  &  2.367  &  1.928  &  1.595  &  1.354 \\
$Z\sol$, 0.54 Mg/Fe\sol  & 3.049  &  3.872  &  4.067  &  3.120  &  1.903  &  1.471  &  1.198  &  1.191 \\
$Z\sol$, Mg/Fe\sol  &  2.995  &  3.823  &  4.056  &  3.185  &  1.929  &  1.483  &  1.209  &  1.190 \\
$Z\sol$, 1.84 Mg/Fe\sol  &  3.426  &  3.729  &  4.029  &  3.297  &  1.986  &  1.510  &  1.230  &  1.214 \\
1.5 $Z\sol$, 0.54 Mg/Fe\sol  &  2.540  &  3.314  &  3.613  &  2.953  &  1.817  &  1.405  &  1.164  &  1.197 \\
1.5 $Z\sol$, Mg/Fe\sol  &  2.495  &  3.264  &  3.591  &  2.994  &  1.844  &  1.419  &  1.174  &  1.192 \\
1.5 $Z\sol$, 1.84 Mg/Fe\sol  &  2.415  &  3.173  &  3.553  &  3.471  &  1.896  &  1.443  &  1.189  &  1.196 \\\\
0.1 $Z\sol$, 0.5 Ne/Fe\sol  &  5.712  &  5.779  &  4.297  &  2.955  &  2.332  &  1.909  &  1.577  &  1.333 \\
0.1 $Z\sol$, Ne/Fe\sol  &  5.679  &  5.815  &  4.368  &  2.981  &  2.357  &  1.918  &  1.584  &  1.341 \\
0.1 $Z\sol$, 2.0 Ne/Fe\sol  &  5.614  &  5.877  &  4.515  &  3.038  &  2.387  &  1.944  &  1.608  &  1.369 \\
$Z\sol$, 0.5 Ne/Fe\sol  &  3.069  &  3.884  &  4.067  &  3.099  &  1.903  &  1.472  &  1.200  &  1.197 \\
$Z\sol$, Ne/Fe\sol  &  2.995  &  3.823  &  4.056  &  3.185  &  1.929  &  1.483  &  1.209  &  1.190 \\
$Z\sol$, 2.0 Ne/Fe\sol  &  2.867  &  3.701  &  4.036  &  3.353  &  2.006  &  1.517  &  1.232  &  1.221 \\
1.5 $Z\sol$, 0.5 Ne/Fe\sol  &  2.555  &  3.321  &  3.613  &  2.946  &  1.819  &  1.409  &  1.165  &  1.202 \\
1.5 $Z\sol$, Ne/Fe\sol  &  2.495  &  3.264  &  3.591  &  2.994  &  1.844  &  1.419  &  1.174  &  1.192 \\
1.5 $Z\sol$, 2.0 Ne/Fe\sol  &  2.395  &  3.155  &  3.542  &  3.474  &  1.892  &  1.439  &  1.187  &  1.194 \\
\enddata
\end{deluxetable}

\begin{deluxetable}{lcccccccc}
\tabletypesize{\scriptsize}
\tablecaption{$\Delta$$T_{eff}$ (K) for each mass and end-member composition for all elements.\label{tab3}}
\tablehead{
  \colhead{Composition}
& \colhead{0.5 M\sol}
& \colhead{0.6 M\sol}
& \colhead{0.7 M\sol}
& \colhead{0.8 M\sol}
& \colhead{0.9 M\sol}
& \colhead{1.0 M\sol}
& \colhead{1.1 M\sol}
& \colhead{1.2 M\sol}
}
\startdata
0.1 $Z\sol$, 0.58 C/Fe\sol  &  1173  &  1231  &  708  &  277  &  150  &  141  &  240  &  229 \\
0.1 $Z\sol$, C/Fe\sol  &  1175  &  1119  &  733  &  281  &  146  &  126  &  221  &  216 \\
0.1 $Z\sol$, 1.72 C/Fe\sol  &  1183  &  1151  &  788  &  292  &  144  &  114  &  203  &  210 \\
$Z\sol$, 0.58 C/Fe\sol  & 745  &  840  &  929  &  771  &  270  &  72  &  -27  &  -116 \\
$Z\sol$, C/Fe\sol  &  717  &  825  &  915  &  758  &  257  &  68  &  -21  &  -86 \\
$Z\sol$, 1.72 C/Fe\sol  &  688  &  798  &  892  &  744  &  249  &  67  &  -16  &  -85 \\
1.5 $Z\sol$, 0.58 C/Fe\sol  &  630  &  721  &  802  &  732  &  298  &  99  &  0  &  -86 \\
1.5 $Z\sol$, C/Fe\sol  &  613  &  706  &  789  &  704  &  275  &  87  &  9  &  -59 \\
1.5 $Z\sol$, 1.72 C/Fe\sol  &  592  &  687  &  766  &  668  &  242  &  78  &  19  &  -59 \\\\
0.1 $Z\sol$, 0.44 O/Fe\sol  &  1160  &  1052  &  633  &  253  &  156  &  203  &  316  &  272 \\
0.1 $Z\sol$, O/Fe\sol  &  1175  &  1119  &  733  &  281  &  146  &  126  &  221  &  216 \\
0.1 $Z\sol$, 2.28 O/Fe\sol  &  1161  &  1175  &  969  &  404  &  170  &  86  &  79  &  118 \\
$Z\sol$, 0.44 O/Fe\sol  &  758  &  868  &  977  &  661  &  192  &  21  &  -49  &  -103 \\
$Z\sol$, O/Fe\sol  &  717  &  825  &  915  &  758  &  257  &  68  &  -21  &  -86 \\
$Z\sol$, 2.28 O/Fe\sol  &  643  &  718  &  813  &  776  &  333  &  108  &  14  &  -98 \\
1.5 $Z\sol$, 0.44 O/Fe\sol  &  658  &  765  &  854  &  708  &  243  &  65  &  -17  &  -72 \\
1.5 $Z\sol$, O/Fe\sol  &  613  &  706  &  789  &  704  &  275  &  87  &  9  &  -59 \\
1.5 $Z\sol$, 2.28 O/Fe\sol  &  576  &  653  &  728  &  859  &  255  &  82  &  -25  &  -80 \\\\
0.1 $Z\sol$, 0.54 Mg/Fe\sol  &  1182  &  1118  &  723  &  277  &  149  &  140  &  236  &  219 \\
0.1 $Z\sol$, Mg/Fe\sol  &  1175  &  1119  &  733  &  281  &  146  &  126  &  221  &  216 \\
0.1 $Z\sol$, 1.84 Mg/Fe\sol  &  1171  &  1125  &  754  &  285  &  143  &  110  &  201  &  200 \\
$Z\sol$, 0.54 Mg/Fe\sol  & 733  &  840  &  934  &  739  &  245  &  61  &  -20  &  -97 \\
$Z\sol$, Mg/Fe\sol  &  717  &  825  &  915  &  758  &  257  &  68  &  -21  &  -86 \\
$Z\sol$, 1.84 Mg/Fe\sol  &  655  &  802  &  889  &  779  &  287  &  83  &  -13  &  -89 \\
1.5 $Z\sol$, 0.54 Mg/Fe\sol  &  619  &  719  &  802  &  697  &  263  &  83  &  1  &  -62 \\
1.5 $Z\sol$, Mg/Fe\sol  &  613  &  706  &  789  &  704  &  275  &  87  &  9  &  -59 \\
1.5 $Z\sol$, 1.84 Mg/Fe\sol  &  604  &  687  &  767  &  906  &  296  &  98  &  17  &  -53 \\\\
0.1 $Z\sol$, 0.5 Ne/Fe\sol  &  1180  &  1114  &  715  &  277  &  156  &  147  &  250  &  225 \\
0.1 $Z\sol$, Ne/Fe\sol  &  1175  &  1119  &  733  &  281  &  146  &  126  &  221  &  216 \\
0.1 $Z\sol$, 2.0 Ne/Fe\sol  &  1168  &  1134  &  787  &  294  &  142  &  94  &  175  &  197 \\
$Z\sol$, 0.5 Ne/Fe\sol  &  740  &  848  &  942  &  739  &  247  &  62  &  -19  &  -96 \\
$Z\sol$, Ne/Fe\sol  &  717  &  825  &  915  &  758  &  257  &  68  &  -21  &  -86 \\
$Z\sol$, 2.0 Ne/Fe\sol  &  687  &  783  &  868  &  782  &  292  &  84  &  -15  &  -86 \\
1.5 $Z\sol$, 0.5 Ne/Fe\sol  &  628  &  722  &  808  &  702  &  266  &  84  &  1  &  -63 \\
1.5 $Z\sol$, Ne/Fe\sol  &  613  &  706  &  789  &  704  &  275  &  87  &  9  &  -59 \\
1.5 $Z\sol$, 2.0 Ne/Fe\sol  &  592  &  675  &  753  &  891  &  290  &  91  &  8  &  -57 \\
\enddata
\end{deluxetable}

\begin{deluxetable}{lllllllll}
\tabletypesize{\scriptsize}
\tablecaption{[d($L$/$L_{ZAMS}$)/dt] for each mass and end-member composition for all elements.\label{tab4}}
\tablehead{
  \colhead{Composition}
& \colhead{0.5 M\sol}
& \colhead{0.6 M\sol}
& \colhead{0.7 M\sol}
& \colhead{0.8 M\sol}
& \colhead{0.9 M\sol}
& \colhead{1.0 M\sol}
& \colhead{1.1 M\sol}
& \colhead{1.2 M\sol}
}
\startdata
0.1 $Z\sol$, 0.58 C/Fe\sol  &  0.083  &  0.149  &  0.193  &  0.224  &  0.277  &  0.336  &  0.395  &  0.464 \\
0.1 $Z\sol$, C/Fe\sol  &  0.081  &  0.141  &  0.189  &  0.217  &  0.269  &  0.324  &  0.381  &  0.448 \\
0.1 $Z\sol$, 1.72 C/Fe\sol  &  0.077  &  0.135  &  0.183  &  0.207  &  0.255  &  0.309  &  0.363  &  0.424 \\
$Z\sol$, 0.58 C/Fe\sol  &  0.029  &  0.055  &  0.091  &  0.122  &  0.123  &  0.146  &  0.174  &  0.232 \\
$Z\sol$, C/Fe\sol  &  0.027  &  0.053  &  0.089  &  0.119  &  0.120  &  0.143  &  0.175  &  0.237 \\
$Z\sol$, 1.72 C/Fe\sol  &  0.025  &  0.049  &  0.084  &  0.114  &  0.114  &  0.138  &  0.174  &  0.240 \\
1.5 $Z\sol$, 0.58 C/Fe\sol  &  0.023  &  0.044  &  0.075  &  0.105  &  0.107  &  0.126  &  0.152  &  0.205 \\
1.5 $Z\sol$, C/Fe\sol  &  0.022  &  0.043  &  0.073  &  0.102  &  0.103  &  0.124  &  0.154  &  0.214 \\
1.5 $Z\sol$, 1.72 C/Fe\sol  &  0.021  &  0.041  &  0.070  &  0.100  &  0.098  &  0.122  &  0.157  &  0.223 \\\\
0.1 $Z\sol$, 0.44 O/Fe\sol  &  0.091  &  0.157  &  0.207  &  0.243  &  0.301  &  0.365  &  0.434  &  0.503 \\
0.1 $Z\sol$, O/Fe\sol  &  0.081  &  0.141  &  0.189  &  0.217  &  0.269  &  0.324  &  0.381  &  0.448 \\
0.1 $Z\sol$, 2.28 O/Fe\sol  &  0.063  &  0.114  &  0.166  &  0.185  &  0.219  &  0.265  &  0.312  &  0.358 \\
$Z\sol$, 0.44 O/Fe\sol  &  0.031  &  0.060  &  0.101  &  0.126  &  0.131  &  0.161  &  0.199  &  0.263 \\
$Z\sol$, O/Fe\sol  &  0.027  &  0.053  &  0.089  &  0.119  &  0.120  &  0.143  &  0.175  &  0.237 \\
$Z\sol$, 2.28 O/Fe\sol  &  0.022  &  0.043  &  0.073  &  0.105  &  0.105  &  0.121  &  0.142  &  0.206 \\
1.5 $Z\sol$, 0.44 O/Fe\sol  &  0.025  &  0.048  &  0.081  &  0.110  &  0.111  &  0.136  &  0.171  &  0.233 \\
1.5 $Z\sol$, O/Fe\sol  &  0.022  &  0.043  &  0.073  &  0.102  &  0.103  &  0.124  &  0.154  &  0.214 \\
1.5 $Z\sol$, 2.28 O/Fe\sol  &  0.019  &  0.038  &  0.066  &  0.106  &  0.093  &  0.110  &  0.153  &  0.201 \\\\
0.1 $Z\sol$, 0.54 Mg/Fe\sol  &  0.082  &  0.142  &  0.190  &  0.219  &  0.270  &  0.326  &  0.385  &  0.452 \\
0.1 $Z\sol$, Mg/Fe\sol  &  0.081  &  0.141  &  0.189  &  0.217  &  0.269  &  0.324  &  0.381  &  0.448 \\
0.1 $Z\sol$, 1.84 Mg/Fe\sol  &  0.079  &  0.138  &  0.187  &  0.214  &  0.264  &  0.320  &  0.377  &  0.444 \\
$Z\sol$, 0.54 Mg/Fe\sol  &  0.028  &  0.054  &  0.091  &  0.120  &  0.121  &  0.146  &  0.178  &  0.241 \\
$Z\sol$, Mg/Fe\sol  &  0.027  &  0.053  &  0.089  &  0.119  &  0.120  &  0.143  &  0.175  &  0.237 \\
$Z\sol$, 1.84 Mg/Fe\sol  &  0.031  &  0.050  &  0.084  &  0.117  &  0.117  &  0.139  &  0.170  &  0.231 \\
1.5 $Z\sol$, 0.54 Mg/Fe\sol  &  0.023  &  0.044  &  0.075  &  0.103  &  0.105  &  0.126  &  0.157  &  0.218 \\
1.5 $Z\sol$, Mg/Fe\sol  &  0.022  &  0.043  &  0.073  &  0.102  &  0.103  &  0.124  &  0.154  &  0.214 \\
1.5 $Z\sol$, 1.84 Mg/Fe\sol  &  0.021  &  0.040  &  0.069  &  0.111  &  0.101  &  0.120  &  0.150  &  0.207 \\\\
0.1 $Z\sol$, 0.5 Ne/Fe\sol  &  0.083  &  0.144  &  0.191  &  0.220  &  0.272  &  0.330  &  0.387  &  0.453 \\
0.1 $Z\sol$, Ne/Fe\sol  &  0.081  &  0.141  &  0.189  &  0.217  &  0.269  &  0.324  &  0.381  &  0.448 \\
0.1 $Z\sol$, 2.0 Ne/Fe\sol  &  0.076  &  0.135  &  0.185  &  0.211  &  0.260  &  0.315  &  0.371  &  0.439 \\
$Z\sol$, 0.5 Ne/Fe\sol  &  0.029  &  0.055  &  0.091  &  0.120  &  0.121  &  0.146  &  0.178  &  0.242 \\
$Z\sol$, Ne/Fe\sol  &  0.027  &  0.053  &  0.089  &  0.119  &  0.120  &  0.143  &  0.175  &  0.237 \\
$Z\sol$, 2.0 Ne/Fe\sol  &  0.025  &  0.048  &  0.083  &  0.116  &  0.117  &  0.138  &  0.169  &  0.230 \\
1.5 $Z\sol$, 0.5 Ne/Fe\sol  &  0.023  &  0.044  &  0.075  &  0.103  &  0.105  &  0.126  &  0.156  &  0.218 \\
1.5 $Z\sol$, Ne/Fe\sol  &  0.022  &  0.043  &  0.073  &  0.102  &  0.103  &  0.124  &  0.154  &  0.214 \\
1.5 $Z\sol$, 2.0 Ne/Fe\sol  &  0.020  &  0.040  &  0.069  &  0.111  &  0.101  &  0.120  &  0.150  &  0.208 \\
\enddata
\end{deluxetable}

\begin{deluxetable}{lcccccccc}
\tabletypesize{\scriptsize}
\tablecaption{Fraction (\%) of radii which enter the HZ after the midpoint of MS (for carbon).\label{tab5}}
\tablehead{
  \colhead{Composition}
& \colhead{0.5 M\sol}
& \colhead{0.6 M\sol}
& \colhead{0.7 M\sol}
& \colhead{0.8 M\sol}
& \colhead{0.9 M\sol}
& \colhead{1.0 M\sol}
& \colhead{1.1 M\sol}
& \colhead{1.2 M\sol}
}
\startdata
0.1 $Z\sol$, 0.58 C/Fe\sol  &  29.10  &  27.92  &  33.75  &  37.69  &  41.40  &  45.09  &  49.15  &  52.93 \\
0.1 $Z\sol$, C/Fe\sol  &  29.18  &  30.39  &  33.91  &  37.94  &  41.68  &  45.44  &  49.49  &  53.35 \\
0.1 $Z\sol$, 1.72 C/Fe\sol  &  29.30  &  30.46  &  34.10  &  38.25  &  42.11  &  45.85  &  49.96  &  53.86 \\
$Z\sol$, 0.58 C/Fe\sol  & 35.89  &  33.86  &  35.56  &  40.56  &  46.33  &  50.89  &  55.21  &  54.86 \\
$Z\sol$, C/Fe\sol  &  36.63  &  34.40  &  36.06  &  41.13  &  46.99  &  51.43  &  55.43  &  55.21 \\
$Z\sol$, 1.72 C/Fe\sol  &  37.58  &  35.17  &  36.78  &  41.98  &  48.02  &  52.34  &  55.96  &  55.08 \\
1.5 $Z\sol$, 0.58 C/Fe\sol  &  38.47  &  35.82  &  36.70  &  41.38  &  47.33  &  52.03  &  56.27  &  55.84 \\
1.5 $Z\sol$, C/Fe\sol  &  39.38  &  36.55  &  37.50  &  42.25  &  48.26  &  52.75  &  56.60  &  55.89 \\
1.5 $Z\sol$, 1.72 C/Fe\sol  &  40.31  &  37.50  &  38.63  &  43.63  &  49.82  &  53.40  &  57.37  &  55.45 \\
\enddata
\end{deluxetable}

\subsection{Location of the Habitable Zone}

We produce complete evolutionary tracks for the position of the HZ as a function of time for all stellar models. With an independent age estimate for the star, as well as measurements for its mass and specific elemental composition, we can fairly accurately predict the future and past location of a given exoplanet, and whether that planet ever inhabited the parent star's HZ; furthermore, we can assess the timeline for when a planet $will$ enter the star's HZ if it hasn't yet, as well as estimate how long the planet has been outside of the HZ if it has already departed. Assuming the aforementioned stellar properties are well measured, the time that an observed exoplanet may exist in the HZ can be estimated to the level of accuracy of the planetary atmosphere models that predict the HZ boundaries.

Generally, we find that a higher abundance of carbon, magnesium, or neon in the host star correlates with a closer-in HZ , because the star is less luminous, at a lower T$_{eff}$, and the MS lifetime is longer. Likewise, a lower elemental abundance value will typically produce shorter overall MS lifetimes with HZ distances that are farther away from the host star, which is the same trend that we observed for oxygen abundance ratios in T15.

\begin{figure}
\centering
\includegraphics[width=0.9\textwidth]{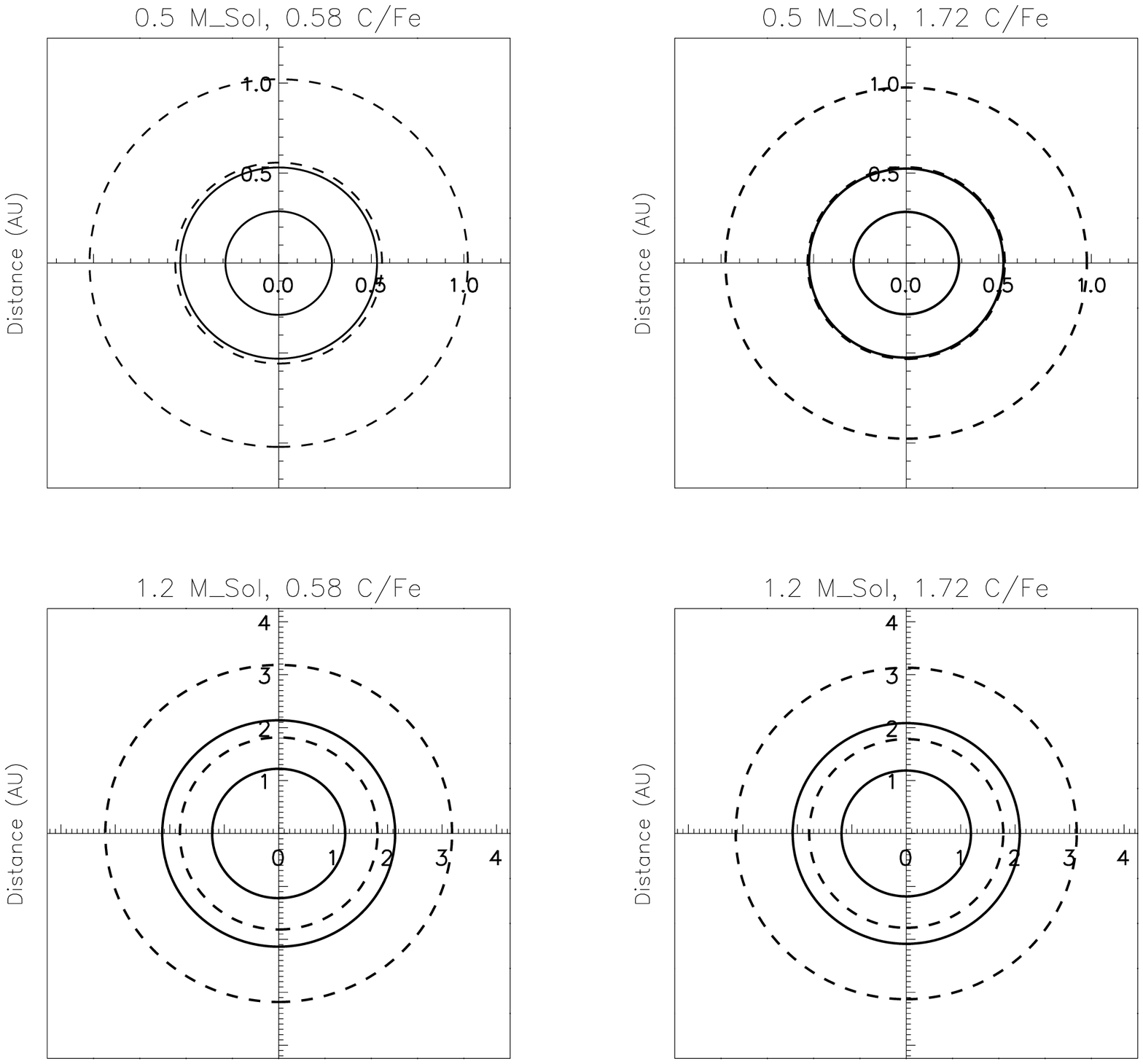}
\caption{Habitable Zone Ranges: 0.5 M\sol (top), 1.2 M\sol (bottom), for 0.58 C/Fe\sol (left), 1.72 C/Fe\sol (right), at Z\sol. HZ shown at ZAMS (solid) and TAMS (dashed). Inner boundaries represent Runaway Greenhouse and outer boundaries represent Maximum Greenhouse.\label{hzpolar_carb}}    
\end{figure}

\begin{figure}
\centering
\includegraphics[width=0.9\textwidth]{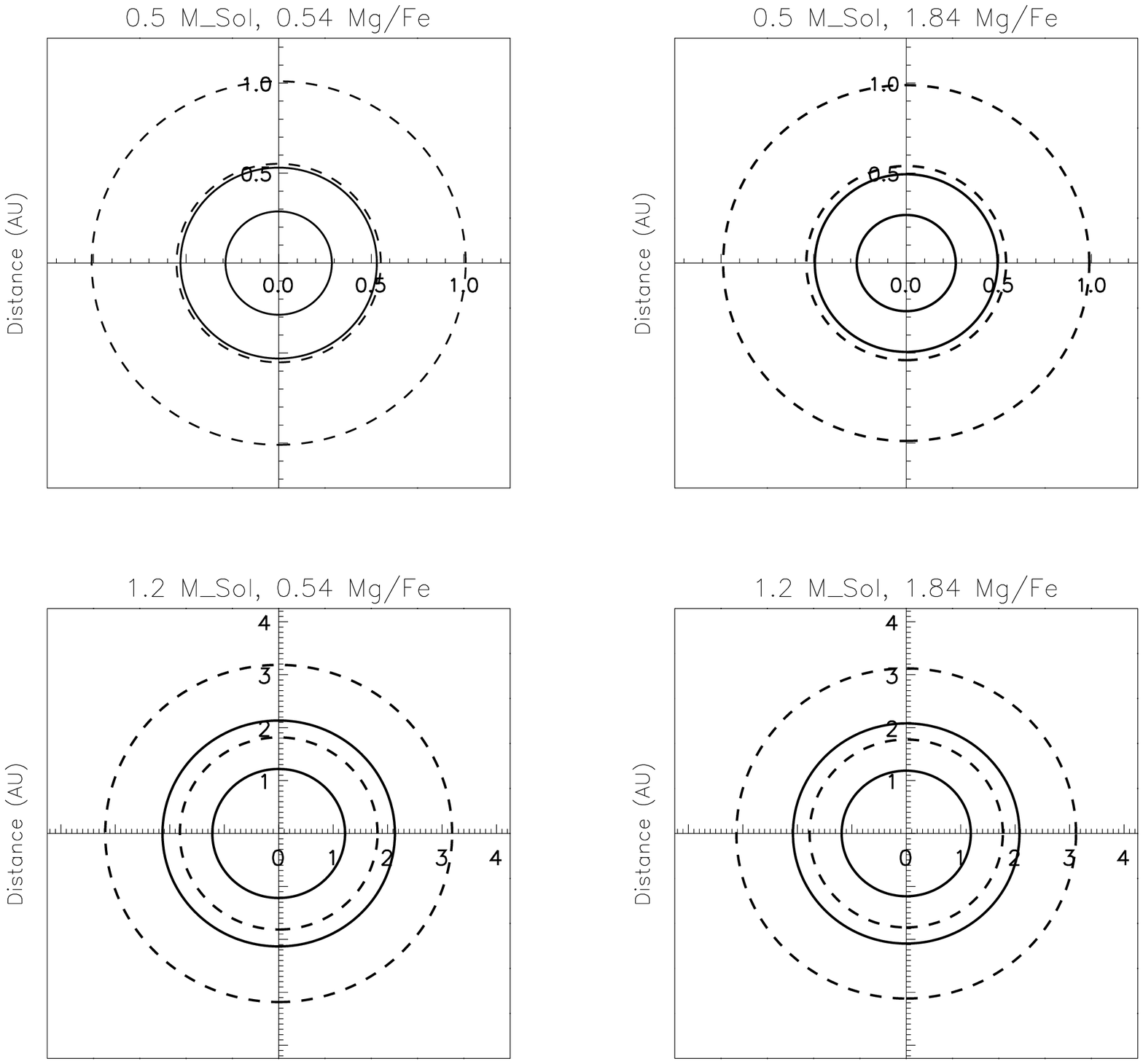}
\caption{Habitable Zone Ranges: 0.5 M\sol (top), 1.2 M\sol (bottom), for 0.54 Mg/Fe\sol (left), 1.84 Mg/Fe\sol (right), at Z\sol. HZ shown at ZAMS (solid) and TAMS (dashed). Inner boundaries represent Runaway Greenhouse and outer boundaries represent Maximum Greenhouse.\label{hzpolar_mag}}    
\end{figure}

\begin{figure}
\includegraphics[width=0.9\textwidth]{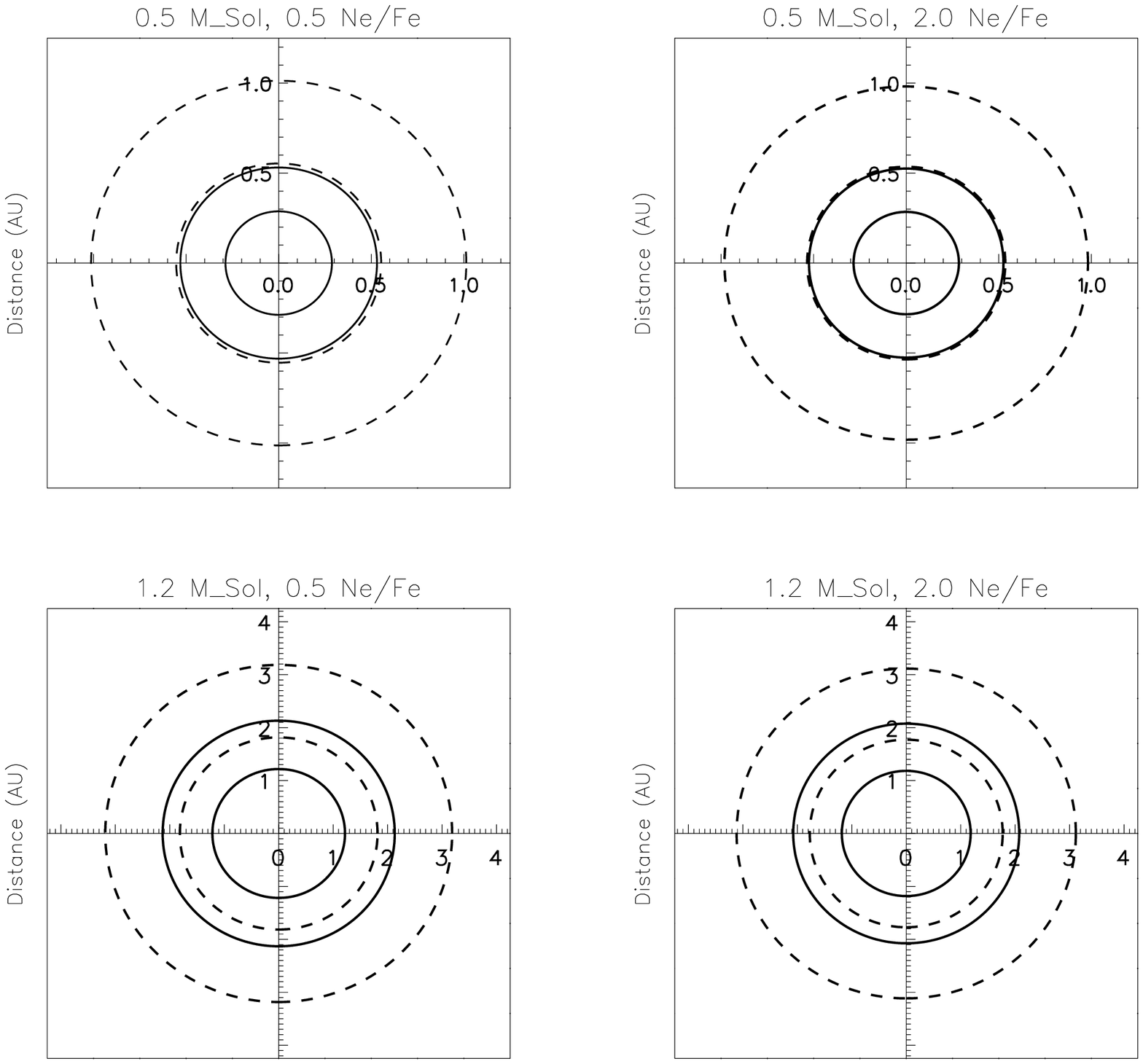}
\caption{Habitable Zone Ranges: 0.5 M\sol (top), 1.2 M\sol (bottom), for 0.5 Ne/Fe\sol (left), 2.0 Ne/Fe\sol (right), at Z\sol. HZ shown at ZAMS (solid) and TAMS (dashed). Inner boundaries represent Runaway Greenhouse and outer boundaries represent Maximum Greenhouse.\label{hzpolar_neon}}    
\end{figure}

Figures~\ref{hzpolar_carb},~\ref{hzpolar_mag}, and~\ref{hzpolar_neon} show polar plots for carbon, magnesium, and neon, respectively. These figures are meant to demonstrate how the HZ varies between different kinds of stars, and what the differences would look like from a perspective perpendicular to that of a hypothetical planet's orbital plane. These figures each include stars of end-member masses 0.5 M\sol star (top) and a 1.2 M\sol star (bottom), at the lowest and highest composition cases for each element of interest. Habitable Zone boundaries are solid for the ZAMS and dashed for the TAMS. The inner and outer HZ boundaries should be clear based on their positions relative to each other. Only the high mass stars exhibit a small degree of overlap between the outer edge at the ZAMS and the inner edge at the TAMS, which might correspond to a ``Continuously Habitable Zone" (see \S 3.3). Given this HZ prescription, it is clear that there are no orbits around the low mass stars that remain within the HZ for the $entire$ MS lifetime. Ultimately, this doesn't matter much in the sense that the low mass stars are sufficiently long-lived that they would still provide a significantly long continuously habitable zone. However, we do eventually need to assess the variation of M-star activity with age, since an extremely long  continuously HZ lifetime would not necessarily be enough to overcome a harsh radiation environment. We will explore these ideas further in a future paper. From the polar figures we also see that the HZ can change substantially over the MS depending on the host star's specific chemical composition. Tables~\ref{tab6} and \ref{tab7} show changes in the location of the HZ radius in AU from ZAMS to TAMS for both the Runaway Greenhouse inner boundary (RGH), and the Maximum Greenhouse outer boundary (MaxGH), respectively, which are the conservative HZ limit cases discussed in \citet{kopp13,kopp14}.

\begin{figure}
\centering
\includegraphics[width=0.4\textwidth]{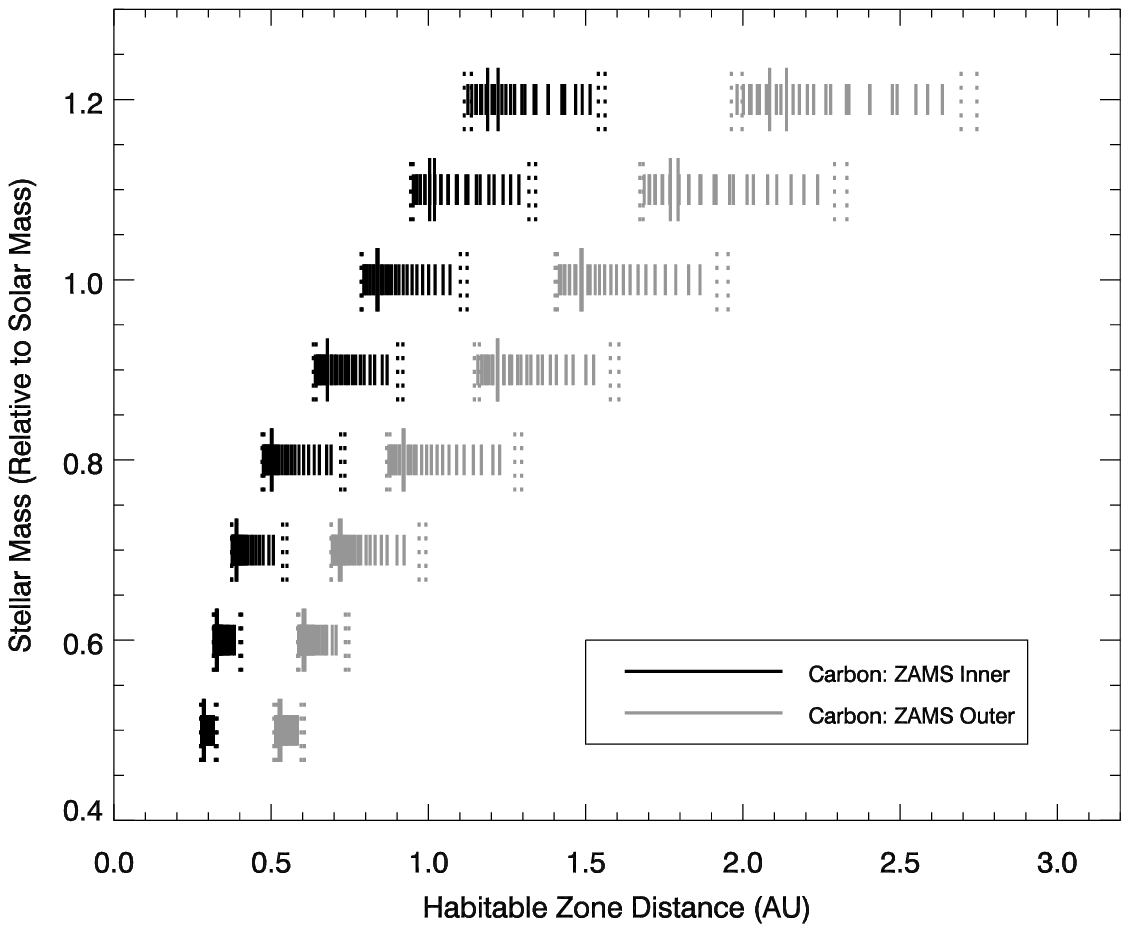}
\includegraphics[width=0.4\textwidth]{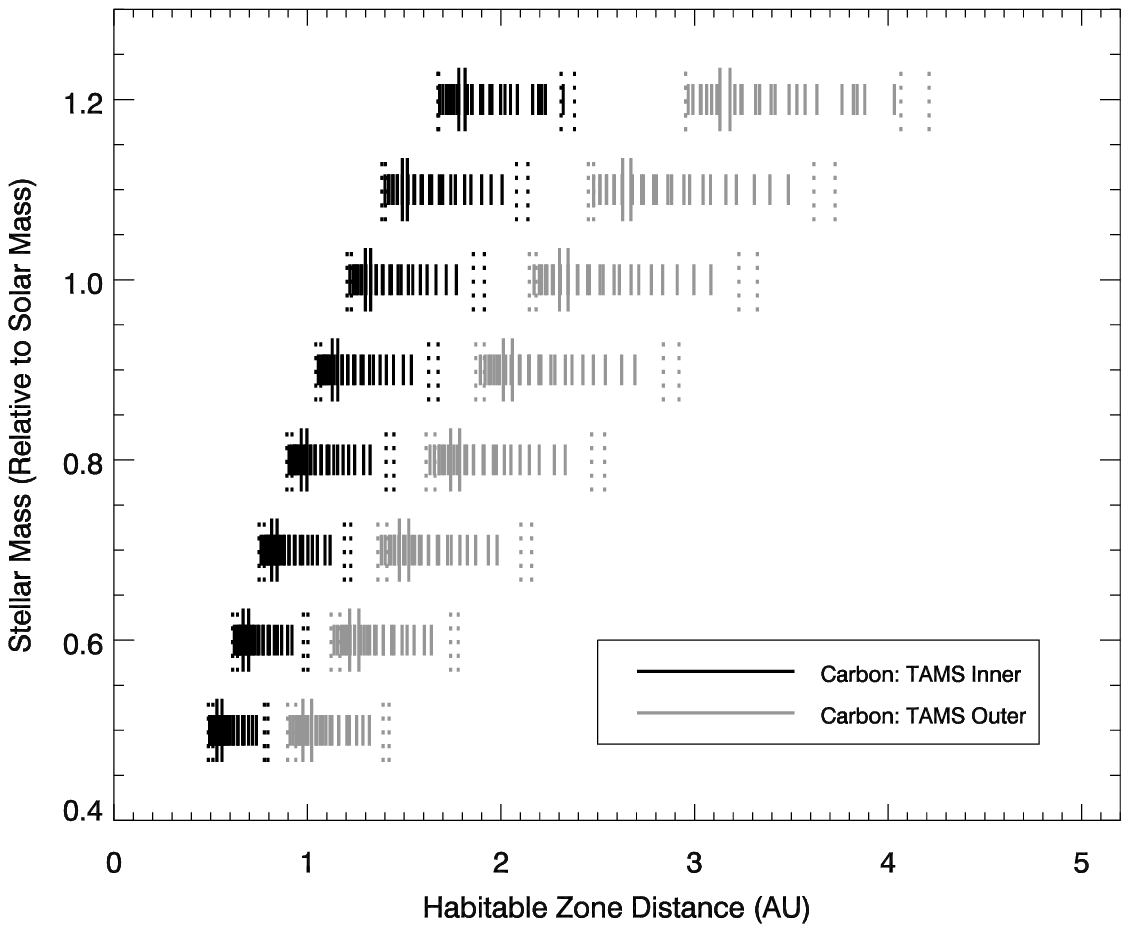}
\includegraphics[width=0.4\textwidth]{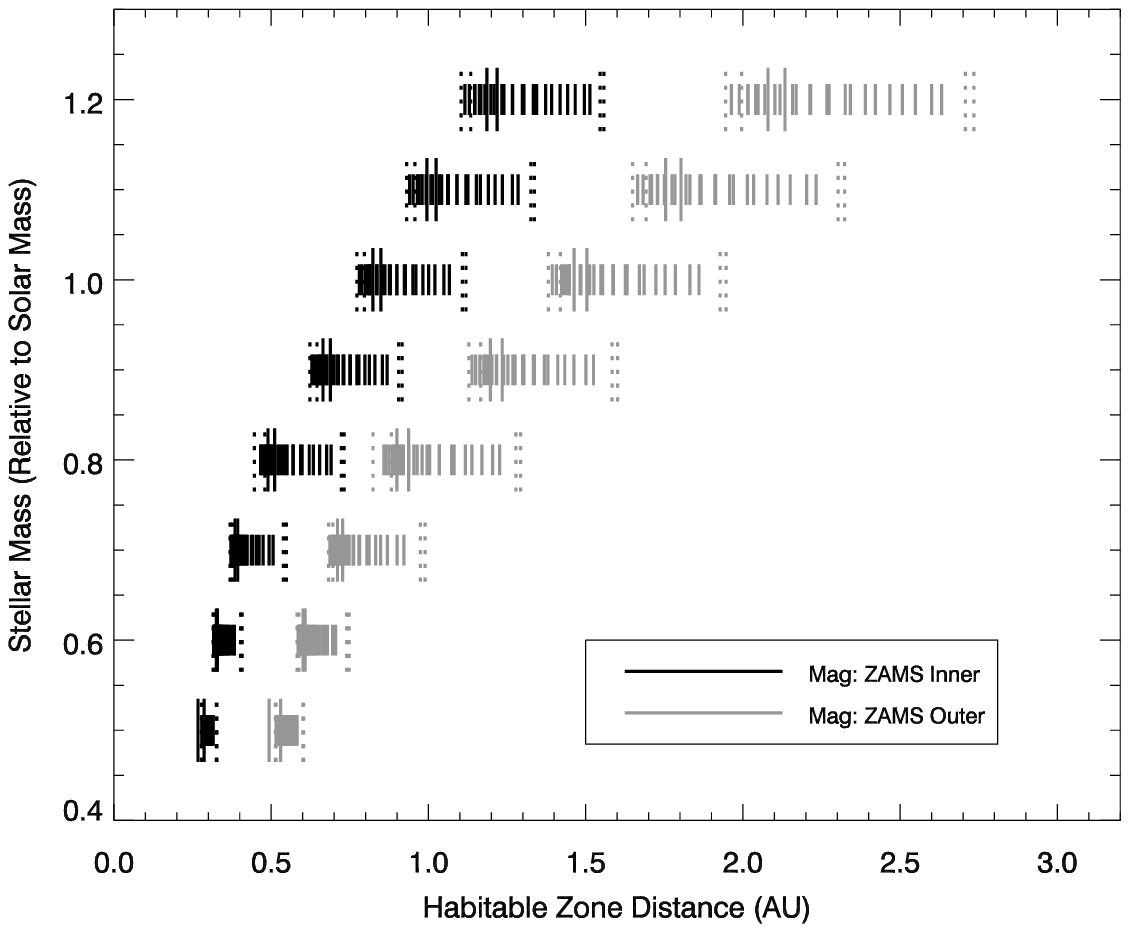}
\includegraphics[width=0.4\textwidth]{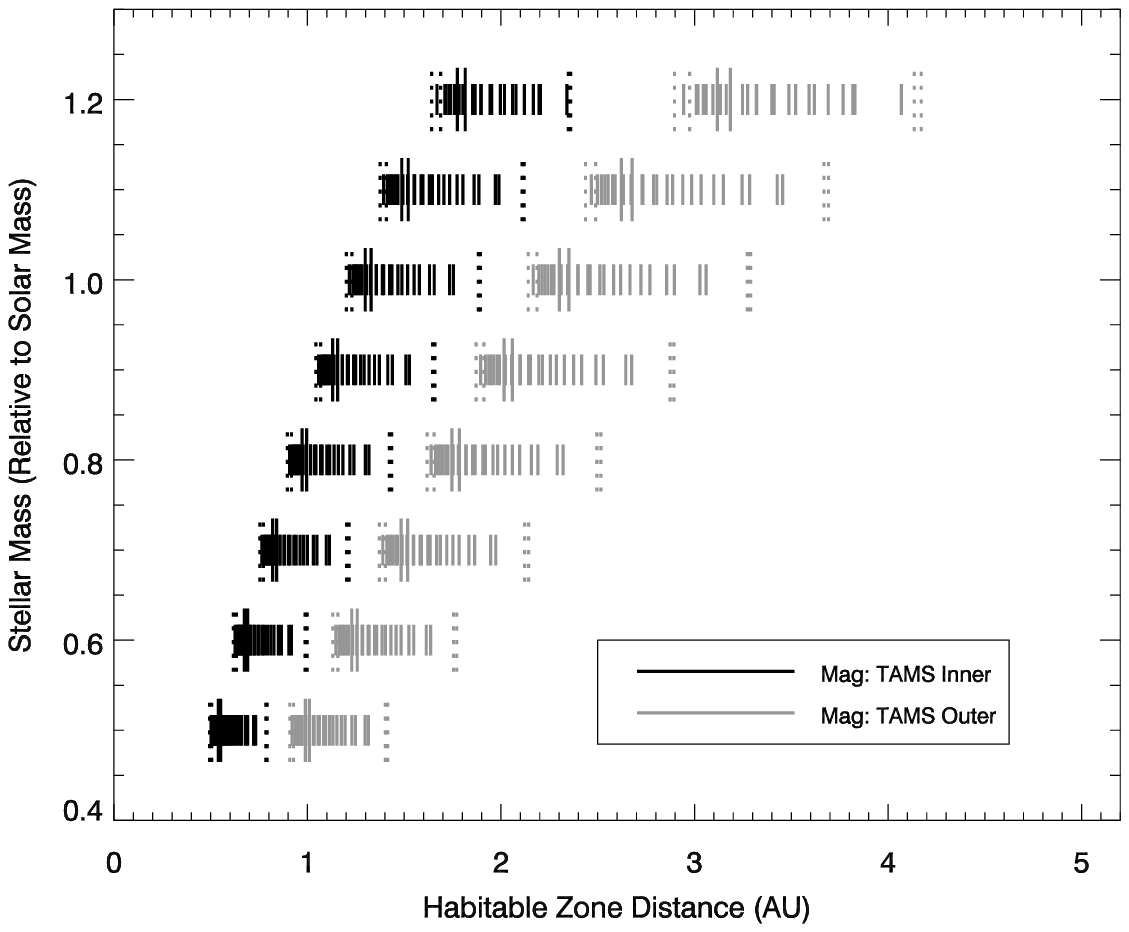}
\includegraphics[width=0.4\textwidth]{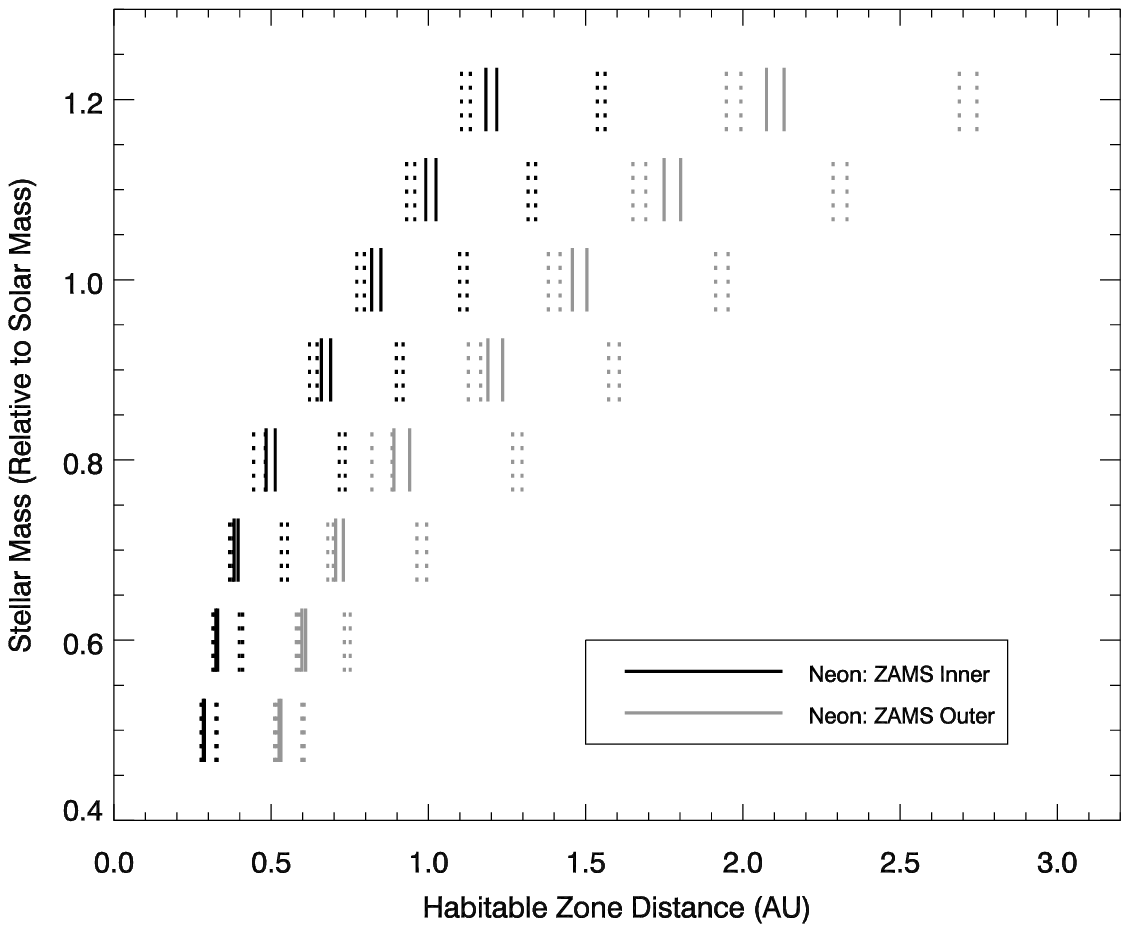}
\includegraphics[width=0.4\textwidth]{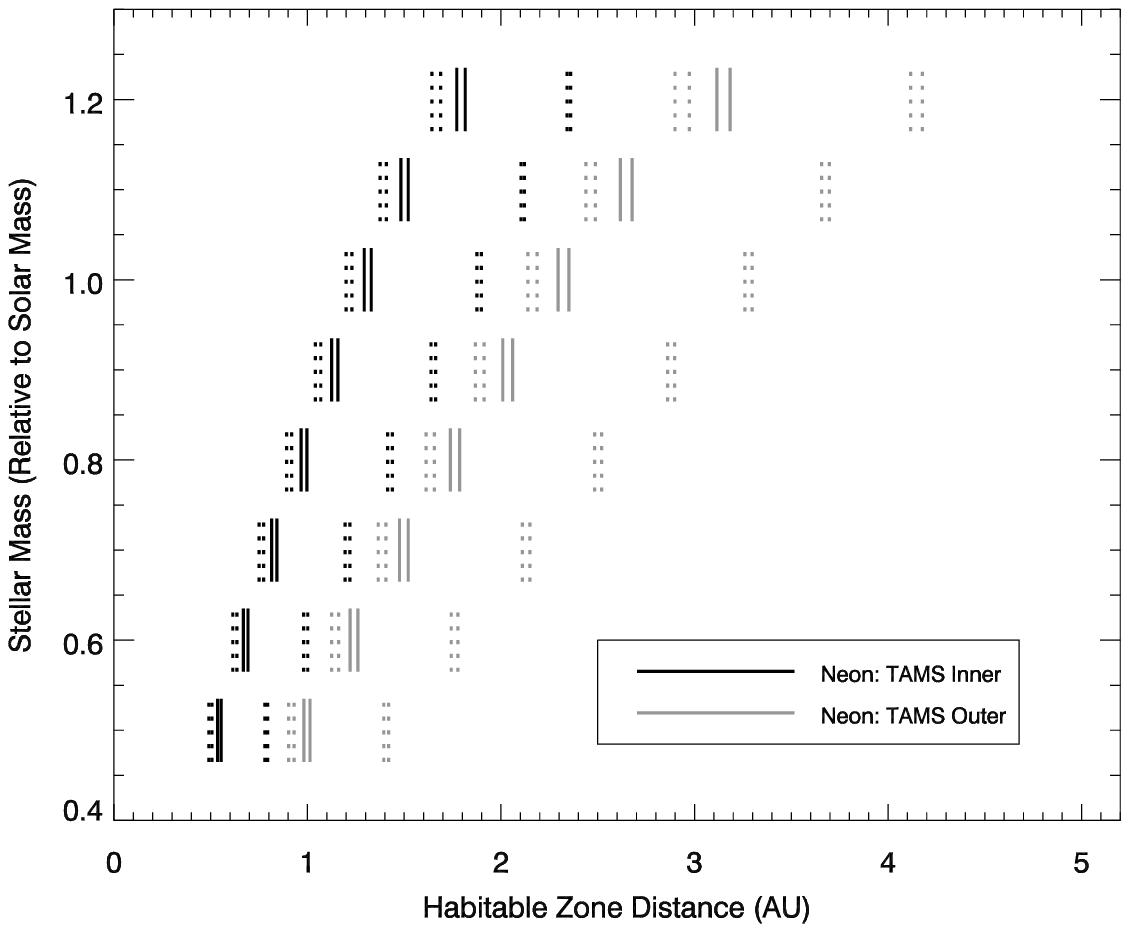}
\caption{Inner and outer HZ boundaries (RGH and MaxGH) for all models at ZAMS (left column) and TAMS (right column). The elongated solid lines in each figure represent the depleted and enriched end-member cases at Z\sol: the top row is carbon, the middle row is magnesium, and the bottom row is neon. The dotted lines represent the end member values for the spread in each elemental abundance value, now at end member Z values (0.1 and 1.5 Z\sol). It is clear that compositional variation has a much larger effect for the outer HZ limit, and for the higher mass stars in particular. Additionally, there exists a more exaggerated spreading trend for the lower mass stars at TAMS than at ZAMS.\label{zams_tams}}
\end{figure}

The lefthand column of Figure~\ref{zams_tams} shows the inner and outer edges of the HZ for each stellar mass for all compositions at the ZAMS, while the righthand column shows the same information for the TAMS. The top row is for carbon, the middle row is magnesium, and the bottom row is neon. There are a smaller number of neon lines included since we only modeled the end-member scenarios for the neon cases. We find that for all elements, a higher overall opacity results in the associated HZ boundaries at radii much closer to the host star. As observed with our original grid of oxygen models, we see that with increasing stellar mass, there seems to be a widening of the overall HZ range, as well as a larger spread in the HZ distances due to compositional variation. The spread in specific abundance ratios for each element of interest (at solar metallicity value) are indicated by the elongated solid lines.

The range in distance of the HZ edges for the each element at Z\sol is clearly smaller than the range that exists for the variations in overall scaled metallicity. This is expected, since the total change in opacity of the stellar material is much larger for a factor of fifteen change in total Z than a factor of about two change in each elemental abundance. This figure also similarly includes the spread in C, Mg, or Ne, calculated at each scaled metallicity value. The elongated dotted lines represent the end-member values for the spread in C, Mg, and Ne abundance (0.58 C/Fe\sol, 1.72 C/Fe\sol, 0.54 Mg/Fe\sol, 1.84 Mg/Fe\sol, 0.5 Ne/Fe\sol and 2.0 Ne/Fe\sol, respectively) calculated at end-member metallicity values (0.1 and 1.5 Z\sol). These models extend the range of HZ distance even further than do the models for elemental abundances at Z\sol alone. The observed difference in HZ location as a function of composition is larger for higher mass stars because the absolute change in $L$ is larger. The outer HZ edge changes more than that of the inner edge because calculation for the Maximum Greenhouse limit is more sensitive to the spectrum of the incoming radiation and $T_{eff}$.

\begin{figure}
\centering
\includegraphics[width=0.325\textwidth]{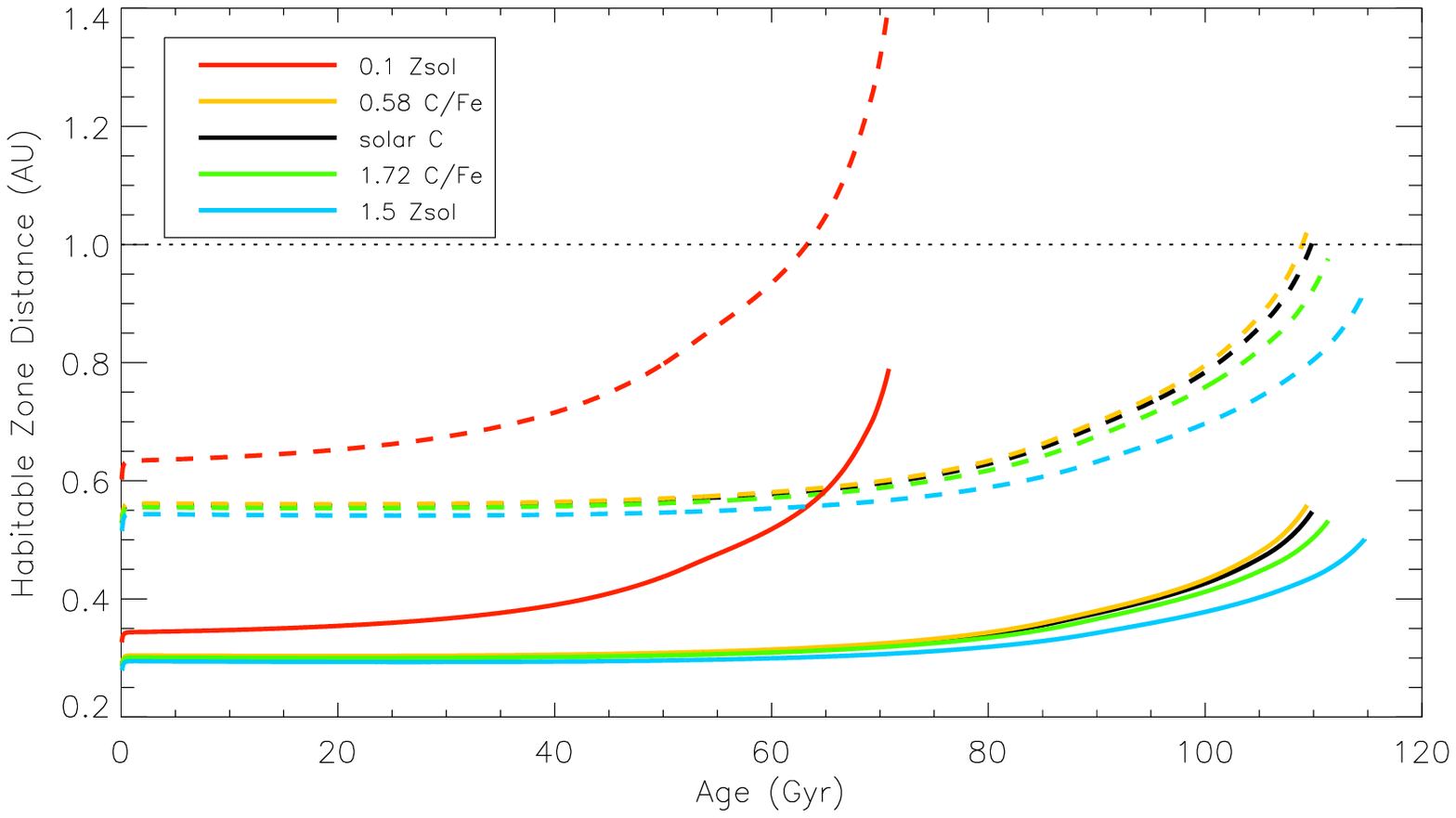}
\includegraphics[width=0.325\textwidth]{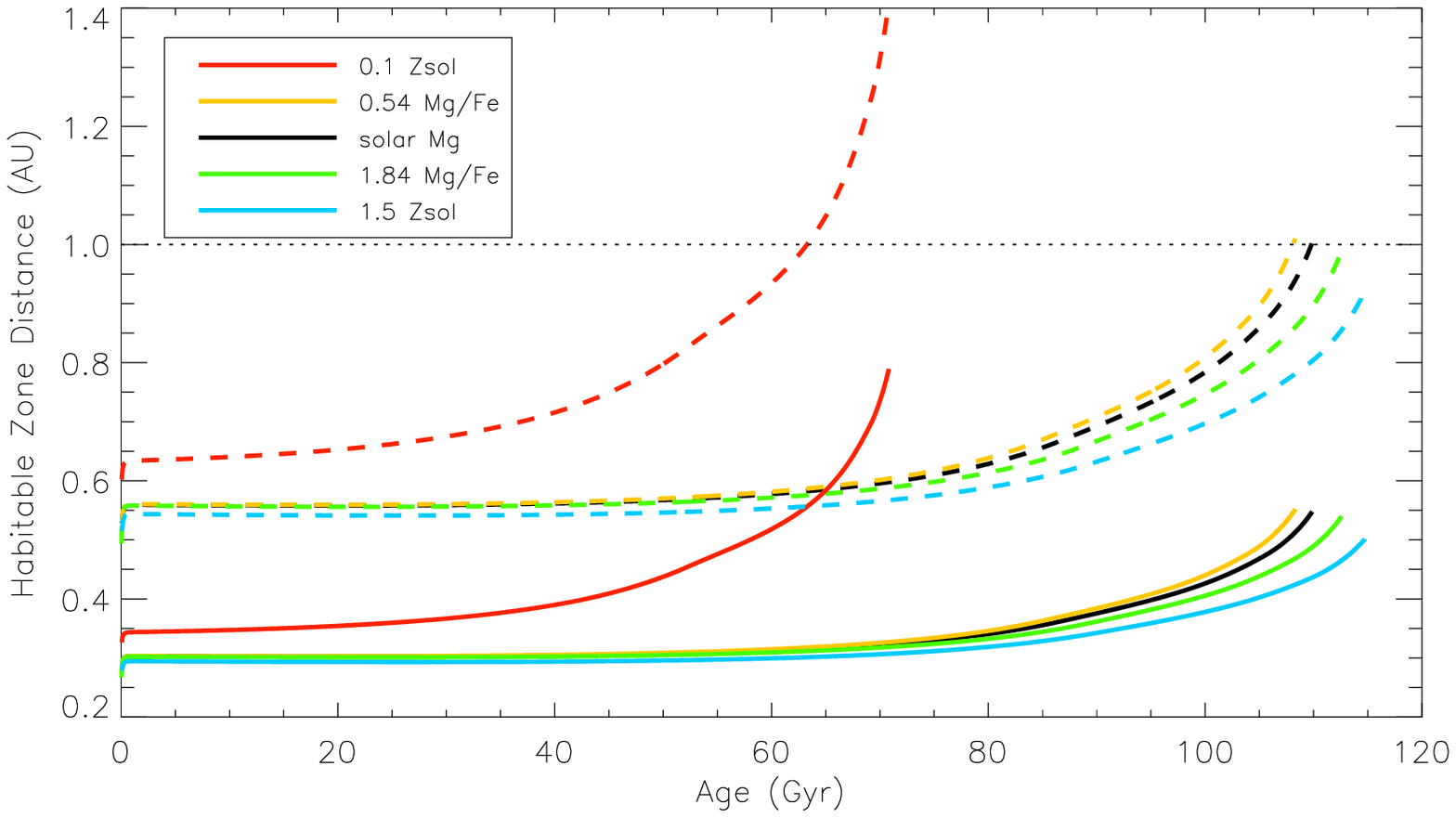}
\includegraphics[width=0.325\textwidth]{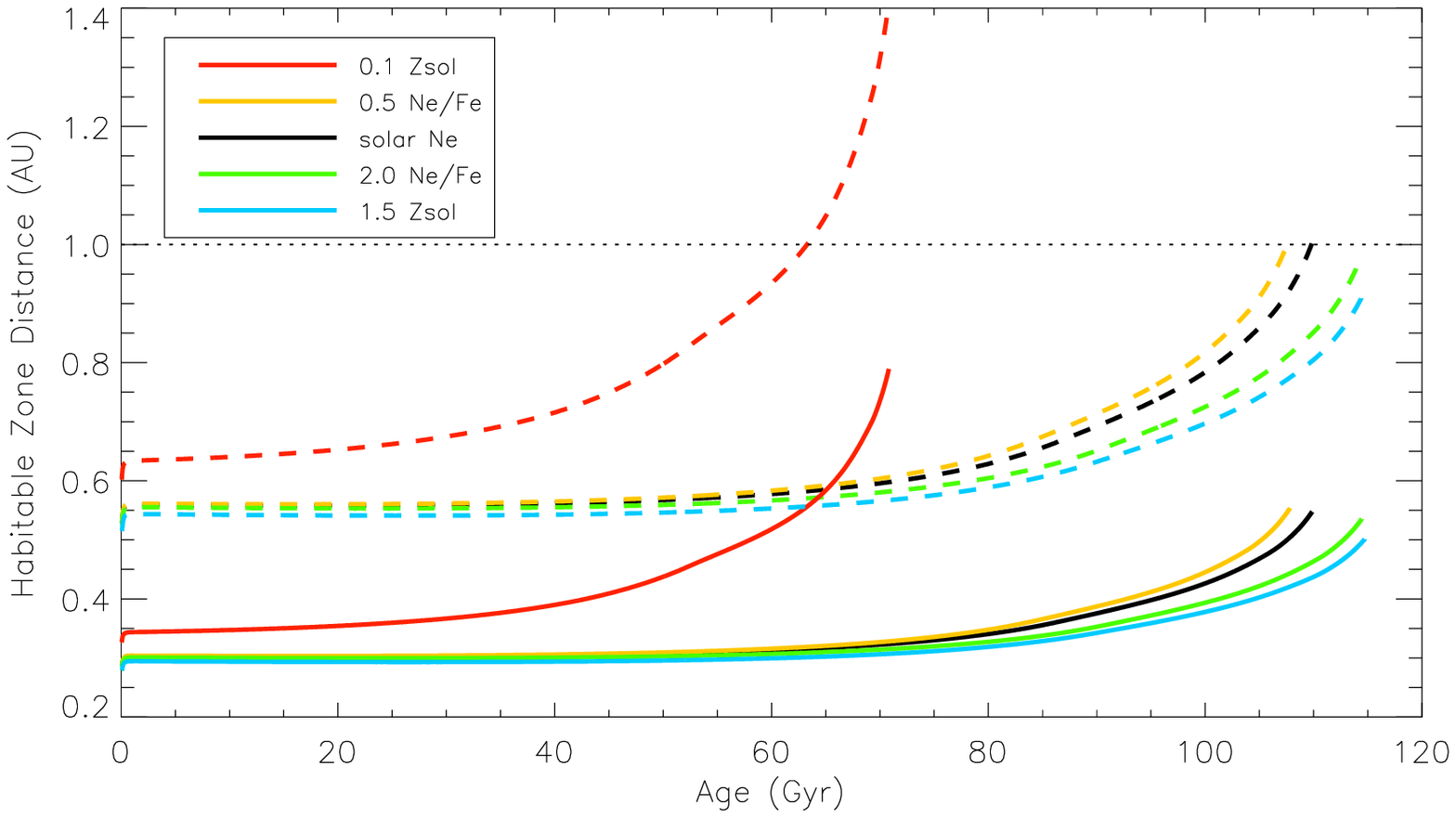}
\includegraphics[width=0.325\textwidth]{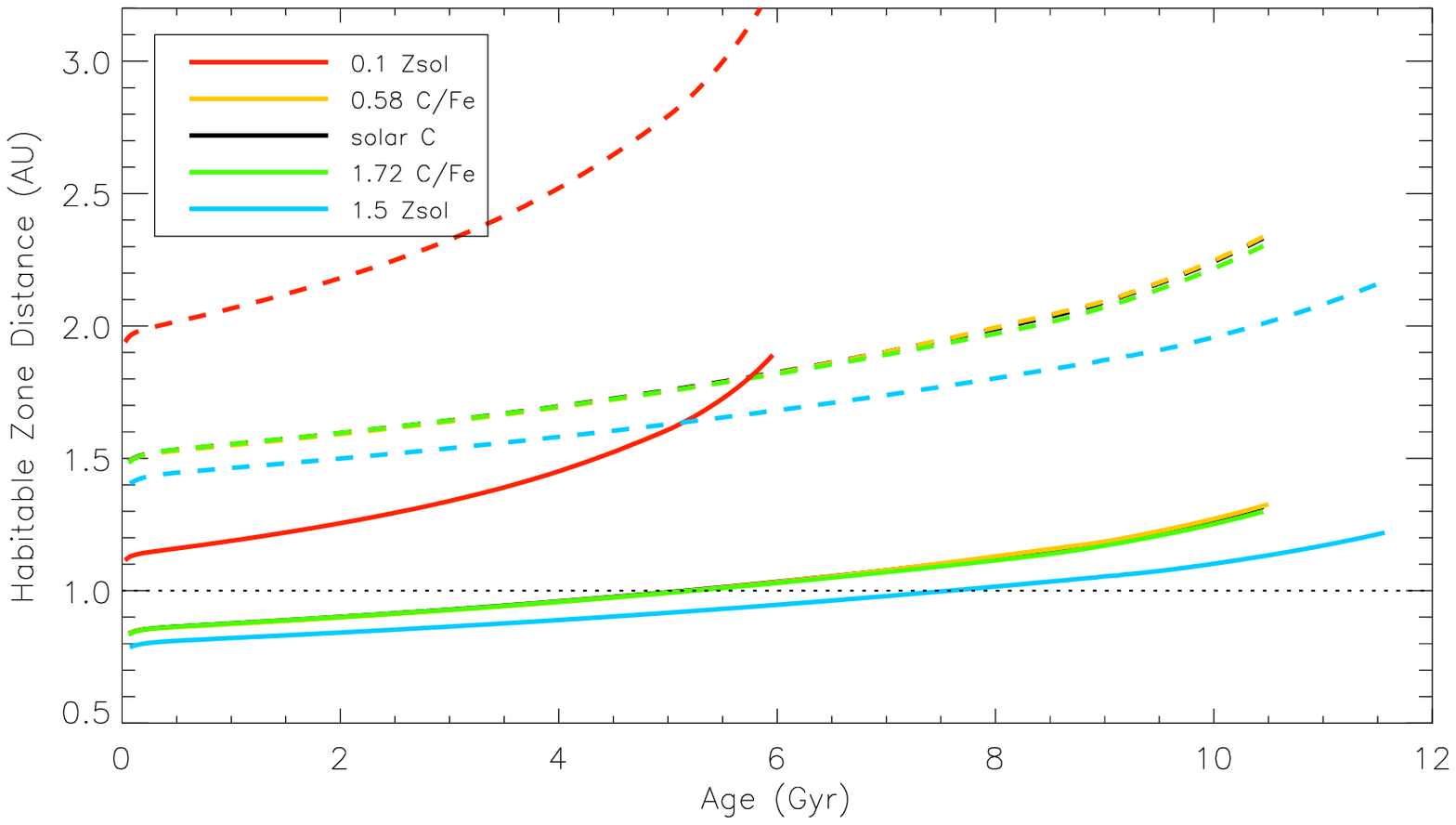}
\includegraphics[width=0.325\textwidth]{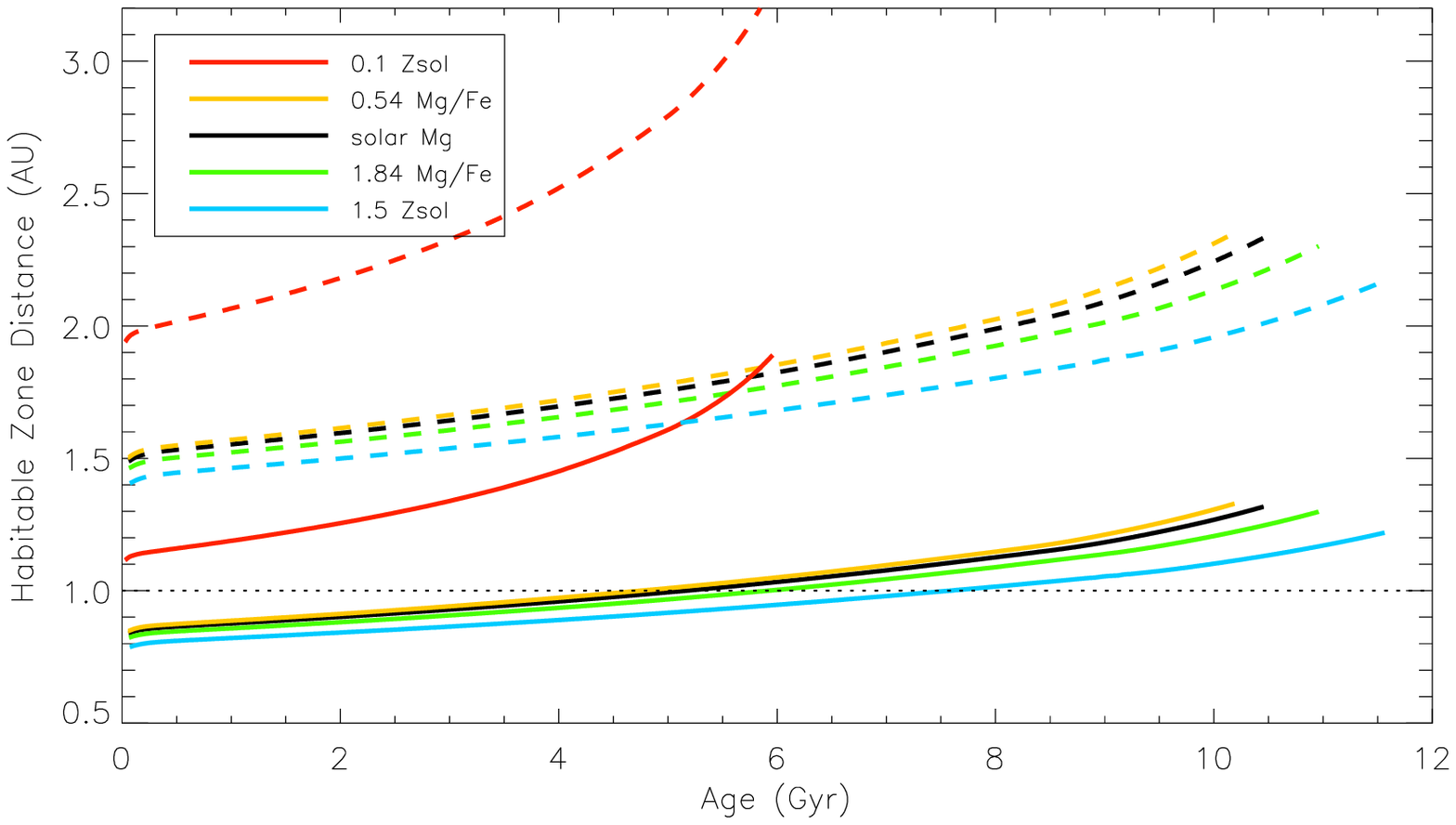}
\includegraphics[width=0.325\textwidth]{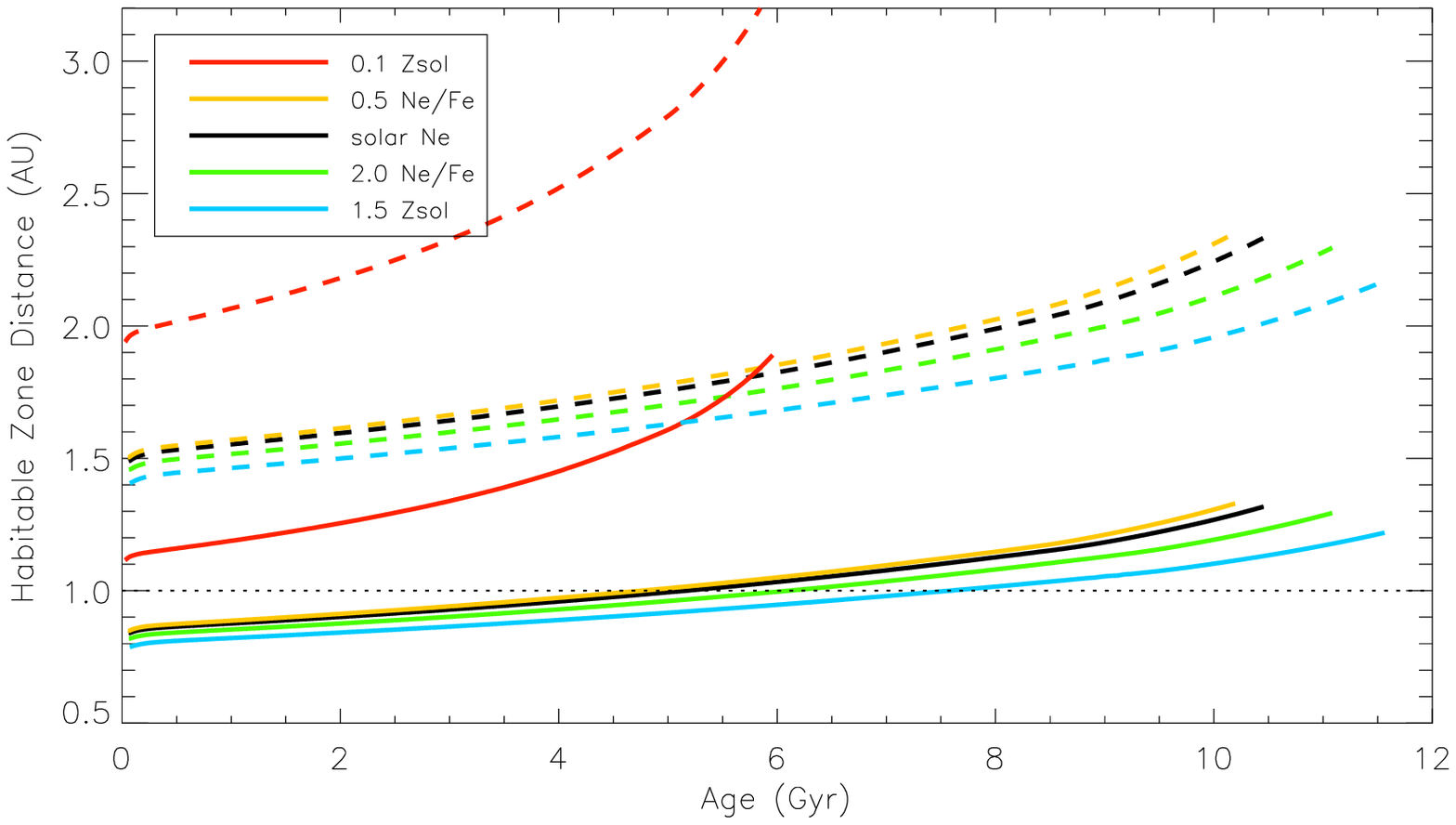}
\includegraphics[width=0.325\textwidth]{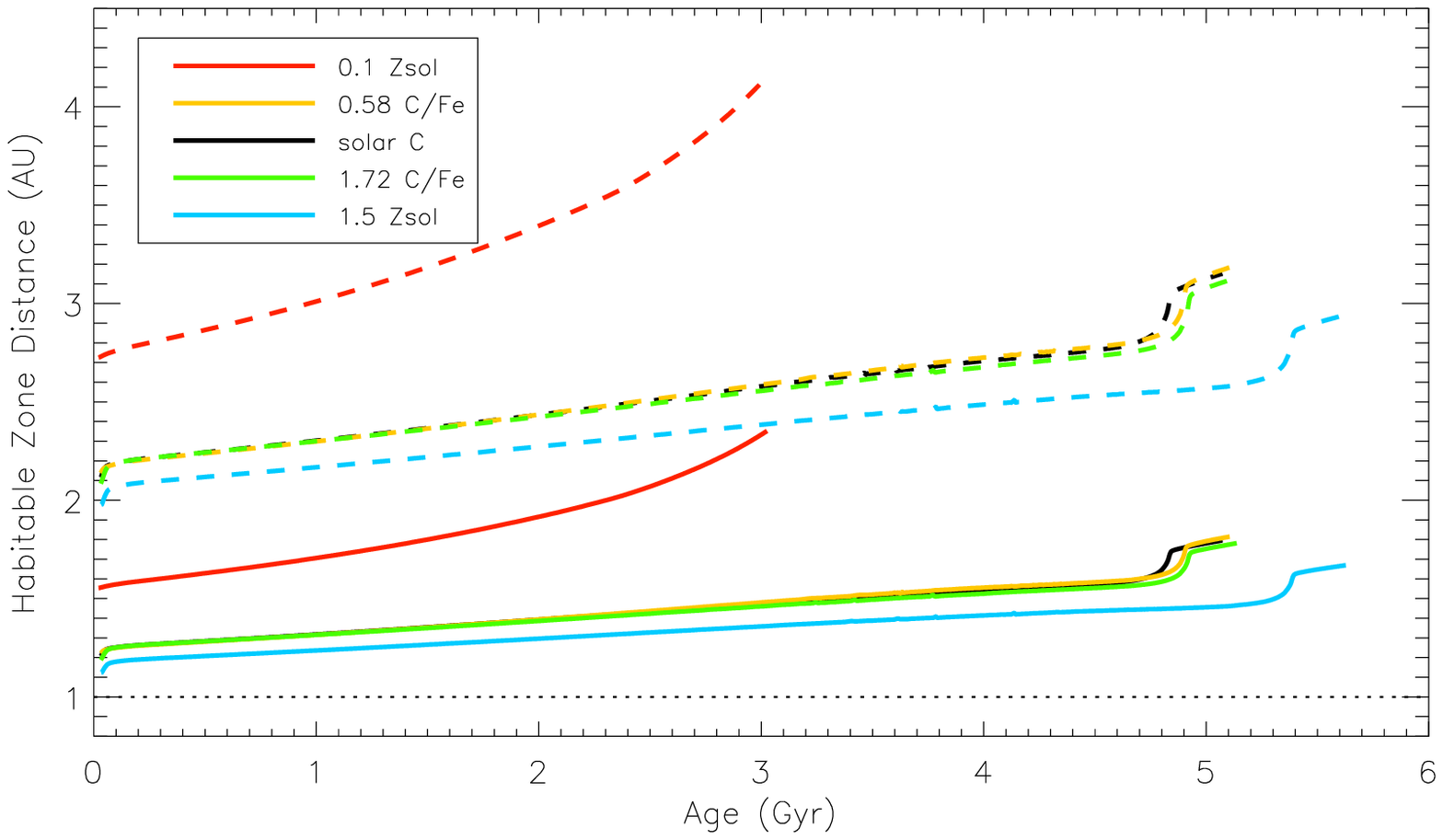}
\includegraphics[width=0.325\textwidth]{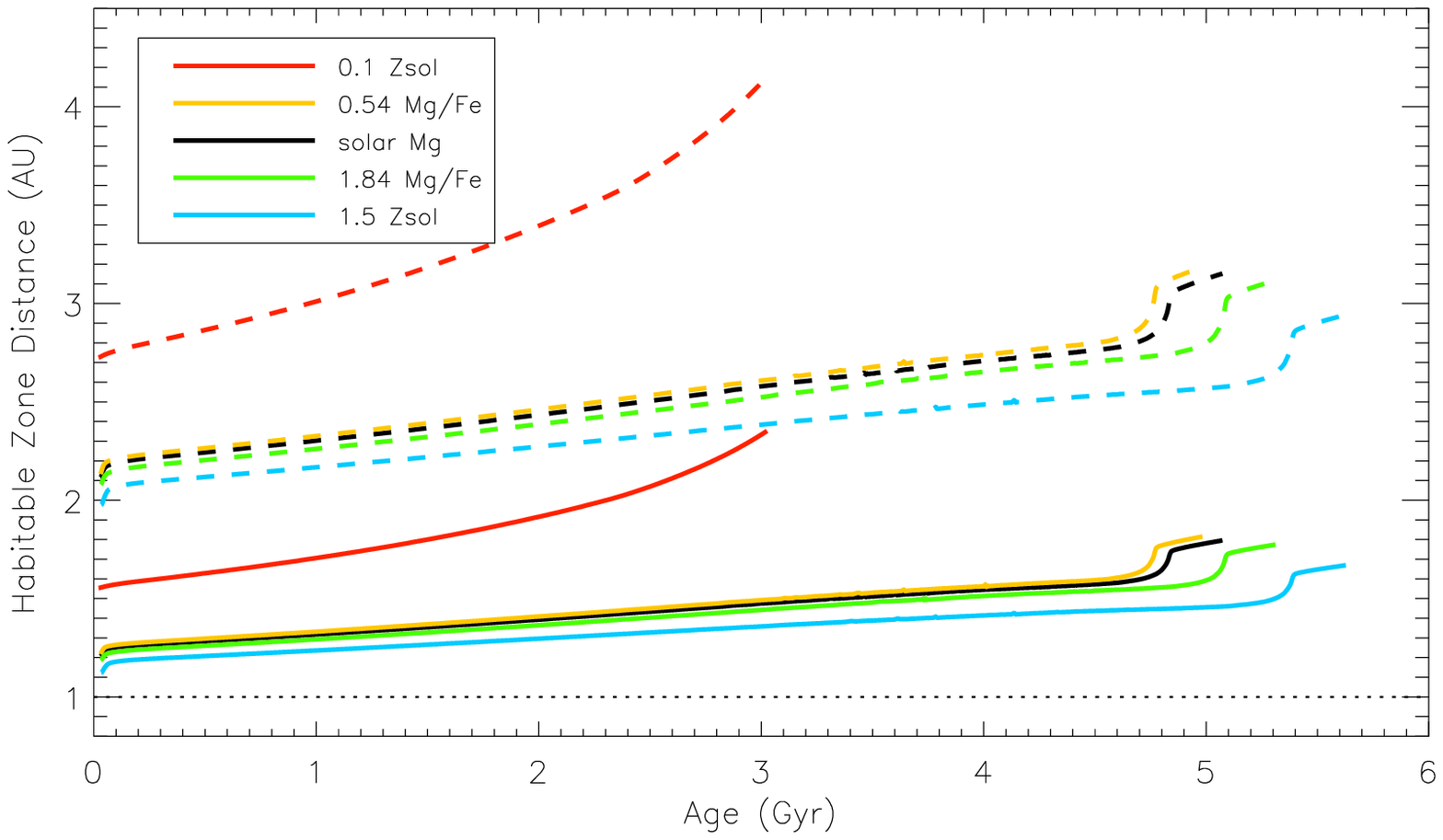}
\includegraphics[width=0.325\textwidth]{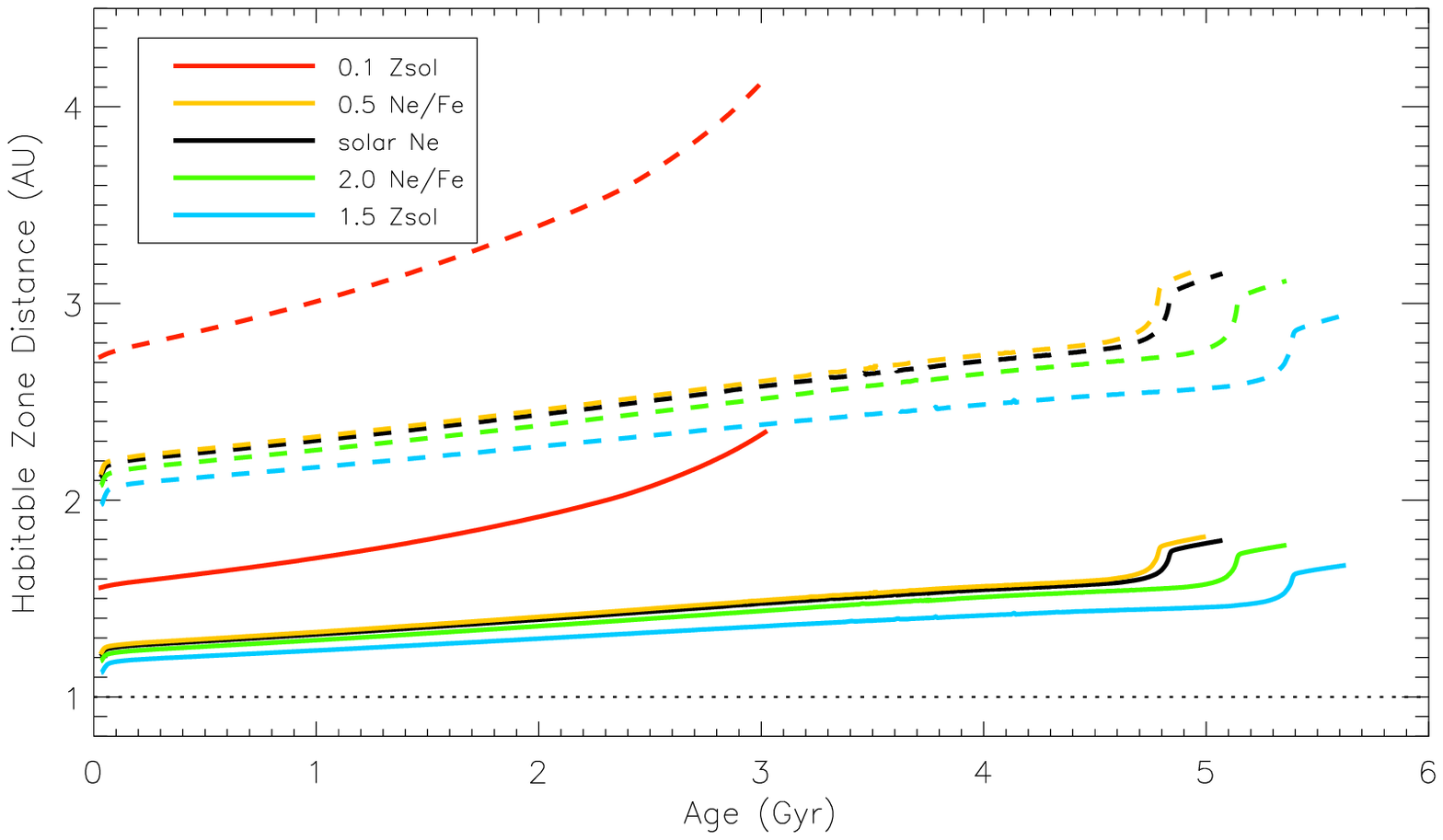}
\caption{Inner (solid) and outer (dashed) edges of the HZ for 0.5 M\sol (top), 1 M\sol (middle), and 1.2 M\sol (bottom), for three elements: carbon (left), magnesium (middle), and neon (right). Each color represents a different abundance value of interest: black is solar, orange is for depleted elemental values (0.58 C/Fe, 0.54 Mg/Fe, 0.5 Ne/Fe), and green is for enriched elemental values (1.72 C/Fe, 1.84 Mg/Fe, 2.0 Ne/Fe). For comparison, the red lines represent 0.1 Z\sol and the blue lines represent 1.5 Z\sol. A 1 AU orbit is also indicated by the dotted line in each frame, for reference. It is clear that abundance variations within a star significantly affect MS lifetime and HZ distance.\label{hzcases_all}}
\end{figure}

A higher specific elemental abundance ratio present in the host star will generally result in a closer HZ, because the star would be less luminous and at a lower effective temperature. Additionally, a star of higher opacity will live significantly longer on the MS than a star of equal mass at lower opacity, due to the higher efficiency of radiation transport in the star. Figure~\ref{hzcases_all} shows the HZ distance as it changes with stellar age, for three stellar mass values (top is 0.5 M\sol, middle is 1.0 M\sol, bottom is 1.2 M\sol), for five different compositions at each element of interest (carbon in left column, magnesium in middle column, neon in right column). Each color represents a different abundance value: black is solar, orange is for depleted elemental values (0.58 C/Fe, 0.54 Mg/Fe, 0.5 Ne/Fe), and green is for enriched elemental values (1.72 C/Fe, 1.84 Mg/Fe, 2.0 Ne/Fe). For comparison, the red lines represent 0.1 Z\sol and the blue lines represent 1.5 Z\sol. A 1 AU orbit is also indicated by the dotted line in each frame, for reference. It is clear that abundance variations within a star significantly affect MS lifetime and HZ distance. As expected, the shortest lifetime corresponds to a star with total metallicity Z = 0.1 Z\sol and the longest lifetime corresponds to Z = 1.5 Z\sol. However, when considering only the variations in the specific elemental abundances, we see that changes in magnesium and neon make the largest difference to the evolution, followed by carbon. The total MS lifetime for a 0.5 M\sol star at end-member neon abundances (at Z\sol) varies by about 7 Gyr, which can also be determined by examining Table~\ref{tab1}. Likewise, the MS varies by about 4 Gyr for magnesium, and only about 1 Gyr for carbon.

\begin{deluxetable}{lcccccccc}
\tabletypesize{\scriptsize}
\tablecaption{$\Delta$AU for each mass and end-member compositions at inner HZ limit (RGH).\label{tab6}}
\tablehead{
  \colhead{Composition}
& \colhead{0.5 M\sol}
& \colhead{0.6 M\sol}
& \colhead{0.7 M\sol}
& \colhead{0.8 M\sol}
& \colhead{0.9 M\sol}
& \colhead{1.0 M\sol}
& \colhead{1.1 M\sol}
& \colhead{1.2 M\sol}
}
\startdata
0.1 $Z\sol$, 0.58 C/Fe\sol  &  0.468  &  0.596  &  0.673  &  0.712  &  0.756  &  0.791  &  0.798  &  0.818 \\
0.1 $Z\sol$, C/Fe\sol  &  0.463  &  0.589  &  0.666  &  0.701  &  0.744  &  0.775  &  0.782  &  0.798 \\
0.1 $Z\sol$, 1.72 C/Fe\sol  &  0.455  &  0.579  &  0.655  &  0.686  &  0.725  &  0.756  &  0.761  &  0.771 \\
$Z\sol$, 0.58 C/Fe\sol  & 0.271  &  0.367  &  0.452  &  0.494  &  0.478  &  0.487  &  0.497  &  0.593 \\
$Z\sol$, C/Fe\sol  &  0.261  &  0.357  &  0.442  &  0.484  &  0.467  &  0.478  &  0.495  &  0.589 \\
$Z\sol$, 1.72 C/Fe\sol  &  0.247  &  0.342  &  0.426  &  0.467  &  0.449  &  0.463  &  0.486  &  0.593 \\
1.5 $Z\sol$, 0.58 C/Fe\sol  &  0.231  &  0.320  &  0.403  &  0.449  &  0.435  &  0.440  &  0.451  &  0.541 \\
1.5 $Z\sol$, C/Fe\sol  &  0.222  &  0.310  &  0.391  &  0.436  &  0.421  &  0.431  &  0.449  &  0.546 \\
1.5 $Z\sol$, 1.72 C/Fe\sol  &  0.213  &  0.297  &  0.375  &  0.416  &  0.400  &  0.415  &  0.441  &  0.560 \\\\
0.1 $Z\sol$, 0.54 Mg/Fe\sol  &  0.465  &  0.590  &  0.666  &  0.703  &  0.743  &  0.774  &  0.783  &  0.8015 \\
0.1 $Z\sol$, Mg/Fe\sol  &  0.463  &  0.589  &  0.666  &  0.701  &  0.744  &  0.775  &  0.782  &  0.798 \\
0.1 $Z\sol$, 1.84 Mg/Fe\sol  &  0.459  &  0.585  &  0.663  &  0.699  &  0.741  &  0.775  &  0.784  &  0.8017 \\
$Z\sol$, 0.54 Mg/Fe\sol  & 0.264  &  0.362  &  0.446  &  0.484  &  0.468  &  0.480  &  0.496  &  0.596 \\
$Z\sol$, Mg/Fe\sol  &  0.261  &  0.357  &  0.442  &  0.484  &  0.467  &  0.478  &  0.495  &  0.589 \\
$Z\sol$, 1.84 Mg/Fe\sol  &  0.272  &  0.349  &  0.435  &  0.482  &  0.465  &  0.475  &  0.492  &  0.589 \\
1.5 $Z\sol$, 0.54 Mg/Fe\sol  &  0.225  &  0.314  &  0.395  &  0.438  &  0.422  &  0.433  &  0.451  &  0.554 \\
1.5 $Z\sol$, Mg/Fe\sol  &  0.222  &  0.310  &  0.391  &  0.436  &  0.421  &  0.431  &  0.449  &  0.546 \\
1.5 $Z\sol$, 1.84 Mg/Fe\sol  &  0.216  &  0.302  &  0.384  &  0.449  &  0.420  &  0.428  &  0.445  &  0.538 \\\\
0.1 $Z\sol$, 0.5 Ne/Fe\sol  &  0.467  &  0.592  &  0.667  &  0.703  &  0.743  &  0.775  &  0.780  &  0.798 \\
0.1 $Z\sol$, Ne/Fe\sol  &  0.463  &  0.589  &  0.666  &  0.701  &  0.744  &  0.775  &  0.782  &  0.798 \\
0.1 $Z\sol$, 2.0 Ne/Fe\sol  &  0.455  &  0.582  &  0.663  &  0.698  &  0.740  &  0.776  &  0.787  &  0.805 \\
$Z\sol$, 0.5 Ne/Fe\sol  &  0.266  &  0.363  &  0.447  &  0.484  &  0.468  &  0.481  &  0.500  &  0.598 \\
$Z\sol$, Ne/Fe\sol  &  0.261  &  0.357  &  0.442  &  0.484  &  0.467  &  0.478  &  0.495  &  0.589 \\
$Z\sol$, 2.0 Ne/Fe\sol  &  0.251  &  0.345  &  0.433  &  0.483  &  0.465  &  0.474  &  0.491  &  0.589 \\
1.5 $Z\sol$, 0.5 Ne/Fe\sol  &  0.227  &  0.315  &  0.396  &  0.438  &  0.423  &  0.434  &  0.451  &  0.555 \\
1.5 $Z\sol$, Ne/Fe\sol  &  0.222  &  0.310  &  0.391  &  0.436  &  0.421  &  0.431  &  0.449  &  0.546 \\
1.5 $Z\sol$, 2.0 Ne/Fe\sol  &  0.214  &  0.300  &  0.382  &  0.448  &  0.419  &  0.427  &  0.445  &  0.538 \\
\enddata
\end{deluxetable}

\begin{deluxetable}{lcccccccc}
\tabletypesize{\scriptsize}
\tablecaption{$\Delta$AU for each mass and end-member compositions at outer HZ limit (MaxGH).\label{tab7}}
\tablehead{
  \colhead{Composition}
& \colhead{0.5 M\sol}
& \colhead{0.6 M\sol}
& \colhead{0.7 M\sol}
& \colhead{0.8 M\sol}
& \colhead{0.9 M\sol}
& \colhead{1.0 M\sol}
& \colhead{1.1 M\sol}
& \colhead{1.2 M\sol}
}
\startdata
0.1 $Z\sol$, 0.58 C/Fe\sol  &  0.816  &  1.032  &  1.166  &  1.240  &  1.315  &  1.372  &  1.396  &  1.469 \\
0.1 $Z\sol$, C/Fe\sol  &  0.808  &  1.021  &  1.153  &  1.221  &  1.294  &  1.345  &  1.366  &  1.428 \\
0.1 $Z\sol$, 1.72 C/Fe\sol  &  0.796  &  1.004  &  1.133  &  1.194  &  1.261  &  1.312  &  1.326  &  1.372 \\
$Z\sol$, 0.58 C/Fe\sol  & 0.490  &  0.659  &  0.801  &  0.864  &  0.840  &  0.858  &  0.877  &  1.045 \\
$Z\sol$, C/Fe\sol  &  0.473  &  0.642  &  0.784  &  0.847  &  0.822  &  0.843  &  0.873  &  1.038 \\
$Z\sol$, 1.72 C/Fe\sol  &  0.451  &  0.616  &  0.757  &  0.820  &  0.792  &  0.818  &  0.859  &  1.047 \\
1.5 $Z\sol$, 0.58 C/Fe\sol  &  0.423  &  0.579  &  0.720  &  0.792  &  0.767  &  0.779  &  0.796  &  0.957 \\
1.5 $Z\sol$, C/Fe\sol  &  0.406  &  0.561  &  0.700  &  0.770  &  0.744  &  0.764  &  0.794  &  0.965 \\
1.5 $Z\sol$, 1.72 C/Fe\sol  &  0.389  &  0.539  &  0.673  &  0.736  &  0.708  &  0.737  &  0.779  &  0.991 \\\\
0.1 $Z\sol$, 0.54 Mg/Fe\sol  &  0.811  &  1.023  &  1.154  &  1.224  &  1.292  &  1.343  &  1.369  &  1.437 \\
0.1 $Z\sol$, Mg/Fe\sol  &  0.808  &  1.021  &  1.153  &  1.221  &  1.294  &  1.345  &  1.366  &  1.428 \\
0.1 $Z\sol$, 1.84 Mg/Fe\sol  &  0.802  &  1.015  &  1.148  &  1.218  &  1.290  &  1.345  &  1.367  &  1.428 \\
$Z\sol$, 0.54 Mg/Fe\sol  & 0.480  &  0.649  &  0.790  &  0.848  &  0.824  &  0.847  &  0.875  &  1.051 \\
$Z\sol$, Mg/Fe\sol  &  0.473  &  0.642  &  0.784  &  0.847  &  0.822  &  0.843  &  0.873  &  1.038 \\
$Z\sol$, 1.84 Mg/Fe\sol  &  0.495  &  0.628  &  0.773  &  0.846  &  0.820  &  0.838  &  0.868  &  1.038 \\
1.5 $Z\sol$, 0.54 Mg/Fe\sol  &  0.412  &  0.569  &  0.707  &  0.772  &  0.746  &  0.766  &  0.797  &  0.978 \\
1.5 $Z\sol$, Mg/Fe\sol  &  0.406  &  0.561  &  0.700  &  0.770  &  0.744  &  0.764  &  0.794  &  0.965 \\
1.5 $Z\sol$, 1.84 Mg/Fe\sol  &  0.395  &  0.548  &  0.688  &  0.794  &  0.743  &  0.759  &  0.787  &  0.951 \\\\
0.1 $Z\sol$, 0.5 Ne/Fe\sol  &  0.814  &  1.026  &  1.154  &  1.223  &  1.291  &  1.345  &  1.366  &  1.434 \\
0.1 $Z\sol$, Ne/Fe\sol  &  0.808  &  1.021  &  1.153  &  1.221  &  1.294  &  1.345  &  1.366  &  1.428 \\
0.1 $Z\sol$, 2.0 Ne/Fe\sol  &  0.796  &  1.011  &  1.147  &  1.216  &  1.289  &  1.347  &  1.370  &  1.428 \\
$Z\sol$, 0.5 Ne/Fe\sol  &  0.482  &  0.651  &  0.791  &  0.846  &  0.824  &  0.848  &  0.875  &  1.053 \\
$Z\sol$, Ne/Fe\sol  &  0.473  &  0.642  &  0.784  &  0.847  &  0.822  &  0.843  &  0.873  &  1.038 \\
$Z\sol$, 2.0 Ne/Fe\sol  &  0.456  &  0.622  &  0.771  &  0.848  &  0.820  &  0.838  &  0.868  &  1.040 \\
1.5 $Z\sol$, 0.5 Ne/Fe\sol  &  0.414  &  0.571  &  0.707  &  0.771  &  0.747  &  0.767  &  0.797  &  0.981 \\
1.5 $Z\sol$, Ne/Fe\sol  &  0.406  &  0.561  &  0.700  &  0.770  &  0.744  &  0.764  &  0.794  &  0.965 \\
1.5 $Z\sol$, 2.0 Ne/Fe\sol  &  0.391  &  0.544  &  0.686  &  0.793  &  0.741  &  0.758  &  0.788  &  0.952 \\
\enddata
\end{deluxetable}

\begin{deluxetable}{lcccccccc}
\tabletypesize{\scriptsize}
\tablecaption{Fraction (\%) of time spent in CHZ${_2}$ vs. the entire MS (for carbon).\label{tab8}}
\tablehead{
  \colhead{Composition}
& \colhead{0.5 M\sol}
& \colhead{0.6 M\sol}
& \colhead{0.7 M\sol}
& \colhead{0.8 M\sol}
& \colhead{0.9 M\sol}
& \colhead{1.0 M\sol}
& \colhead{1.1 M\sol}
& \colhead{1.2 M\sol}
}
\startdata
0.1 $Z\sol$, 0.58 C/Fe\sol  &  80.88  &  78.31  &  74.16  &  68.68  &  63.16  &  56.83  &  49.88  &  0.00\tablenotemark{a} \\
0.1 $Z\sol$, C/Fe\sol  &  80.84  &  78.84  &  74.52  &  69.42  &  63.85  &  57.31  &  50.65  &  0.00 \\
0.1 $Z\sol$, 1.72 C/Fe\sol  &  81.12  &  79.19  &  75.43  &  70.11  &  64.90  &  58.76  &  51.95  &  0.00 \\
$Z\sol$, 0.58 C/Fe\sol  & 86.48  &  85.84  &  85.19  &  83.05  &  79.41  &  75.60  &  70.20  &  61.06 \\
$Z\sol$, C/Fe\sol  &  86.90  &  86.22  &  85.46  &  83.25  &  79.43  &  75.39  &  69.82  &  61.17 \\
$Z\sol$, 1.72 C/Fe\sol  &  87.51  &  86.88  &  86.15  &  83.87  &  79.91  &  75.31  &  69.44  &  60.76 \\
1.5 $Z\sol$, 0.58 C/Fe\sol  &  87.38  &  86.86  &  86.33  &  84.88  &  81.48  &  77.68  &  72.62  &  63.85 \\
1.5 $Z\sol$, C/Fe\sol  &  87.81  &  87.29  &  86.69  &  85.08  &  81.52  &  77.35  &  71.91  &  63.53 \\
1.5 $Z\sol$, 1.72 C/Fe\sol  &  88.11  &  87.87  &  85.19  &  85.45  &  81.82  &  77.09  &  71.00  &  61.94 \\
\enddata
\tablenotetext{a}{No orbits are continually habitable for 2 Gyr as a result of the short MS lifetime.}
\end{deluxetable}

\subsection{Continuously Habitable Zones}

It should now be abundantly clear that it is an extremely useful pursuit to quantify how long any given planet would remain in a star's HZ as a function of its orbital distance; however, the instantaneous habitability of a planet alone is insufficient to determine the likelihood that it actually hosts extant life, or whether any life present would even be detectable. Groups at the Virtual Planet Laboratory at the University of Washington have worked on this problem from the perspective of viewing the Earth as an exoplanet, in order to determine the current technological limits of what ``biosignatures'' might be measurable in the atmospheres of real exoplanets (e.g. \citet{har15,kriss16}). This kind of information plays an integral role in determining how we think about habitability; indeed, with more sophisticated planetary atmosphere models and a broader understanding of what might be directly observable about them \citep{kast14}, we will have a better idea of how to apply the data from our stellar evolution tracks to paint a more complete picture of HZ evolution around different types of stars. 

\begin{figure}
\centering
\includegraphics[width=0.4\textwidth]{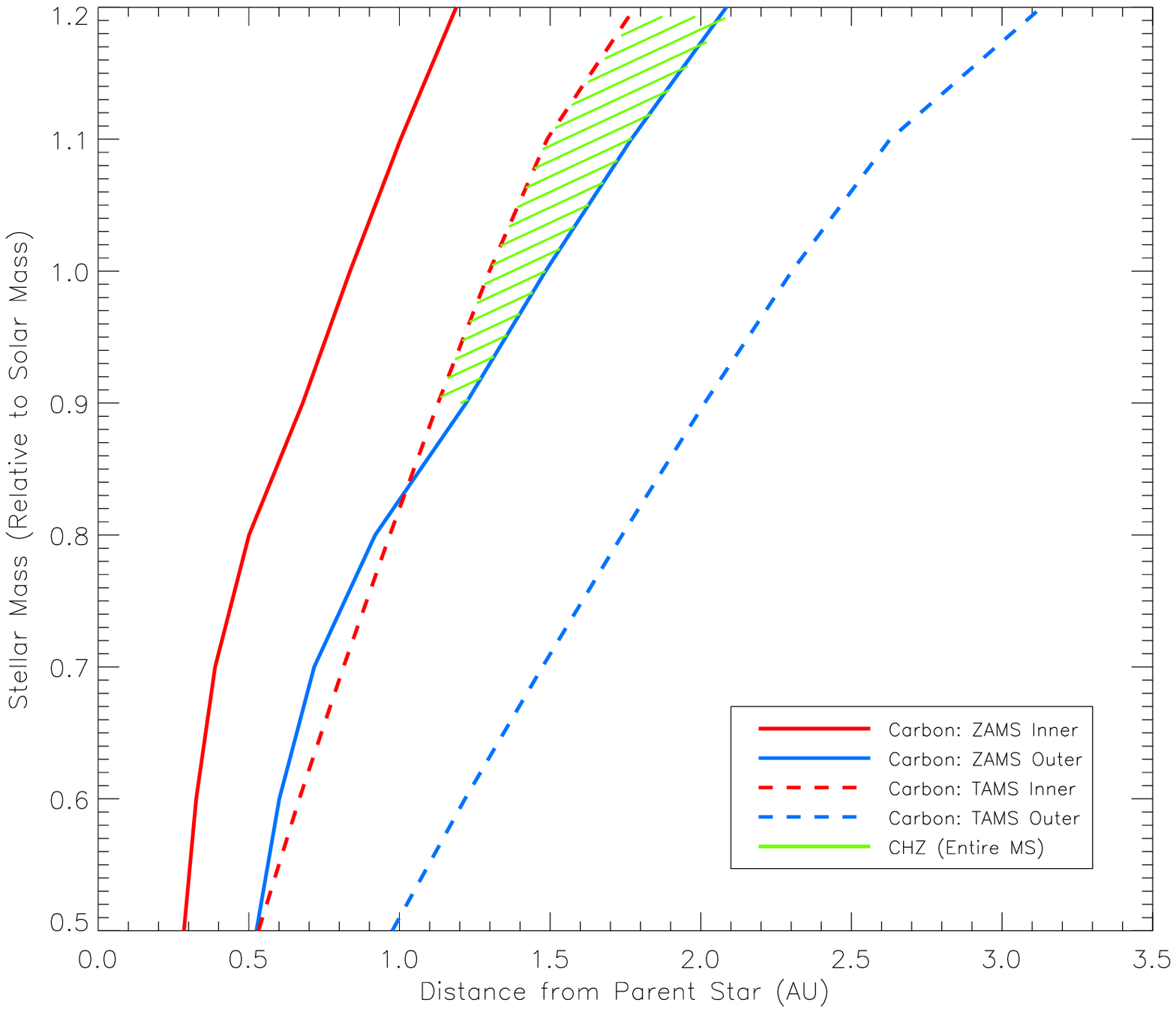}
\includegraphics[width=0.4\textwidth]{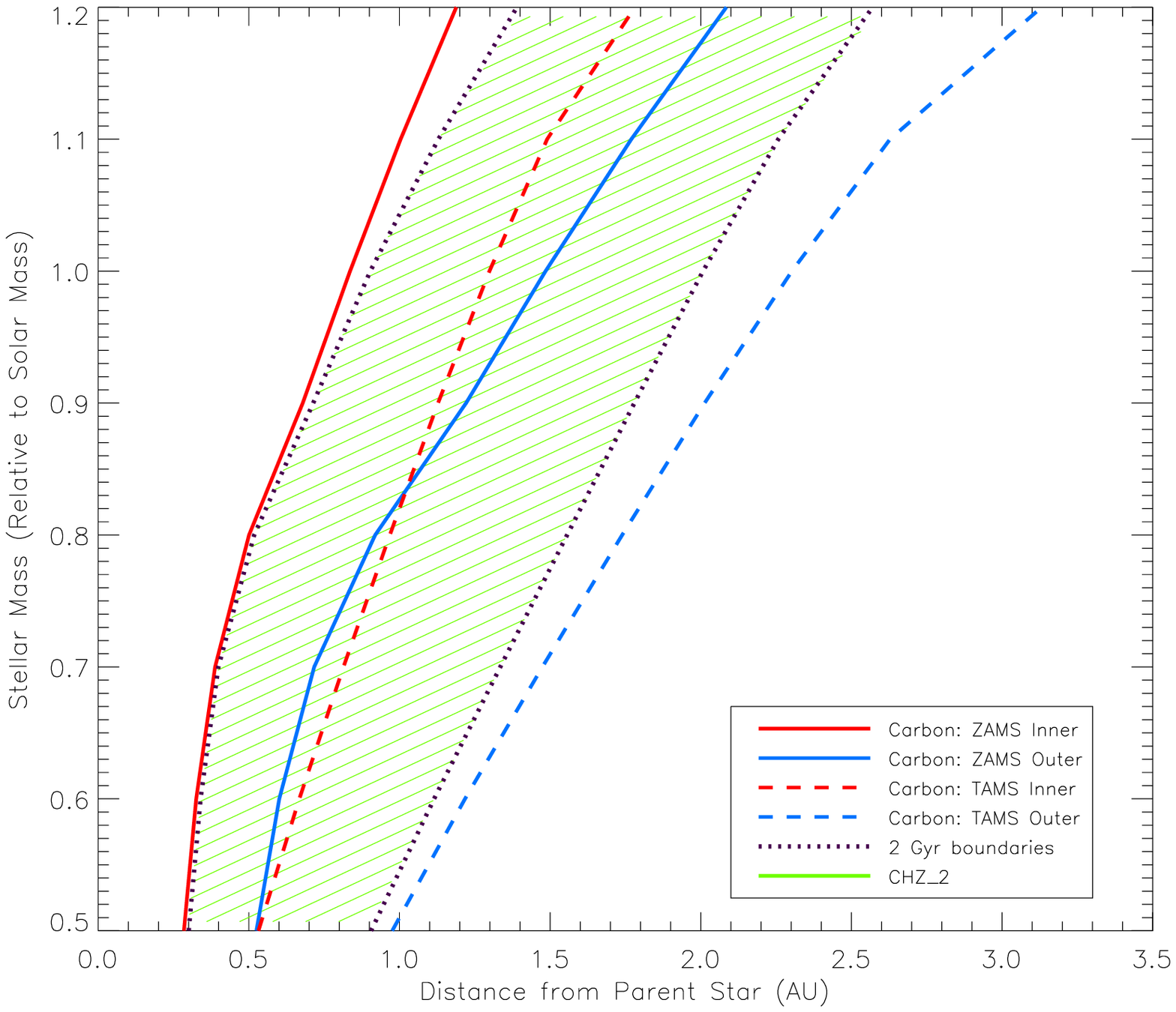}
\includegraphics[width=0.4\textwidth]{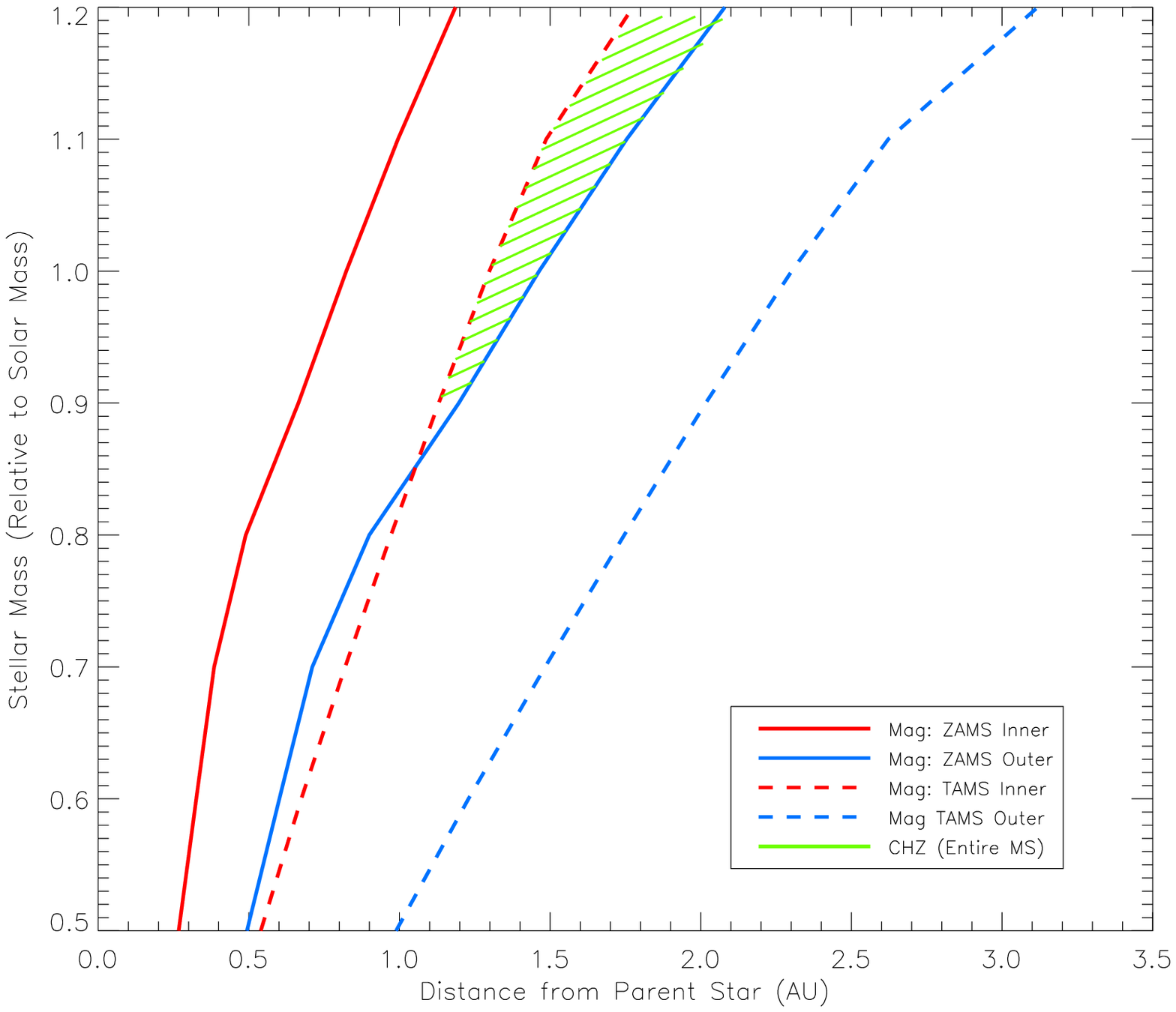}
\includegraphics[width=0.4\textwidth]{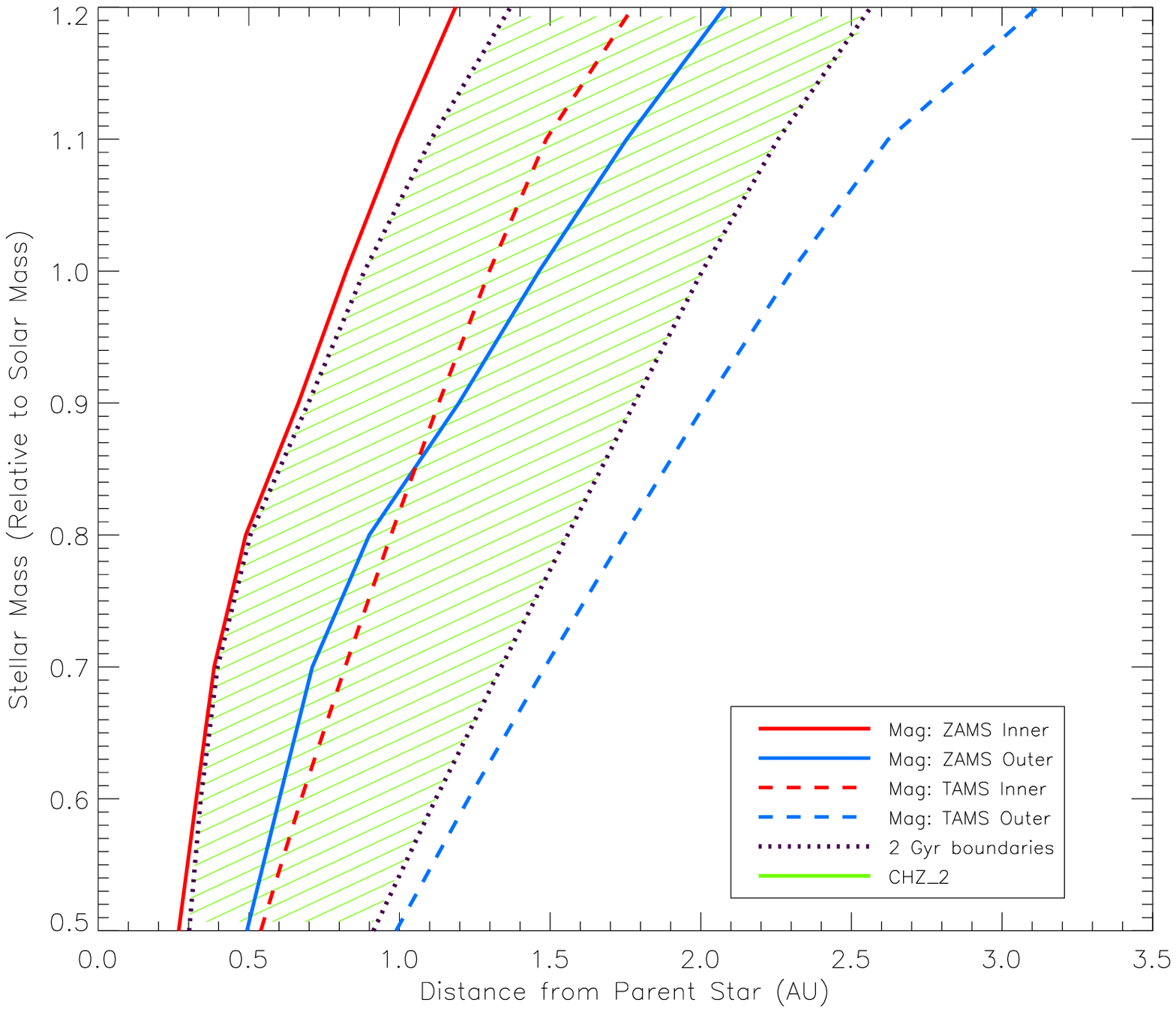}
\includegraphics[width=0.4\textwidth]{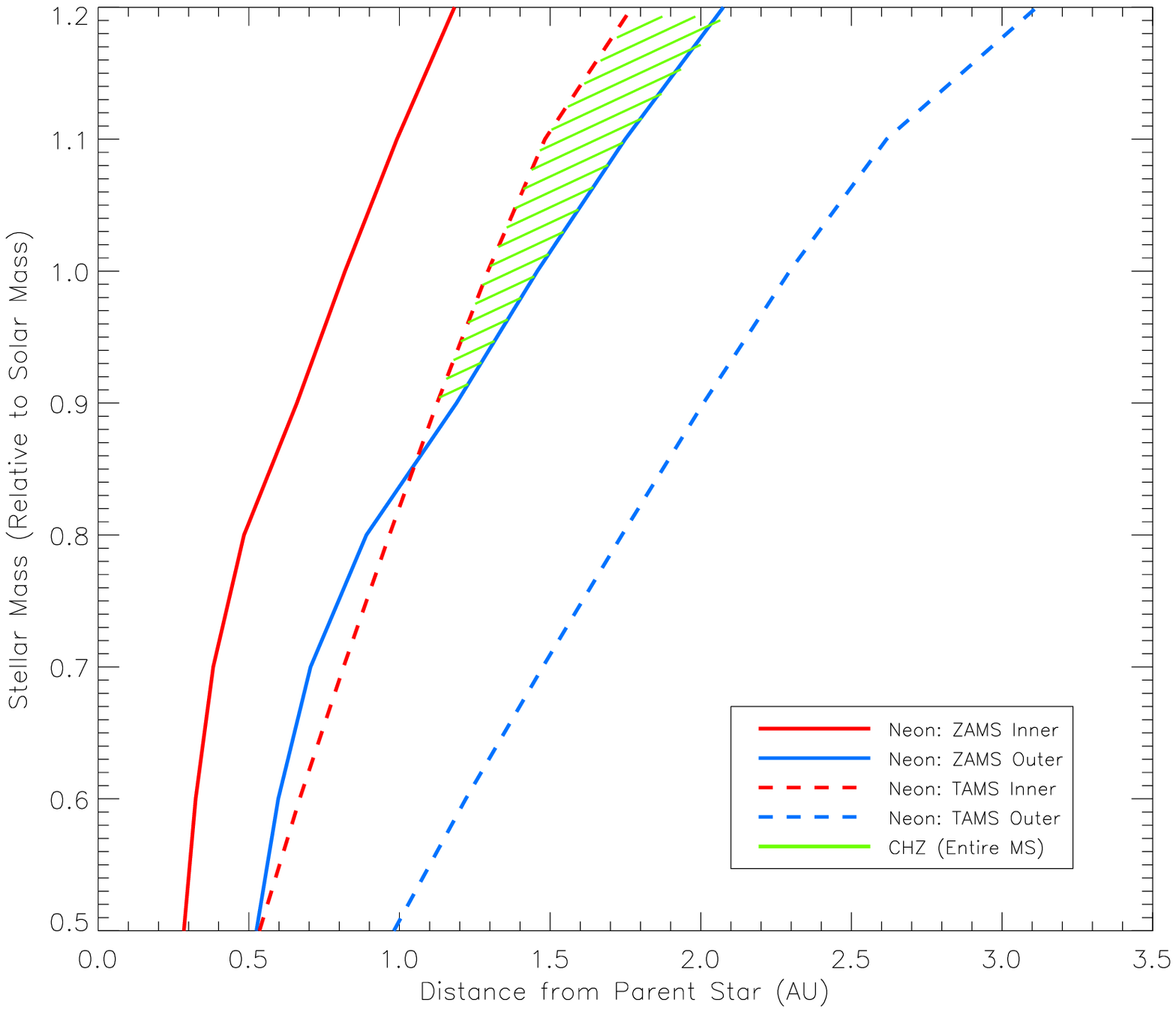}
\includegraphics[width=0.4\textwidth]{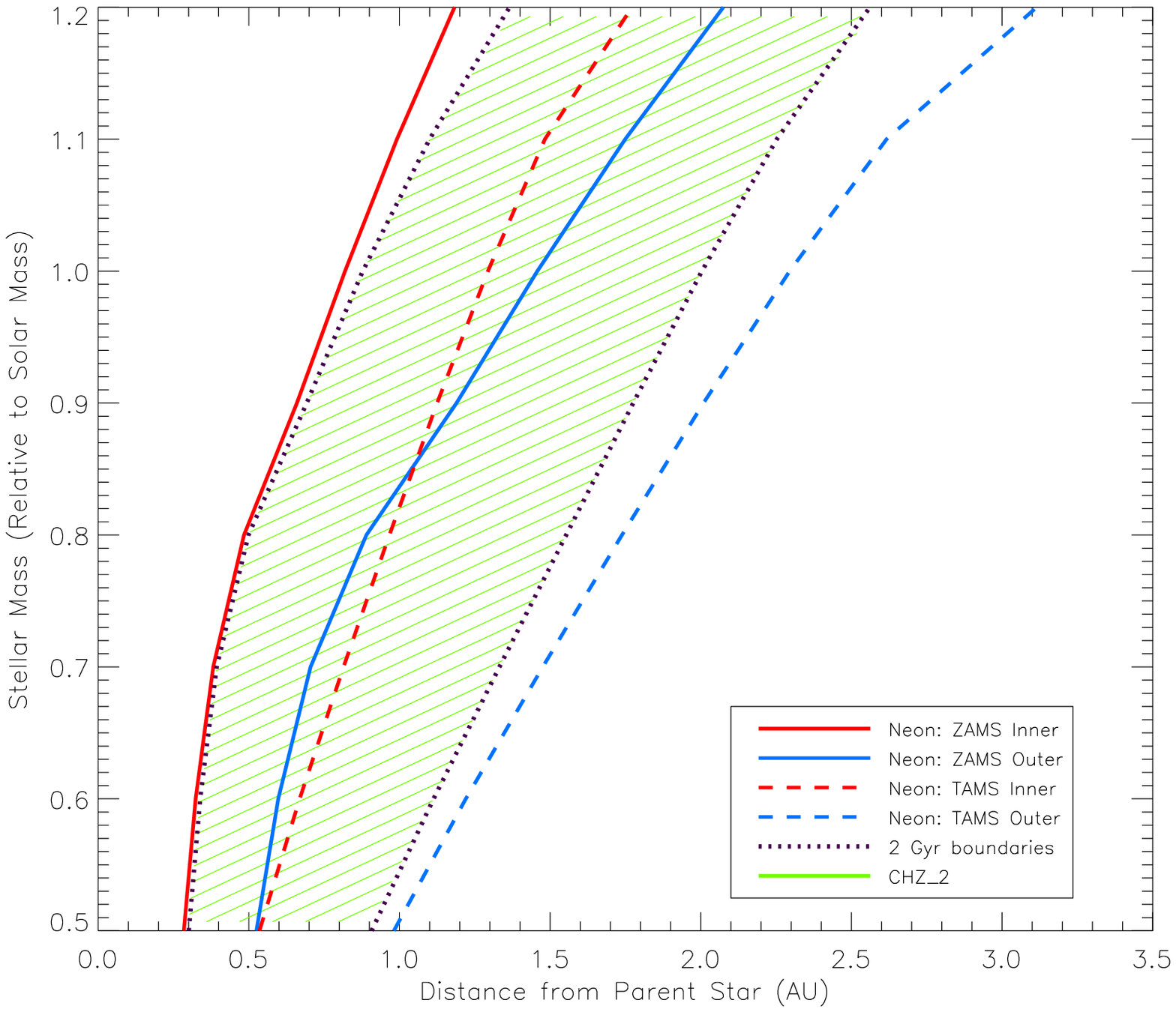}
\caption{Inner (red) and outer (blue) boundaries of the HZ at the ZAMS (solid) and TAMS (dashed). This is for stars at each mass in our range, at a composition of solar metallicity and enriched elemental values (carbon/top, magnesium/middle, neon/bottom). In the left column, the green shaded region is the CHZ, where an orbit would remain in the HZ for the star's entire MS lifetime. In the right column, the inner edge 2 Gyr after the ZAMS and the outer edge 2 Gyr before the TAMS are indicated by dotted purple lines, and the shaded region is the CHZ$_2$, in which an orbiting planet would remain in the HZ for at least 2 Gyr. For conservative HZ limits (RGH and MaxGH), low-mass stars have no CHZ, and the fraction of the total habitable orbits in the CHZ$_2$ is higher.\label{chz_all}}
\end{figure}

In T15, our initial goal was to estimate a continuously habitable zone (CHZ) for each star in our catalog, which would simply include a range of orbital radii that remain in the HZ for the entire MS. The CHZ is rather straightforwardly defined by considering the boundary overlap between ZAMS and TAMS. The lefthand column of Figure~\ref{chz_all} shows the CHZ for stars of all masses in our grid, at a composition of solar metallicity and enriched abundance values for each element of interest. The top row is for carbon, the middle is magnesium, and the bottom is neon. It is clear that the low mass stars have no CHZ for the conservative HZ limits (at all elements), which would seem to indicate that low mass stars would have a low statistical likelihood to host a long-term habitable planet; however, that is somewhat misleading due to the extremely long MS lifetimes of low mass stars. Thus, we have defined a much more useful 2 Gyr CHZ (the CHZ$_{2}$), which is the range of orbital radii that would be continuously habitable for at least 2 billion years. We use this time because it is estimated that life on Earth took approximately 2 Gyr to produce a measurable chemical change in the atmosphere \citep{sum99,kc03,holl06}. Of course, the CHZ$_{2}$ assumes that Earth's timescale for the evolution of life with the capability to modify the entire planetary atmosphere is representative of the norm. This is not meant to imply that other suggested timescales are unreasonable (e.g. \citet{rush13}), but in order to narrow down the large pool of potentially habitable exoplanets to the ones with the highest potential for both long-term habitability $and$ detectability, there is an advantage in using Earth's history as a starting point. In addition, we provide a robust consideration of the HZ evolution because we incorporate detailed stellar properties. It is fairly clear that for this particular consideration, the variations in the elemental abundances of interest don't show a significantly large difference between them, though the difference is more pronounced when compared to the associated figures from T15 that were created for the oxygen cases.

The righthand column of Figure~\ref{chz_all} shows the orbits that remain habitable for at least 2 Gyr (CHZ$_{2}$). This is determined from the inner edge of the HZ at 2 Gyr after the beginning of the MS and the outer edge 2 Gyr before the TAMS. Now we see that the significantly longer MS lifetimes of the low-mass stars create a higher proportion of the HZ that is included in the CHZ$_2$ than for the basic CHZ.  Based on these results, we would be less likely to find a planet that has been in the CHZ for at least 2 Gyr orbiting a more massive star; at the very least, we would be less confident that a planet located outside of the CHZ$_2$ would produce detectable biosignatures than one within the CHZ$_2$. Table~\ref{tab8} shows the fraction of time a planet would spend in the CHZ$_2$ vs. time it would spend in the HZ over its entire MS lifetime. Comparing stars of interest with chemical compositions will inform which ones we should focus on in the continued search for detectable inhabited exoplanets.

As in T15, our consideration of HZ evolution and the CHZ must also address the issue of cold starts. Our discussion until now has assumed that any planets initially beyond the boundaries of the HZ could easily become habitable as soon as the host star's HZ expanded outward to engulf them; indeed, the albedos used in the planetary atmosphere models of \citet{kopp14} are relatively low, which assumes a planet could fairly easily become habitable upon entering the HZ. However, it may be unlikely that a completely frozen planet (a ``hard snowball'') entering the HZ late in the host star's MS lifetime would receive enough energy in the form of stellar radiation to reverse a global glaciation, especially if the planet harbors reflective CO$_2$ clouds \citep{calkast92,kwr93}.

\begin{figure}
\centering
\includegraphics[width=0.325\textwidth]{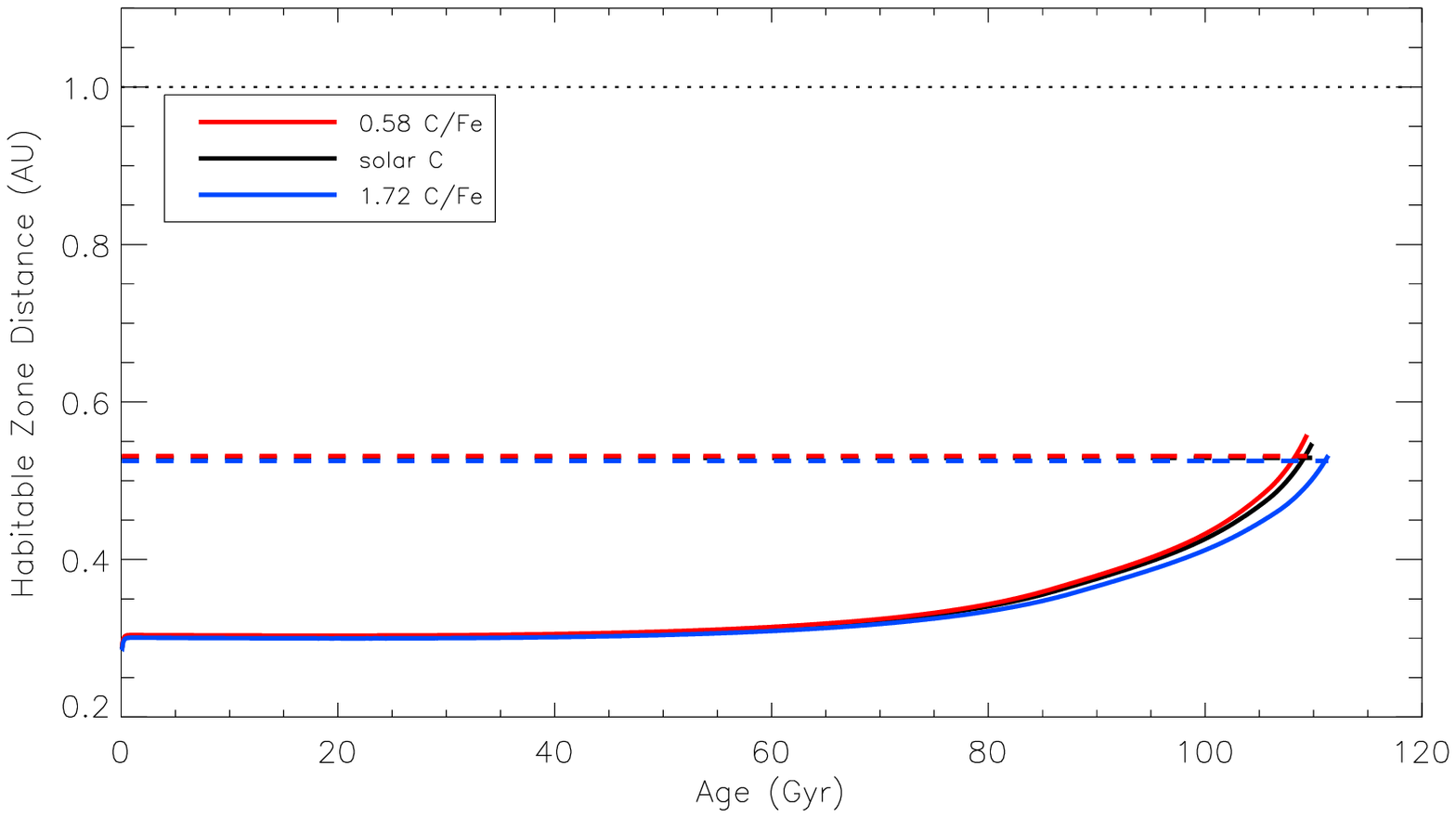}
\includegraphics[width=0.325\textwidth]{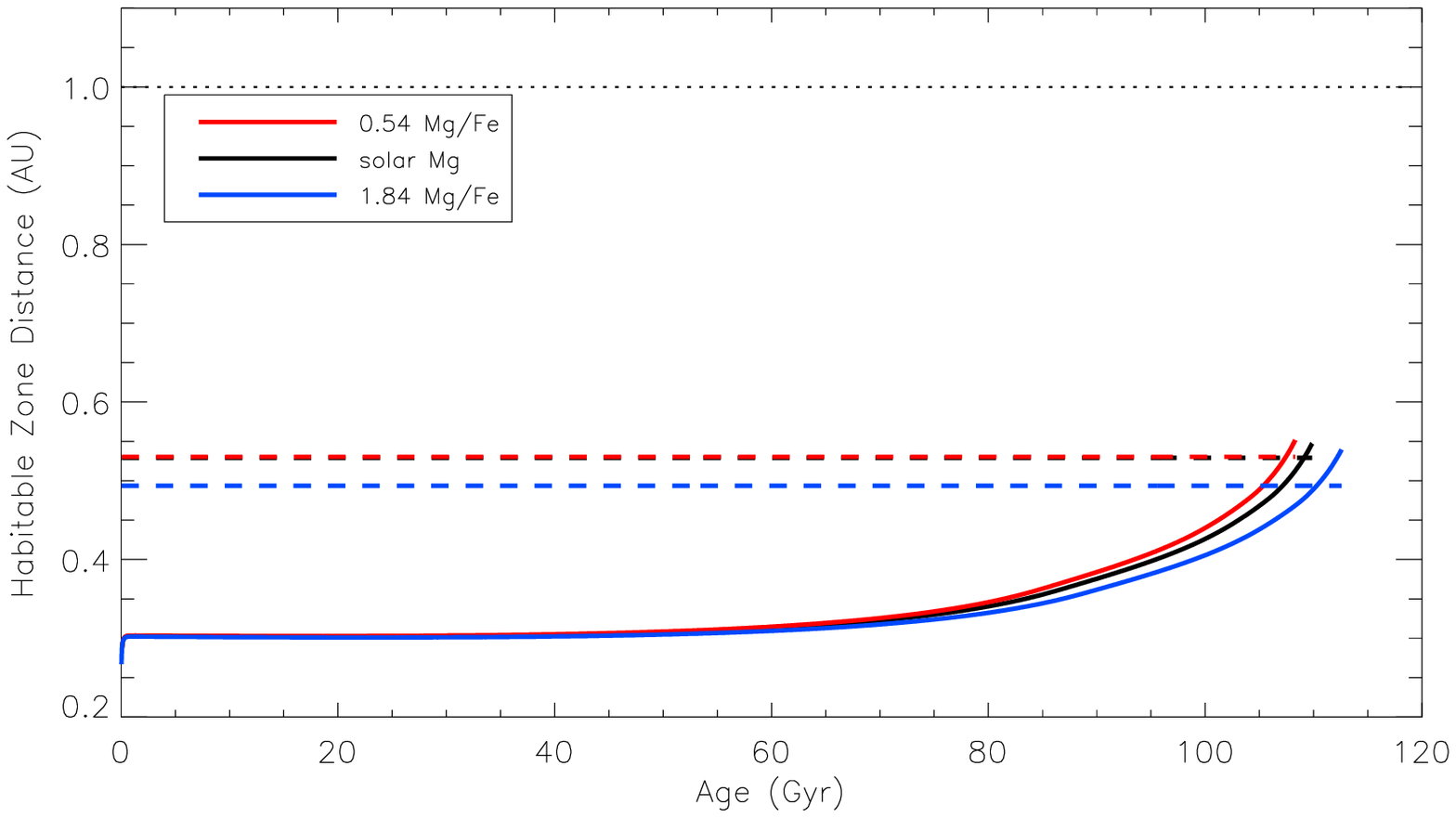}
\includegraphics[width=0.325\textwidth]{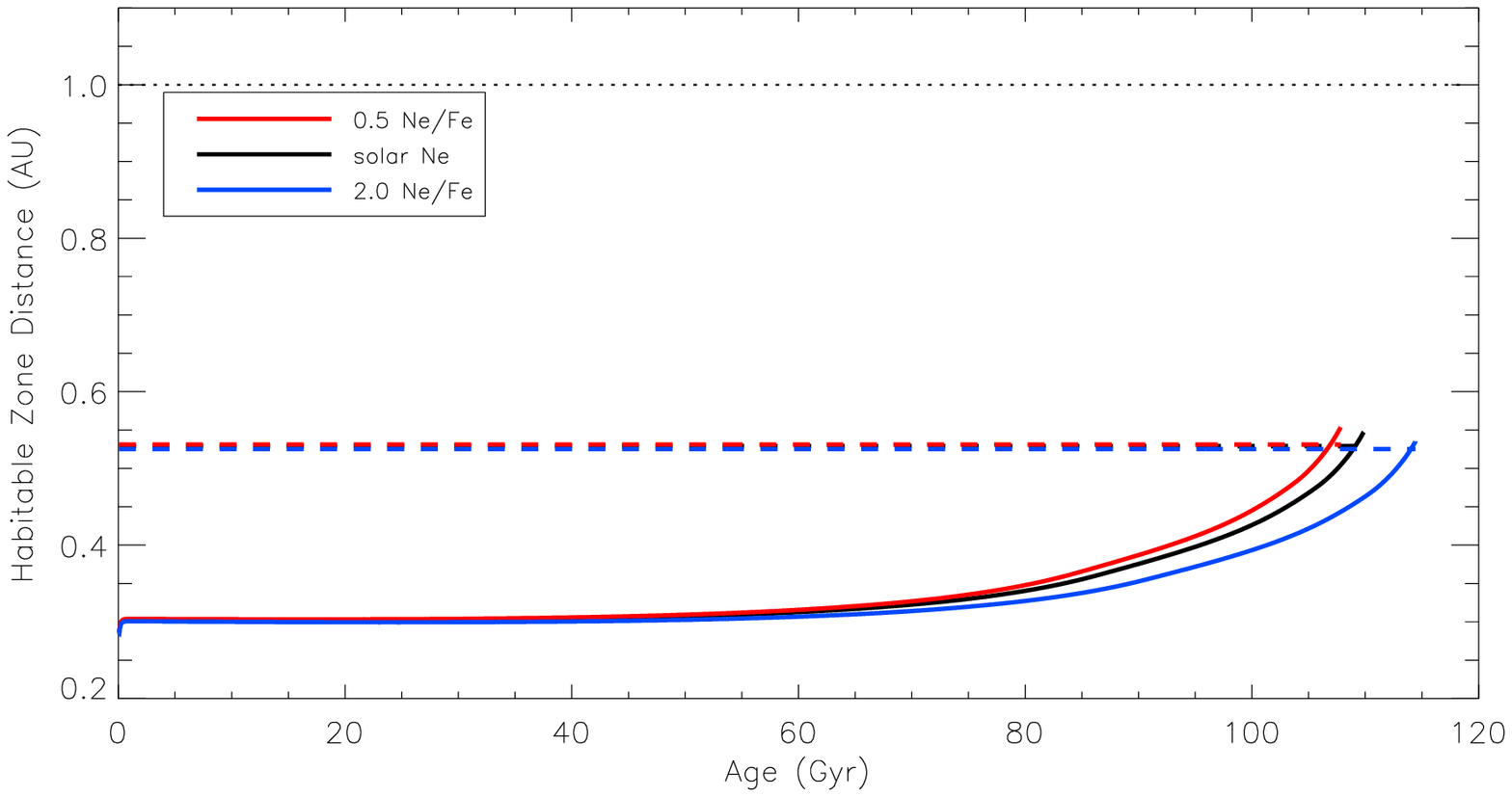}
\includegraphics[width=0.325\textwidth]{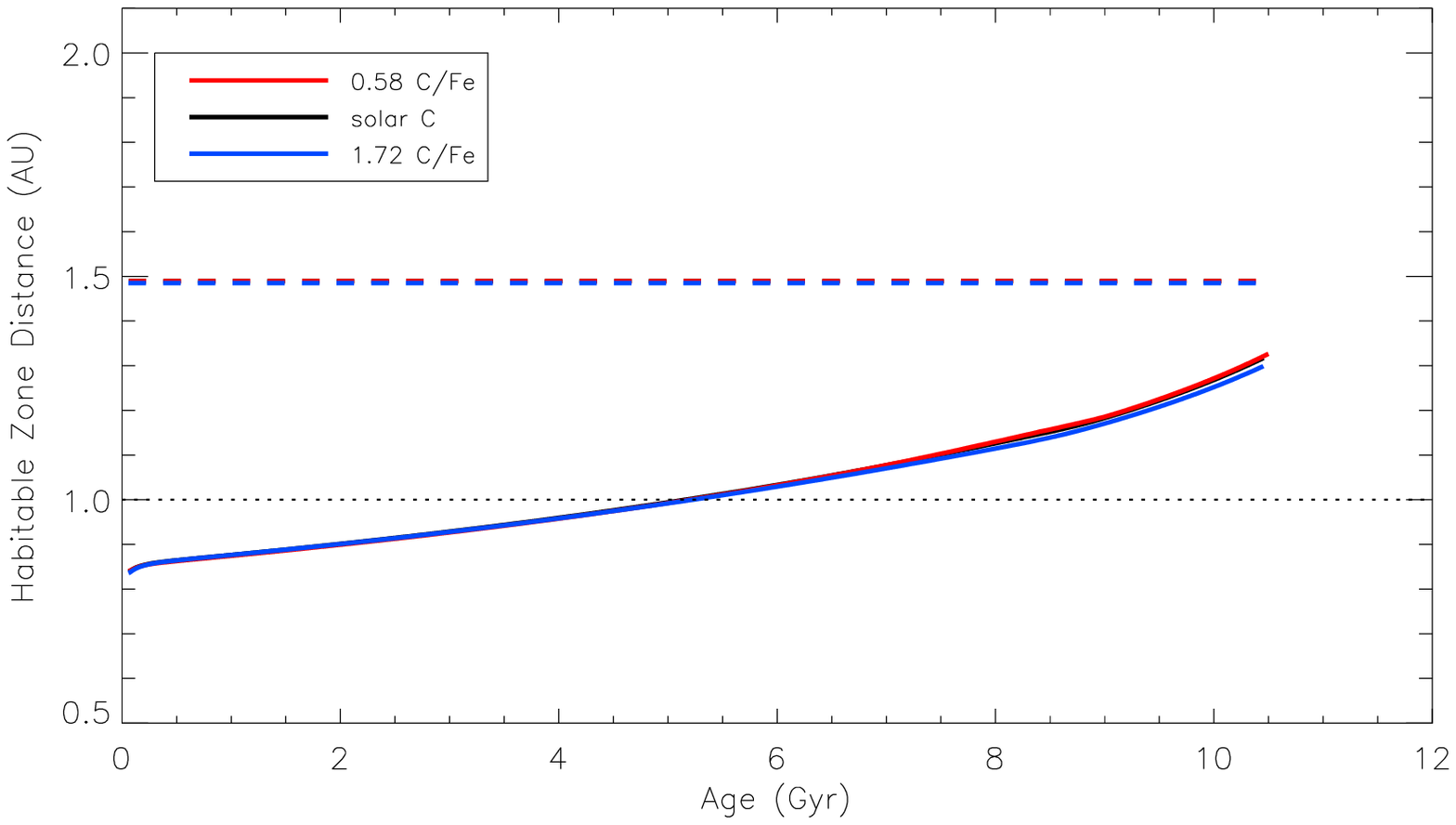}
\includegraphics[width=0.325\textwidth]{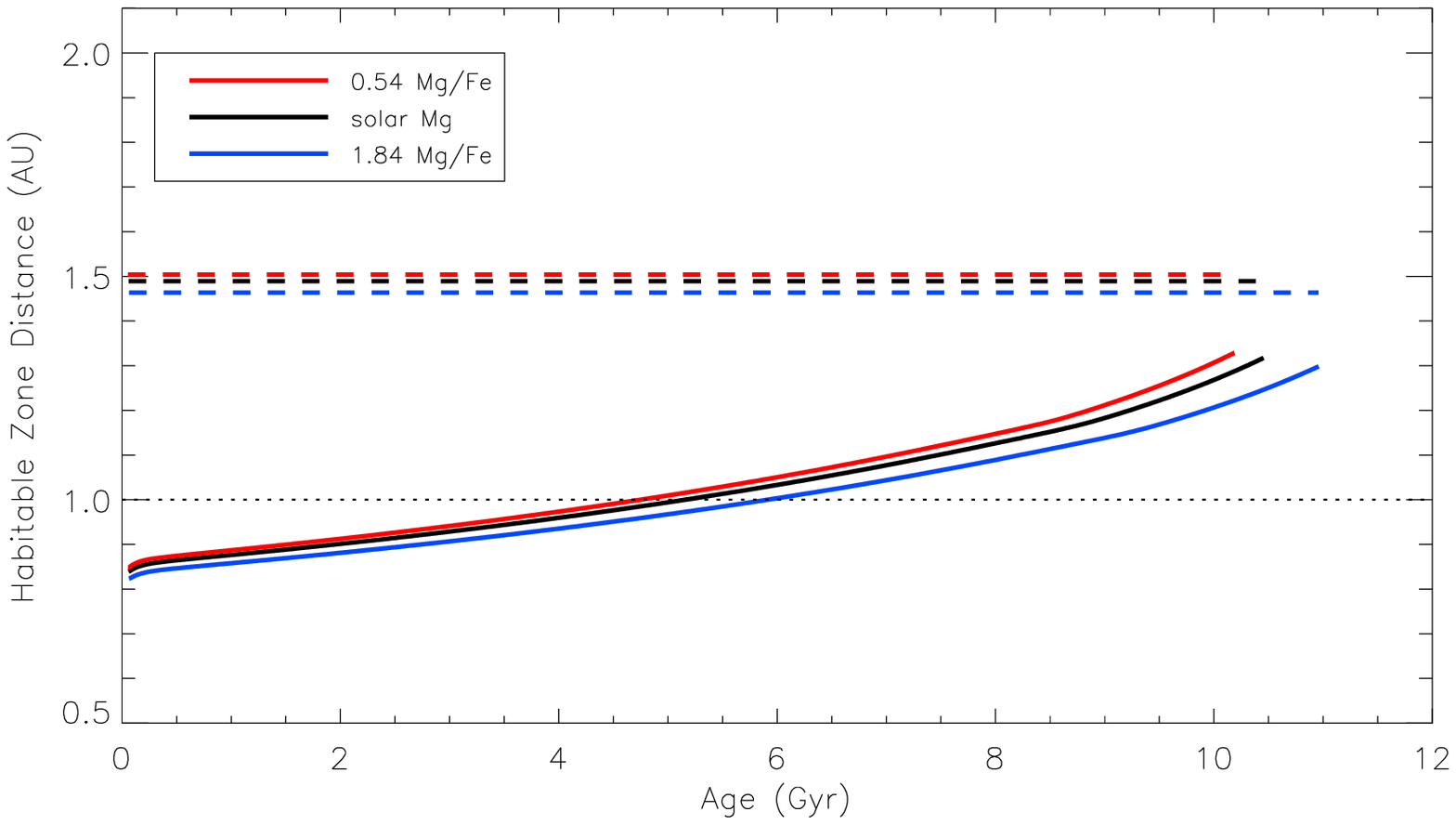}
\includegraphics[width=0.325\textwidth]{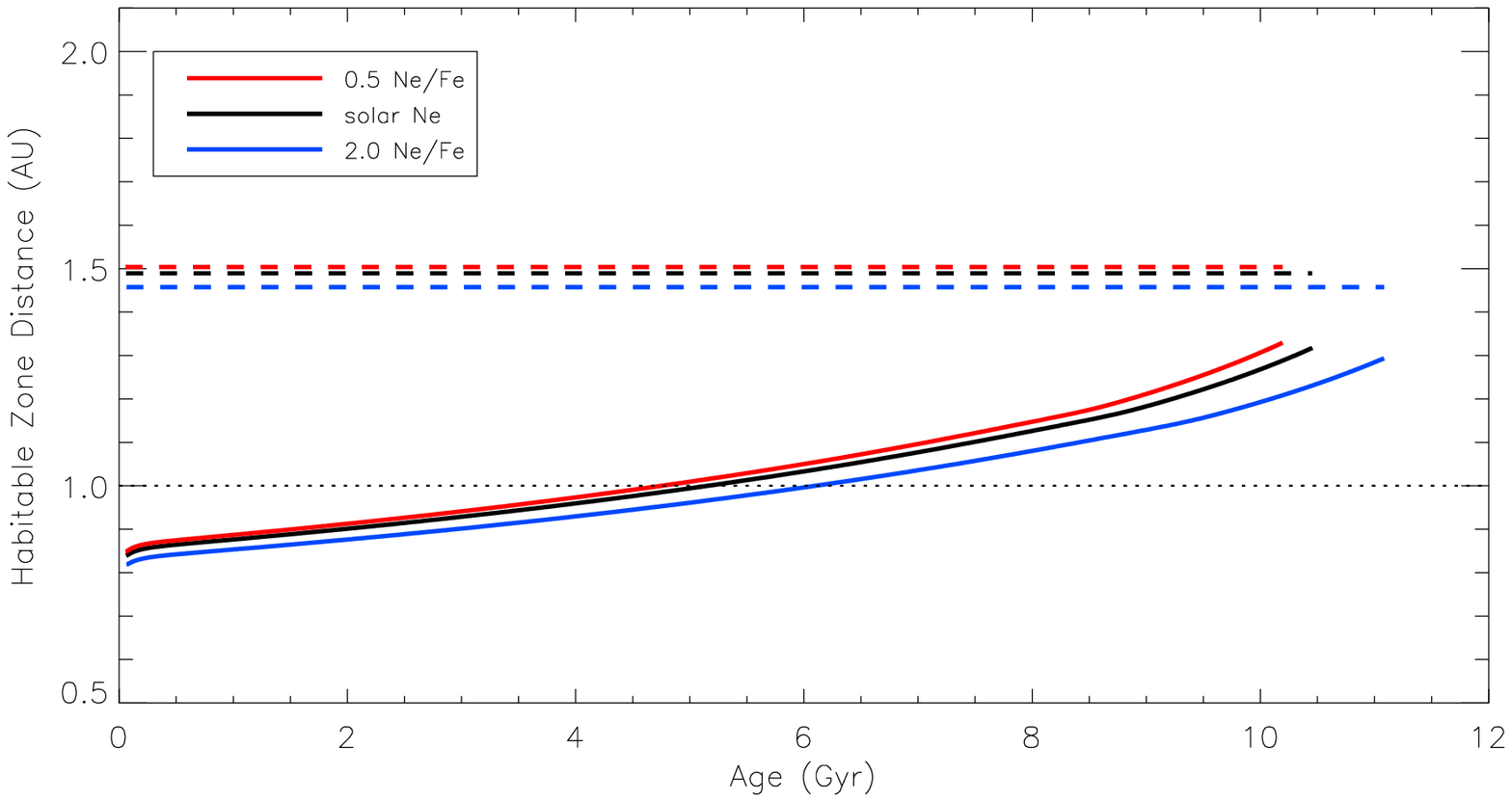}
\includegraphics[width=0.325\textwidth]{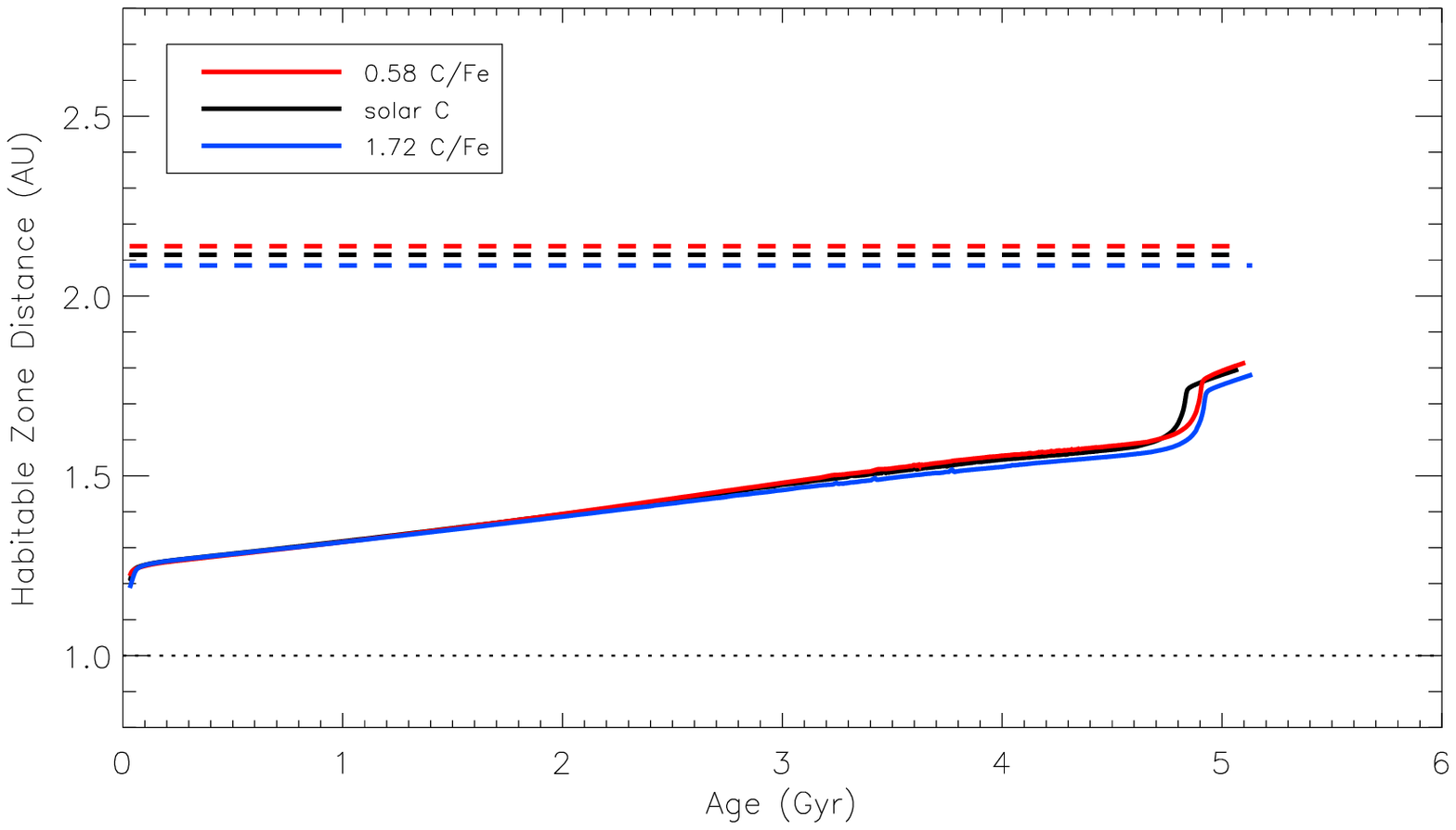}
\includegraphics[width=0.325\textwidth]{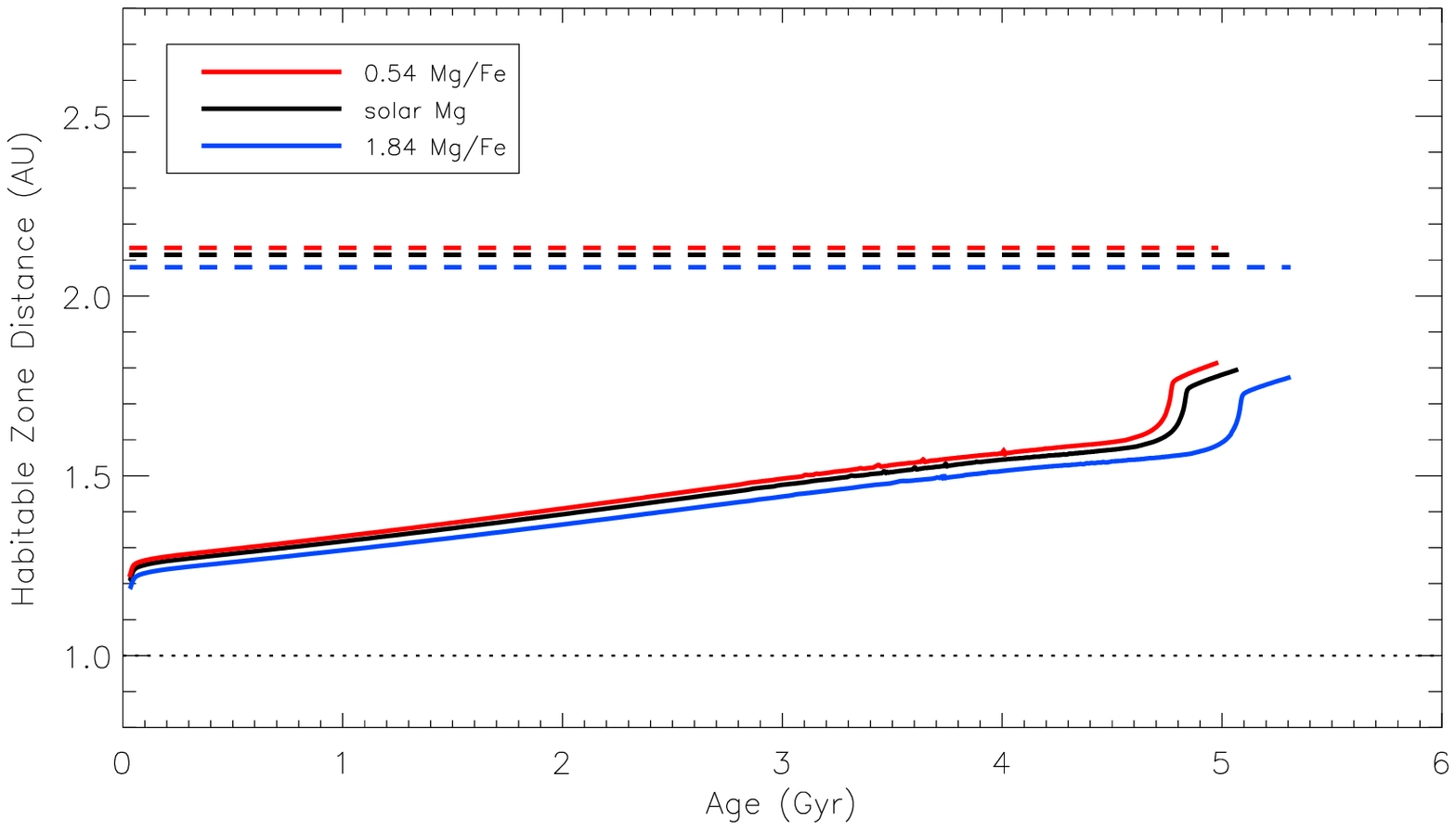}
\includegraphics[width=0.325\textwidth]{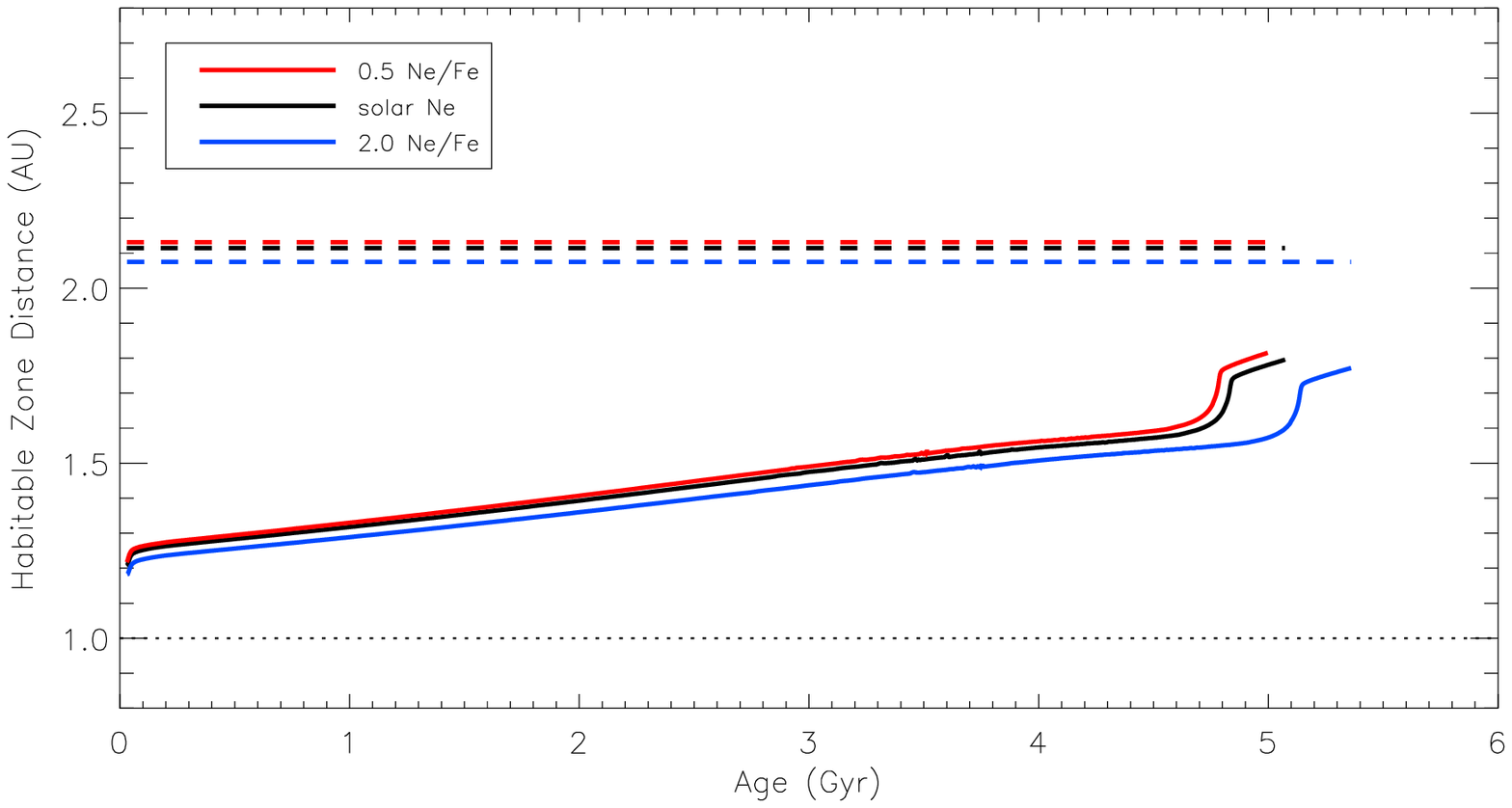}
\caption{Inner (solid) and outer (dashed) edges of the HZ for 0.5 M\sol (top), 1 M\sol (middle), and 1.2 M\sol (bottom), for three elements: carbon (left), magnesium (middle), and neon (right). Each line represents a different abundance value of interest: black is solar, red is for depleted elemental values (0.58 C/Fe, 0.54 Mg/Fe, 0.5 Ne/Fe), and blue is for enriched elemental values (1.72 C/Fe, 1.84 Mg/Fe, 2.0 Ne/Fe). A 1 AU orbit is also indicated by the light dotted line in each frame, for reference. It is clear that abundance variations within a star significantly affect MS lifetime and HZ distance. The inner radius is Runaway Greenhouse and the outer edge is the Maximum Greenhouse, at ZAMS value.\label{hzcold_all}}
\end{figure}

Figure~\ref{hzcold_all} offers an alternative scenario to that of Figure~\ref{hzcases_all}, wherein we presumed a cold start would be possible and thus allowed the outer HZ limit to expand with time. Instead, we now treat the outer boundary of the HZ at ZAMS as a hard limit that does not co-evolve with the star, so a particular planet would be required to exist in the HZ from the beginning of the host star's MS in order to be considered habitable over the long-term. Obviously, a planet that is in a star's HZ from very early times would not face the problem of a cold start. We defer discussion of the cold start problem as applied to the CHZ to our previous paper.

\section{CONCLUSIONS}

As we have discussed at length here and in T15, the stellar evolution depends strongly on elemental composition. It is highly important to distinguish between the metallicity of a star as measured by [Fe/H] and the specific abundances of individual elements. Though we have discussed the concept of overall scaled ``metallicity'' as the abundance of all heavy elements scaled relative to solar value, the term is often used interchangeably with [Fe/H], which actually only indicates the amount of iron relative to the Sun. The typical approach to stellar modeling is to simply measure the iron in a star and assume that every individual element scales in the same proportions (relative to iron) as observed in the Sun, though the abundances in actual stars can vary significantly.

We have also provided additional evidence that the HZ distance can be substantially affected even when only abundance ratios are changed. Evaluating habitability potential by modeling the co-evolution of stars and HZs requires models that span a range of variation in abundance ratios, as well as total scaled metallicity. For the same reason, characterizing a system requires measurements of multiple elemental abundances, not just [Fe/H].

In this paper, we discussed the new models we have created for inclusion in our catalog of stellar evolution profiles. We have now considered variation in the abundance ratios for carbon (from 0.58 - 1.72 C/Fe\sol), magnesium (from 0.54 - 1.84 Mg/Fe\sol), and neon (0.5 - 2.0 Ne/Fe\sol) and have investigated how each of these elements affects the co-evolution of stars and habitable zones. Though carbon is the most abundant of the three, we actually find that magnesium provides the largest contribution to stellar opacity and thus exhibits the largest effects in terms of MS lifetime, $L$, and $T_{eff}$. For this set of elements the effects on lifetimes and luminosities are smaller than typical observational uncertainties. This situation, however, will change dramatically with the release of Gaia data beginning in late 2016. High precision distances will allow for a much more accurate determination of luminosities, which is the largest source of error in stellar age determinations. Gaia will acquire distance measurements of our nearest stellar neighbors to an accuracy of 0.001\% and will provide parallaxes and proper motions with accuracy ranging from 10 to 1000 microarcseconds for over one billion stars. For an unreddened K giant at 6 kpc, it will provide a distance measurement accurate to 15\% and the transverse velocity to an accuracy of about 1 km/s. Even stars near the center of the galaxy (approximately 30,000 light-years away) will have distance measurements to within an accuracy of 20\% (e.g. \citet{bj2008}). For nearby stars, the luminosity uncertainty attributable to distance error will be of order 0.3\%, and the dominant source of error will be bolometric corrections. The effect of C abundance ratios will still likely be undiscernable, but Mg should be taken into account. Ne would also produce a detectable change, but since it is in most cases unobservable, its effects should be included in uncertainty estimates for quantities derived from stellar models.

Since many targets of radial velocity planet searches have high quality spectra that can be used to determine fairly precise stellar abundances, it should be standard practice to compare stellar models with more accurate compositions as long as such models and abundance measurements exist. To this end, we have updated the online database\footnote{http://bahamut.sese.asu.edu/$\sim$payoung/AST\_522/Evolutionary\_Tracks\_Database.html} of stellar evolution models and predicted HZs to include the 528 new models discussed in this paper. The library will be extended in the future to include a comprehensive grid for very low mass M-dwarf stars and evolved stars as well.

{\bf Acknowledgements} The results reported herein benefitted from collaborations and/or information exchange within NASA's Nexus for Exoplanet System Science (NExSS) research coordination network sponsored by NASA's Science Mission Directorate.

\clearpage

\clearpage

\clearpage

\clearpage

\clearpage

\clearpage

\clearpage

{}

\end{document}